\documentclass[bibyear]{aa} 
\usepackage{natbib}

\usepackage{xcolor}

\usepackage[pdfpagelabels=false]{hyperref}
\hypersetup{colorlinks = true, linkcolor = blue, citecolor = blue, filecolor = blue, urlcolor = blue}

\usepackage{graphicx}
\usepackage[labelfont=bf]{caption}
\usepackage{array}
\newcolumntype{H}{>{\setbox0=\hbox\bgroup}c<{\egroup}@{}}

\usepackage{txfonts}
\usepackage{lipsum}
\usepackage{subcaption}         
                                
\usepackage{lscape}            
\usepackage{placeins}  
\usepackage{longtable}

\begin{document}

   \title{Unexpected Plateau: The Constant $ \rm \log \epsilon (Th/Eu)$ Ratio of Actinide-Boost Stars}

   \author{J. Shejeelammal\inst{1,2}\corrauth{shejeelammal.jameela@unifi.it}        
        \and {\'A}sa Sk{\'u}lad{\'o}ttir\inst{1,2}\email{asa.skuladottir@unifi.it}
        \and Marta Molero\inst{3,4}\email{}
        \and Heitor Ernandes\inst{5,6,7}\email{}
        \and Jan Kuske\inst{3}\email{}
        \and Almudena Arcones\inst{3,8,9}\email{}
        \and Raghubar Singh \inst{1,2}
        }

   \institute{Dipartimento di Fisica e Astronomia, Universit{\'a} degli Studi di Firenze, Via G. Sansone 1, I-50019 Sesto Fiorentino, Italy
   \and INAF/Osservatorio Astrofisico di Arcetri, Largo E. Fermi 5, I-50125 Firenze, Italy
   \and Institut für Kernphysik, Technische Universität Darmstadt, Schlossgartenstr. 2, Darmstadt 64289, Germany
   \and Konkoly Observatory, Research Centre for Astronomy and Earth Sciences, HUN-REN, Konkoly Thege Miklös üt 15-17, Budapest H-1121, Hungary
   \and Lund Observatory, Department of Earth and Environmental Sciences, S\"olvegatan 12, SE-223 62 Lund, Sweden
   \and Lund Observatory, Department of Geology, S\"olvegatan 12, SE-223 62 Lund, Sweden
   \and Nicolaus Copernicus Astronomical Center, Polish Academy of Sciences, ul. Bartycka 18, 00-716 Warsaw, Poland
   \and GSI Helmholtzzentrum für Schwerionenforschung GmbH, Planckstr. 1, Darmstadt 64291, Germany
   \and Max-Planck-Institut für Kernphysik, Saupfercheckweg 1, 69117 Heidelberg, Germany}

   \date{Received October XX, 2026}

\abstract{
   The astrophysical site of the rapid neutron-capture process ($r$-process) is still one of the major open questions in astronomy. A key mystery is the origin of the peculiar actinide-to-lanthanide ratio found in the so called actinide-boost stars, with $ \rm \log \epsilon (Th/Eu)>-0.35$.
   Here we present the largest sample of Th measurements in stars formed in an external galaxy, Gaia Sausage Enceladus. From this sample, we present the three most metal-rich actinide-boost stars discovered to date, $\rm -1.6<[Fe/H]<-1.1$. In addition, we measured Th in a comparison sample of Milky Way stars formed in situ, and compiled Th measurements from the literature which do not account for non-local thermodynamic equilibrium (NLTE) effects. After applying NLTE corrections to the combined sample, a bimodality in the $\rm\log\epsilon (Th/Eu)$ ratio is revealed. Unexpectedly, actinide-boost stars lie on a very narrow $(\sigma=0.02)$ plateau of $\rm \log\epsilon(Th/Eu)=-0.2$, clearly separated from actinide-normal stars. Most (if not all) of the actinide-boost stars in our sample originate from outside the Milky Way.
   Chemical evolution models show that such a plateau can arise, even at [Fe/H] = $-$1, when a single actinide-rich event dominates the r-process content of a small or gas-depleted system.
   Combined with other evidence, our results suggest that mergers with neutron stars are the most promising astrophysical site for actinide-boost stars.}

\keywords{Nuclear reactions, nucleosynthesis, abundances -- stars: abundances -- Galaxies: interactions -- Galaxy: evolution -- Galaxy: halo -- Galaxy: abundances}

\maketitle
\nolinenumbers

\section{Introduction} \label{section_introduction}

The astrophysical site of the rapid neutron-capture process ($r$-process) is one of the most persistent open questions in astronomy \citep[e.g.][]{Schatz2022}. It has long been established that elements heavier than Zn ($Z>30$) are primarily created through a series of neutron-captures, and consequent $\beta$-decays \citep{B2FH}. Around half of the abundances of these heavy elements come from the $r$-process, yet its origin remains uncertain. 

Several sites have been proposed to have the physical conditions for sufficiently high neutron-flux. The favoured sites for the $r$-process have long been mergers including neutron stars \citep[e.g.][]{Lattimer1974}, and the detailed observations of GW170817 showed that a merger of two neutron stars (NSM) can indeed host the $r$-process \citep[e.g.][]{Pian2017,Watson2019}. In addition, sites associated with massive stars have also been proposed. These include rare classes of core-collapse supernovae (ccSN), such as magneto-rotational supernovae (mrSN) or collapsars \citep{Nishimura2017,Mosta2018,Siegel2019,Reichert2021, Reichert2023}; and magnetar giant flares \citep{Cehula2024,Patel2025a,Patel2025b}. Currently, none of these prompt $r$-process sites are confirmed observationally, making  their role in the creation of $r$-process elements very uncertain.

A remarkable property of the $r$-process is that relative abundances from the second peak
through the lanthanides ($56 \leq Z \leq 71$) appear to be universal, from the most metal-poor stars \citep[e.g.][]{Sneden2003} to the Sun \citep[e.g.][]{Burris2000}. However, this robustness breaks down at the low-mass ($Z < 56$) and high-mass ($76 \leq Z \leq 92$) ends \citep[e.g.][]{Hill2002}.
In particular, the so-called actinide-boost stars show an enhancement of $ \rm \log \epsilon (Th/Eu) > -0.35$ \citep[e.g.][]{Holmbeck2018,Lin2025}. The underlying cause of these high actinide abundances is still unknown, but might suggest a different nucleosynthetic
site (e.g.~NSM vs. mrSN) and/or different physical conditions \citep[e.g.][]{Eichler2019, Holmbeck_2019, Wanajo2024}.

One of the most promising ways of gaining insights into the $r$-process is through Galactic Archaeology. The chemical evolution of the $r$-process elements can be followed by observations of individual stars in the Milky Way and its satellite systems. These studies suggest that at least two $r$-process sites are needed to explain the data: a delayed site (typically associated with NSM) and a prompt source, whose origin is still debated \citep{cote_2019,skuladottir_2020, Molero_2023,Kobayashi_2023}.

Dynamical studies have associated $r$-rich stars with groups that are remnants of accreted dwarf galaxies \citep{Roederer_2018a, Yuan_2020, Gudin_2021, Hattori_2023}. These stars have also been found in several dwarf galaxies, ranging from ultra-faint dwarfs \citep[e.g.][]{Ji_2016,HansenT_2017} to large dwarf spheroidals such as Fornax \citep{Reichert_2021b,Lucchesi_2024}. Going towards even more massive galaxies, \cite{Ernandes_2024} found 10 r-II stars ([Eu/Fe] $>$ 1.0 and [Ba/Eu] $<$ 0; \citealt{Beers_2005}) that are kinematically part of the Gaia Sausage Enceladus (GSE), which has been dubbed the last major merger of the Milky Way, occurring $\sim10$\,Gyr ago \citep{Nissen_2010,Helmi_18}. Consistent with this picture, a cosmological simulation of a Milky Way analogue finds that most r-II stars originate from dwarf galaxies that are later accreted. The predicted accreted fraction is 100\% at [Fe/H] $= -3$ and 97\% at [Fe/H] $= -2$ and it decreases with metallicity \citep{Hirai_2022, Hirai_2025}. 

Among the $r$-process enhanced stars, about 30\% present actinide-boost, with an over-abundance of [Th/Eu] \citep{Holmbeck_2020}. \cite{Lin2025} reported the first confirmed actinide-boost star in GSE, with $\log\epsilon(\mathrm{Th}/\mathrm{Eu})=-0.22$, and found that a magnetorotationally driven jet-supernova model best reproduces its abundance pattern. Their kinematic analysis also indicates that approximately two-thirds of the actinide-boost stars they consider are ex situ. Together, these results suggest that low-mass accreted systems may provide conditions in which a rare actinide-boost $r$-process event can strongly enrich a poorly mixed gas reservoir.

In this paper, we measure the actinide-to-lanthanide ratio in 23 stars formed in GSE, and in 27 Milky Way stars formed in situ. 
Thus, we present the first systematic study of the actinide-to-lanthanide ratio in stars from an external galaxy.

\section{Stellar sample} \label{section_sample}
\subsection{Primary sample} \label{section_primary_sample}

The primary sample used in this study is a sub-sample of 50 stars selected from \cite{Nissen_2010, Nissen_2011} and \cite{Nissen_2024}, NS sample hereafter. 
The stellar sample consists of nine thick disk stars and 41 halo stars (V$\rm _{tot}$ $>$ 180 km s$^{-1}$) encompassing both high- and low-$\alpha$ populations.
The high-resolution (R $\geq$ 40,000) and high $S/N$ ($\geq$ 80 pix$^{-1}$ at 4000 \AA) optical spectroscopic data for 44 stars in the sample come from new observations via two ESO VLT/UVES programs (PI: Sk{\'u}lad{\'o}ttir, ESO IDs 109.22 and 110.240). For the remaining six stars, raw echelle spectra of comparable quality were adopted from the UVES archive\footnote{\url{https://archive.eso.org/cms.html}}. These spectra were acquired with the same UVES instrumental configurations as our observations, ensuring structural homogeneity across this dataset. 

All spectra were reduced using the standard UVES pipeline \citep{Freudling_2013}. 
Individual exposure frames were then Doppler-corrected, co-added to maximize $S/N$, and carefully 
continuum-normalized. The detailed procedure of data processing is discussed in our previous paper \citep{Ernandes_2025}. 
We have directly adopted the high-precision stellar atmospheric parameters and Mg abundances of the program stars 
from \cite{Ernandes_2025} that originally come from \cite{Nissen_2011} and \cite{Nissen_2024}. This sample consists of 
unevolved dwarf stars ($\rm \log\, g > 3.8$) with metallicities in the range $\rm -1.62 < [Fe/H] < -0.44$.

\subsection{Literature sample} \label{section_literature_sample}

To expand our statistical leverage across a wider metallicity range and test the broader universality of the thorium-to-europium ratio, we have compiled from the literature an additional sample of 60 stars (primarily giants with $\rm \log\, g \leq 2.9$ and metallicity down to $\rm[Fe/H]=-3.5$) in the Milky Way with reported high-resolution spectroscopic abundances for Eu and Th. Only stars with detections are selected and the ones with upper limits are excluded. We have also adopted the Mg abundances as representative of the $\alpha$-elements for this sample. To ensure consistency, we made sure that these stars were selected exclusively from studies employing line parameters and methodologies similar to our program star analysis, establishing a homogeneous 110-star baseline with a metallicity range $-$3.5 $\leq$ [Fe/H] $\leq$ $-$0.44.  

\section{Analysis} \label{section_analysis}

\subsection{Abundance determination} \label{section_abundance_determination}
In this work, we present new measurements of Eu and Th for the NS sample. 
The elemental abundances are computed with the \texttt{MOOG} code (\citealt{Sneden_1973, Sneden_2012}; version 2019), using 
ATLAS9 Kurucz grid of model atmospheres \citep{Castelli_2003}, assuming Local Thermodynamic Equilibrium (LTE) and 1D. 
The line parameters are taken from the \texttt{linemake}\footnote{\url{https://github.com/vmplacco/linemake}} atomic and molecular line database \citep{Placco_2021}. 

The europium abundances are derived from the spectral synthesis of the Eu II 4129.725 and 4205.042 {\rm \AA} lines. 
The hyper-fine structures (hfs) and isotopic components are included in the analysis, assuming
the solar isotopic ratio $^{151}$Eu : $^{153}$Eu = 47.8 : 52.2 \citep{Anders_1989}. 
 
The abundances of Th are derived from the spectral synthesis of the dominant Th~II 4019.129 {\rm \AA} feature. 
This is the only Th line that is strong enough to be detected in our spectra. This spectral feature belongs to a crowded region 
and is blended with overlapping features of other elements. To isolate the pure Th~II contribution, these blends were carefully identified and 
calibrated using the solar atlas and synthetic spectrum comparisons. 

\begin{figure*}[ht!]
\centering
\includegraphics[scale=0.4]{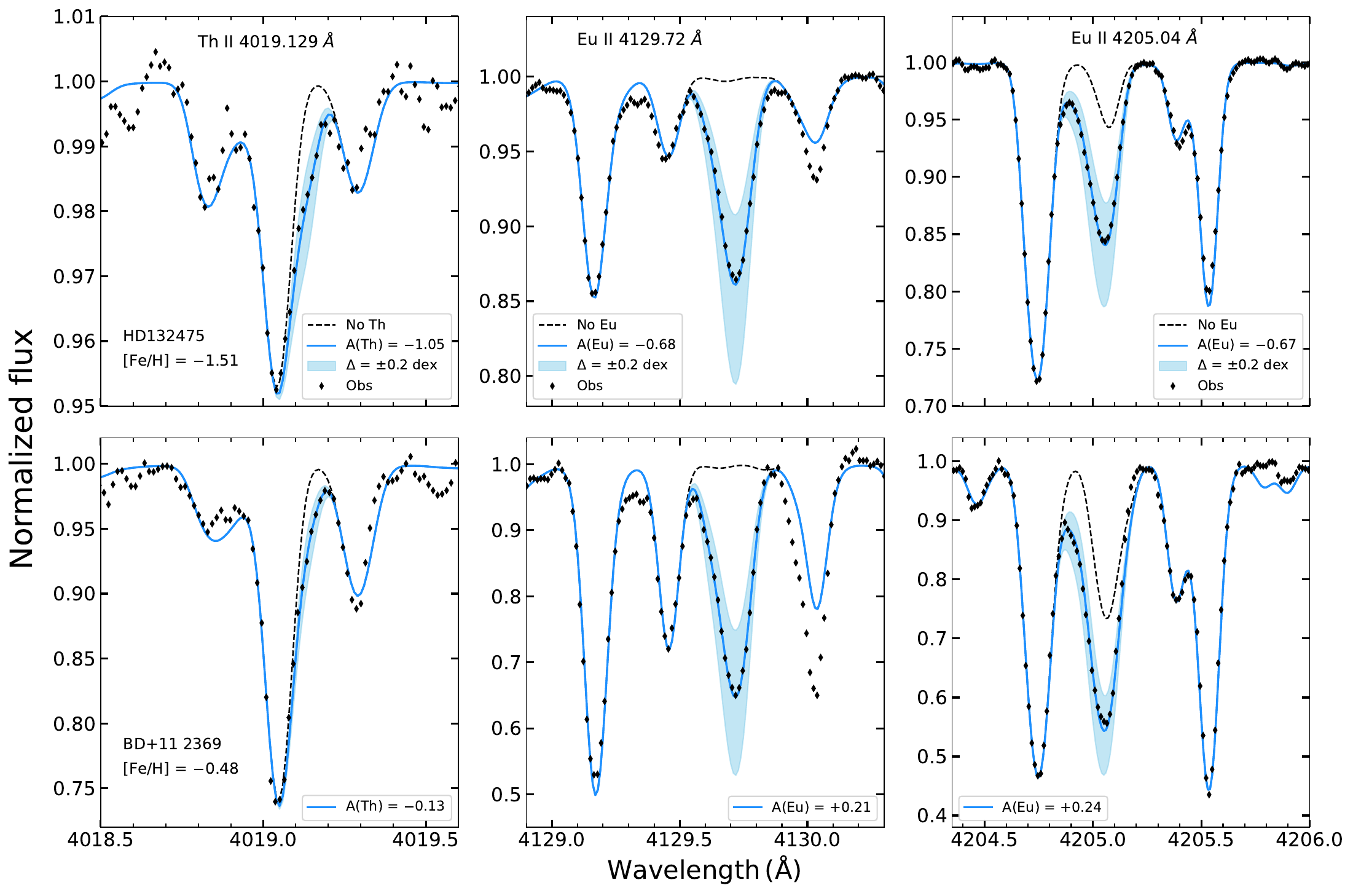}
\caption{Spectral synthesis fits and the derived abundances for Th and Eu for a metal-poor (HD 132475, $\rm[Fe/H]=-1.5$; upper panel) and a metal-rich (BD+11 2369, $\rm[Fe/H]=-0.5$; lower panel) star in our sample. Observed spectra are shown as black dots, along with the best fits (blue lines) and $\pm$0.2 dex deviations (shaded region). Dashed lines show the absence of the element of interest.} \label{figure_spectral_synth}
\end{figure*}

Abundance ratios are calculated for both program sample and the literature sample by adopting the solar abundance values from \cite{Asplund_2021}. 
The spectral synthesis fits for the Eu and Th lines of interest for a metal-poor and a metal-rich star in our primary sample are 
shown in Fig.~\ref{figure_spectral_synth}. 
The atmospheric parameters and abundances of the primary and literature sample are given in Tables~\ref{Table_results_abundances} and \ref{Table_literature_sample}, respectively.

\begin{figure*}[ht!]
\centering
\includegraphics[scale=0.6]{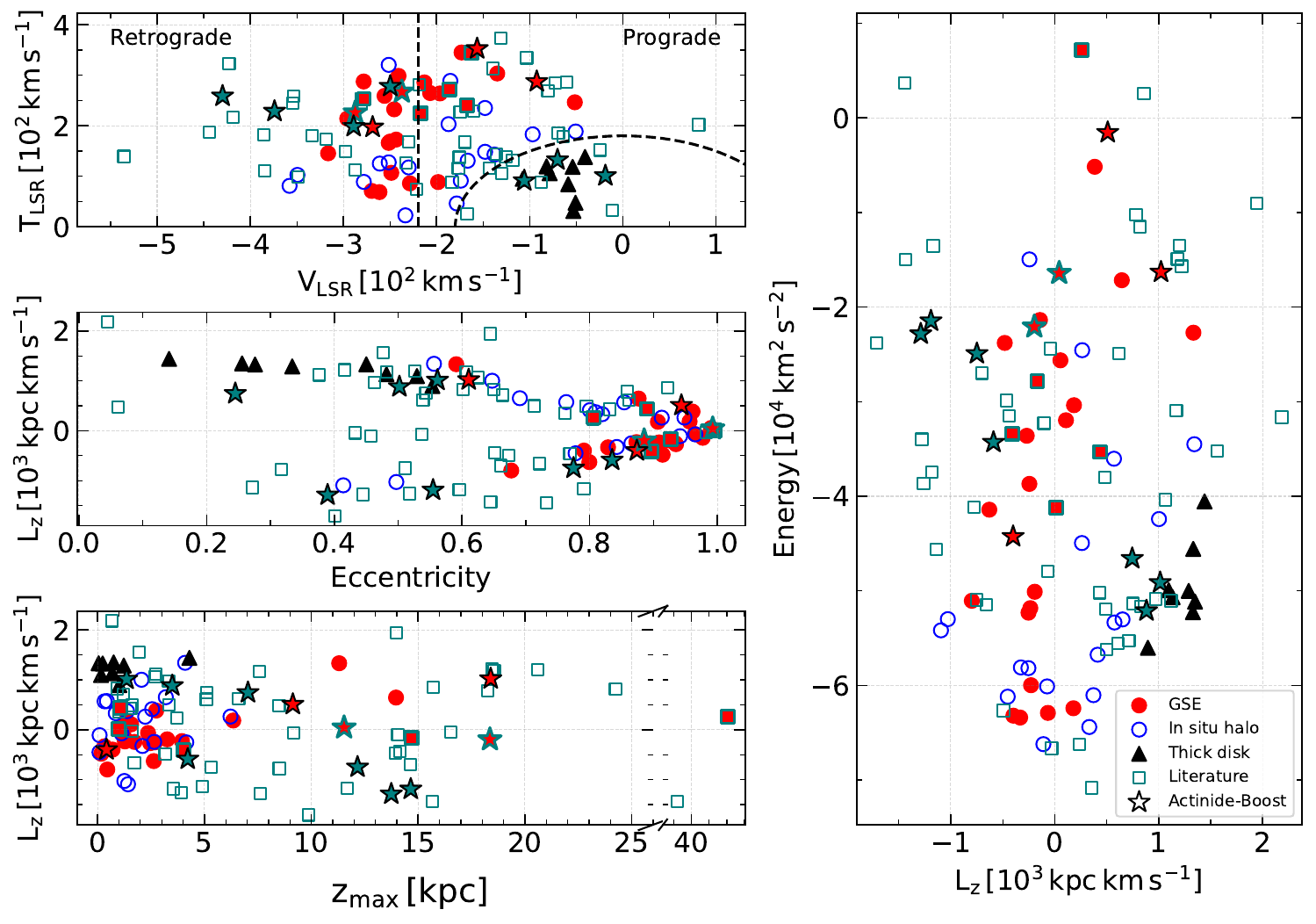} 
\caption{Orbital properties of the primary sample and literature sample in different dynamical spaces. Top left panel: Toomre diagram showing the Galactic velocity components, where the dashed curve separates the disk and the halo regions. Middle left panel: Angular momentum $\rm L_{z}$ as a function of orbital eccentricity. Bottom left panel: Angular momentum as a function of the maximum height above the Galactic plane $\rm Z_{max}$. Right panel: Orbital energy as a function of angular momentum. Red circles show the GSE stars, blue open circles in-situ halo stars, black triangles thick disk stars, teal open squares denote the literature sample, and star symbols are the actinide-boost stars.}
\label{figure_kinematics}
\end{figure*}

\subsection{Uncertainty in the abundances} \label{section_uncertainty}
The typical uncertainties in the adopted atmospheric parameters are 
$\Delta$ (T$_{\rm eff}$, $\rm \log\, g$, v$\rm_{mic}$, [Fe/H]) $\approx$ $\pm$(50 K, 0.10 dex, 0.10 km s$^{-1}$, 0.05 dex) \citep{Nissen_2010}. Given the high quality of our spectra, these errors are the dominant source of the abundance uncertainty. 
The deviations in the elemental abundances from the best-fit values are estimated by changing 
the atmospheric parameters individually by their respective uncertainties. The total uncertainty in the 
derived abundances is then estimated by adding these contributions in quadrature as explained in \cite{Shejeelammal_2020}, 
resulting in a typical uncertainty of $\sim$0.05 dex in $\rm \log \epsilon(X)$ for both Eu and Th. An uncertainty of $\sim$0.04 dex in [Mg/Fe]
is adopted from \cite{Nissen_2010}.

\subsection{Non-LTE corrections}  \label{section_NLTE}
Our combined data set spans two distinct evolutionary stages - dwarfs in our program sample 
and giants in the literature sample - and most of them have metallicities [Fe/H] $<$ $-$ 0.8 
($\approx$ 90\%). Applying NLTE correction ($\Delta$$_{\rm NLTE}$) across our combined sample of 110 
metal-poor stars is essential to eliminate evolutionary offsets and to ensure abundance trends reflect true physical yield variations. Even though several 1D/3D NLTE studies are available 
for various Eu II lines (e.g. \citealt{Mashonkina_2000, Storm_2024, Guo_2025}), \cite{Mashonkina_2012} 
is the only NLTE study (1D) available for Th II. Hence, for consistency, 1D NLTE corrections are 
employed in this study, as described below.

\textbf{Eu II:} The NLTE corrections for the Eu II 4129.725 and 4205.042\,\AA\ lines in our primary
NS sample are adopted from \cite{Storm_2025}. Both lines have identical corrections. Since the 
literature sample comes from multiple high-resolution spectroscopic studies, the reported mean LTE 
abundances were derived from various Eu II lines (predominantly, Eu II 3907.107, 4129.725, 4205.042, 
6437.640, and 6645.064 {\rm \AA}).
To calculate the Eu II NLTE 
corrections of individual stars in this literature sample, we have used the online tool \texttt{NLiTE} \citep{Koutsouridou_2025}\footnote{\url{https://nlite.pythonanywhere.com/}}, optimized for metal-poor 
stars using state-of-the-art publicly available NLTE grids, in this case from \cite{Mashonkina_2000}. The $\Delta$$_{\rm NLTE}$(Eu) of the combined 
sample ranges from +0.06 to +0.27 dex (Tables~\ref{Table_results_abundances} - \ref{Table_literature_sample}). 

\textbf{Th II:} The NLTE corrections for all 110 stars are adopted from \cite{Mashonkina_2012}. All literature studies are based on LTE abundances of Th. Even though $\Delta$$_{\rm NLTE}$ 
is negligible ($\sim$ 0.01 dex) for dwarfs, it becomes significant at lower surface gravities for a 
given temperature, reaching up to +0.21\,dex (Th II 4019.129\,{\rm \AA}) for the $\rm T_{\rm eff}/\log g/[Fe/H]=4000\,K/0.0/-2$ 
model \citep{Mashonkina_2012} (Tables~\ref{Table_results_abundances} - \ref{Table_literature_sample}). 

Depending on the stellar parameters, especially $\rm \log g$ and [Fe/H], the $\Delta$$_{\rm NLTE}$ has 
varying impact on the thorium-to-europium ratio. For instance, extremely metal-poor (EMP) evolved giants
show similar corrections for both Eu and Th, and hence the ratio remains almost unaffected 
(e.g.~HD 108577, 2MASS J20050670$-$3057445). Meanwhile the Eu correction is higher compared to Th 
for dwarfs (Table~\ref{Table_results_abundances}) and less-evolved EMP giants (e.g. HD 10837), which leads to 
non-negligible corrections of the ratio. In the combined sample, $\Delta$$_{\rm NLTE}$(Th/Eu) ranges from 
$-$0.19 to +0.07 dex.

\section{Astrometric and Kinematic data} \label{section_kinematics}

In order to better trace the dynamical origin of our sample, we adopted Gaia astrometric parameters for the whole sample. For that we query the Gaia DR3 IDs using the \texttt{ASTROQUERY.SIMBAD} \citep{Ginsburg_2019} Python package where we got 108 stars that have parallaxes and radial velocity from Gaia Radial Velocity Spectrometer (RVS). Based on the Gaia astrometric data from Gaia DR3 \citep{Gaia_2016b, Gaia_2023} and the zero-point corrected parallaxes, using the prescription by \cite{Lindegren_2021}, we calculated the kinematics and orbital parameters (Table \ref{Table_kinematics}) using the Python packages \texttt{ASTROPY} \citep{astropy13, astropy18, astropy22} and \texttt{GALPY}\footnote{\url{https://github.com/jobovy/galpy}}
\citep{Bovy_2015}.

The kinematic of both the primary and literature sample are shown in Fig.~\ref{figure_kinematics}. We followed the classification of \cite{Ernandes_2025} 
to separate the halo stars in the NS sample into in-situ (blue) and GSE (red) populations. 
For the homogeneity of the literature sample, here the GSE stars are selected based on the high-purity kinematic 
selection criteria given by \cite{Feuillet_2021}.
GSE stars are shown in red throughout this paper, unless stated otherwise.

As shown in Fig.~\ref{figure_kinematics}, all actinide-boost stars in the GSE sample belong to the high-energy class defined by \cite{Skuladottir_2025}, with $\rm E>-4.5\times 10^{4}$\,km$^{2}$\,s$^{-2}$. According to their interpretation, these stars are therefore more likely to have formed in the outer regions of GSE, rather than the more efficiently star-forming inner regions.

Most of the actinide-boost stars in the literature have been associated to an accreted system \citep[e.g.][]{Gudin_2021}. More generally, at the lowest metallicities ($\rm [Fe/H]<-2.5$) the large majority of stars in the Milky Way halo is predicted to be of accreted origin \citep[e.g.][]{Koutsouridou_2026,Hirai_2025}. We note that three actinide-boost (r-II) stars, 2MASS J14244532-2542471 ([Fe/H] = $-$3.39), BPS CS 31078-018 ([Fe/H] = $-$2.85), and BPS CS 30306-132 ([Fe/H] = $-$2.42), belong to the 
disk region in the Toomre diagram, and are in general consistent with disk-like kinematics (see Fig.~\ref{figure_kinematics}). However, two of these stars, 
BPS CS 31078-018 and BPS CS 30306-132, have been identified as belonging to the dynamically tagged groups of \cite{Hattori_2023}, that are ideal candidates for remnants of disrupted $r$-process-enhanced dwarf galaxies that merged with the ancient Milky Way. The third star with disk-like kinematics, 2MASS J14244532-2542471, is an EMP star and therefore likely to have an ex-situ origin; however, the original paper was not able to associate it to any known structure, so it remains possible that this is an in-situ-formed star \citep{Placco_2021}. The other actinide-boost stars in the literature sample exhibit halo kinematics and several previous studies found them to be members of dynamically tagged groups that have likely accreted origin \citep[e.g.][]{Yuan_2020, Hattori_2023}.

\begin{figure}[ht!]
\centering
\includegraphics[width=\columnwidth]{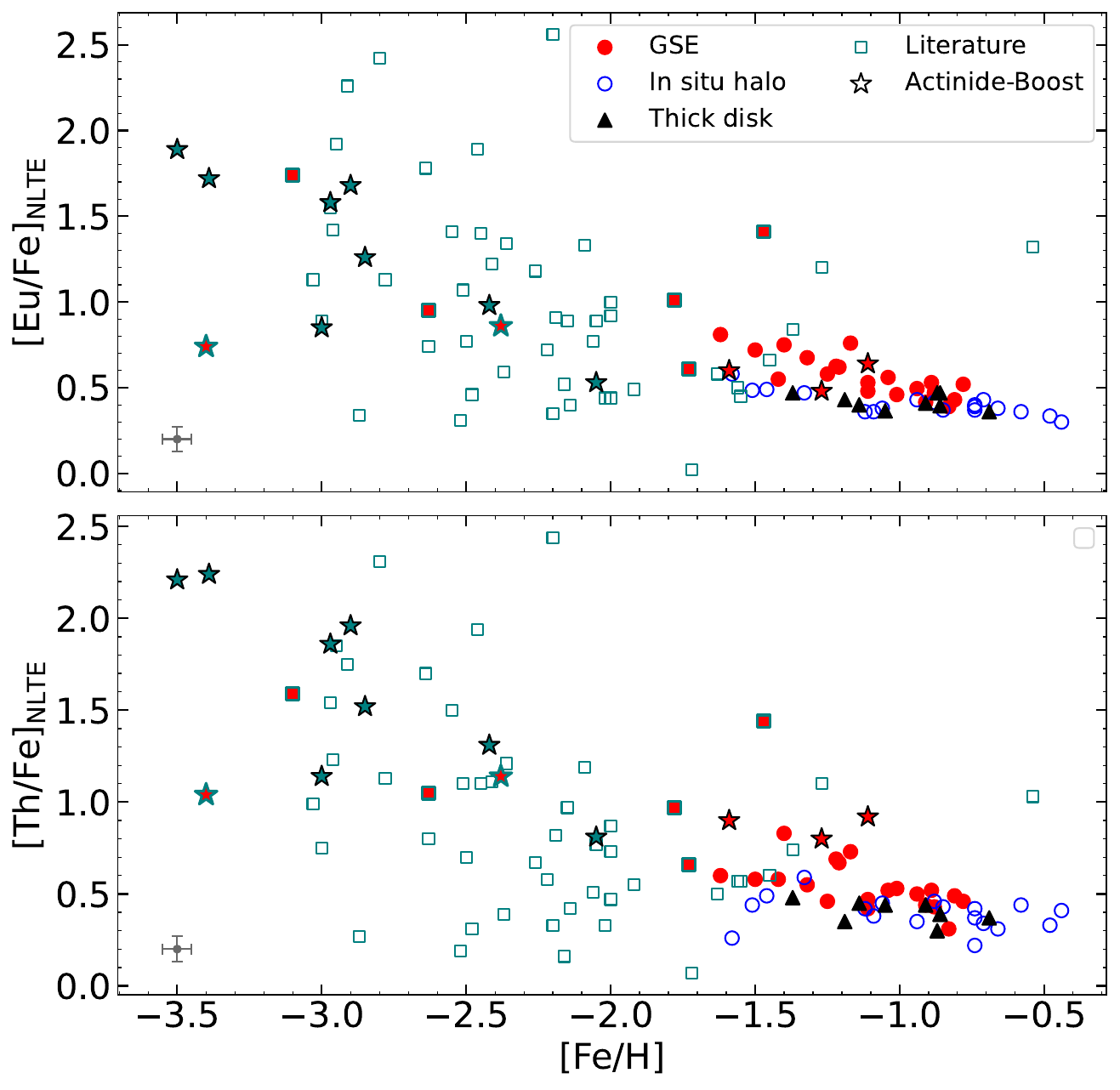}
\caption{Observed [Eu/Fe] and [Th/Fe] ratios of the combined sample with respect to [Fe/H]. Symbols as in Fig. \ref{figure_kinematics}.}\label{figure_Th_Eu}
\end{figure}

\begin{figure*}[ht!]
\centering
\includegraphics[scale=0.5]{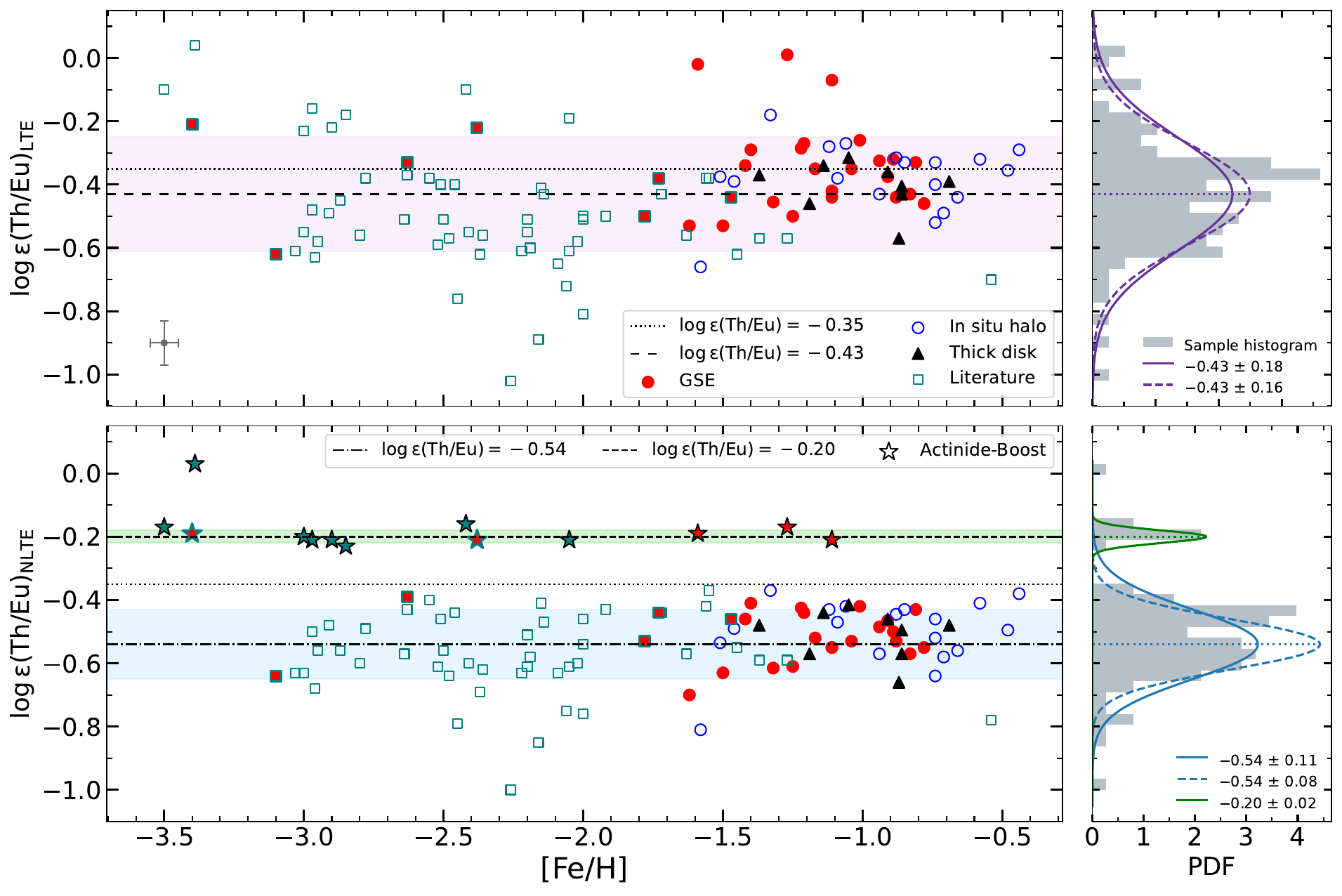}
\caption{Distribution of $ \rm \log \epsilon (Th/Eu)$ with respect to [Fe/H], assuming LTE (top) and NLTE (bottom). Dotted lines at $ \rm \log \epsilon (Th/Eu)$ = $-$0.35 represent the lower boundary of actinide-boost region. 
Probability distribution functions are shown on the right side panel, both the observed (solid) and intrinsic (dashed lines).
The mean values are shown as horizontal lines (see labels), along with the $\rm 1\sigma$ as shaded regions in the main panels.
} \label{figure_bimodality_GMM}
\end{figure*} 

\section{Abundances of Eu and Th} \label{section_abundance_results}
 
We identified three actinide-boost ($ \rm \log \epsilon (Th/Eu)$ $>$ $-$0.35) stars in our NS sample: G 112$-$43, HD 284248, 
and G 87$-$13 with $ \rm \log \epsilon (Th/Eu)$$_{\rm NLTE}$ values $-$0.17, $-$0.19, and $-$0.21\,dex, respectively. These are the most metal-rich ([Fe/H] $>$ $-$1.6) actinide-boost stars discovered to date, and the only main-sequence stars identified 
to belong to this class. One of these actinide-boost stars, HD\,284248, has available spectra from the Hubble Space Telescope in the ultra-violet. The analysis of this spectra shows that this actinide-boost star also has unusual abundances in other heavy elements ($Z>70$), but this will be discussed further in Singh et al. in prep.

\subsection{Abundance ratios of [Eu/Fe] and [Th/Fe]} \label{section_Eu_Th_abundances}
Fig.~\ref{figure_Th_Eu} shows the distribution of [Eu/Fe]$_{\rm NLTE}$ and [Th/Fe]$_{\rm NLTE}$ for our sample. At low metallicities, the scatter is very large with a spread of over $2$\,dex, while at higher metallicities the stars converge. This behavior of the n-capture elements is well known in the literature \citep[e.g.][]{Francois_2007,Cescutti_2015}. 

However, the GSE stars in our primary sample ($\rm [Fe/H]\gtrsim-1.5$, red circles) are on average higher both in [Eu/Fe]$_{\rm NLTE}$ (see also e.g.~\citealt{Fishlock_2017}) and [Th/Fe]$_{\rm NLTE}$ compared to the Milky Way in-situ halo and thick disk stars. In particular, we note that all three actinide-boost stars in our primary sample belong to GSE (3/23: $13^{+13}_{-7}\%$), while none are found among the Milky Way stars selected in the same way (0/27: $0^{+7}\%$). Based on the literature sample, the fraction at low-metallicities, $\rm -3<[Fe/H]<-2$, is 7/43: 16$^{+9}_{-6}$\% (uncertainties estimated using \citealt{Gehrels_1986}).

Finally we note that actinide-boost stars (star symbols in Fig.~\ref{figure_Th_Eu}), are not necessarily $r$-process-rich, that is high in [Eu/Fe], but instead follow a distribution very similar to actinide-normal stars.

\subsection{Bimodality in the $\rm \log \epsilon (Th/Eu)$ ratio} \label{section_bimodality}

The $ \rm \log \epsilon (Th/Eu)$ ratios of the sample in LTE and NLTE are shown 
in Fig.~\ref{figure_bimodality_GMM}.
When evaluated under LTE assumption, the ratios of the sample exhibit a broad, continuous distribution
with a mean value at $\rm \log \epsilon (Th/Eu) = -$0.43 (Fig.~\ref{figure_bimodality_GMM}, upper panel). 

However, when the NLTE corrections are applied, the picture changes substantially. Many of the stars previously categorized as actinide-boost stars are now revealed to be actinide-normal. This is especially true for the sample of metal-rich main-sequence stars, where 17 out of the 20 actinide boost-stars in LTE are actually actinide-normal in NLTE. On the other hand, only 1 star out of the metal-poor giants ($\rm[Fe/H]<-2$) changes from actinide-boost to actinide-normal. Furthermore, the difference seen in LTE between main sequence stars, $\langle\rm\log\epsilon(Th/Eu)\rangle_{\rm MS}=-0.36$, and giant stars, $\langle \rm\log\epsilon(Th/Eu)\rangle_{\rm giants}=-0.48$, disappears when the NLTE corrections are applied.

Thanks to the more accurate NLTE abundances and our three newly discovered actinide-boost stars at high metallicity ($\rm[Fe/H]>-1.6$), the combined sample is split into two distinct populations, separated by a clean observational gap (Fig.~\ref{figure_bimodality_GMM}, lower panel):

{\bf \noindent 1. Actinide-boost stars:} Of the thirteen stars in our sample, twelve belong to a very narrow plateau with a constant 
value of $ \rm \log \epsilon (Th/Eu)$ $\approx$ $-$0.20, and very low $\sigma\sim0.02$. The only star that deviates from this plateau is 2MASS J14244532-2542471 (J1424 hereafter; \citealt{Placco_2023}). 

{\bf \noindent 2. Actinide-normal stars:} Most of the stars (N = 97) in the sample belong to this 
broader population  centered at $ \rm \log \epsilon (Th/Eu)$ = $-$0.54 and $\sigma\sim0.11$.

{\bf \noindent 3. Observational gap:} The NLTE corrections lead to a clean observational gap 
between $\rm -0.35 < \log \epsilon (Th/Eu) < -0.25$, where no stars in our sample lie.

The plateau of actinide-boost stars and the observational gap persist throughout the wide metallicity range of the sample, $\rm -3.5<[Fe/H]<-1$.  
This indicates that the bimodality of the $\rm \log \epsilon (Th/Eu)$ ratio is a genuine and robust physical feature.

\begin{figure*}[ht!]
\centering
\includegraphics[scale=0.6]{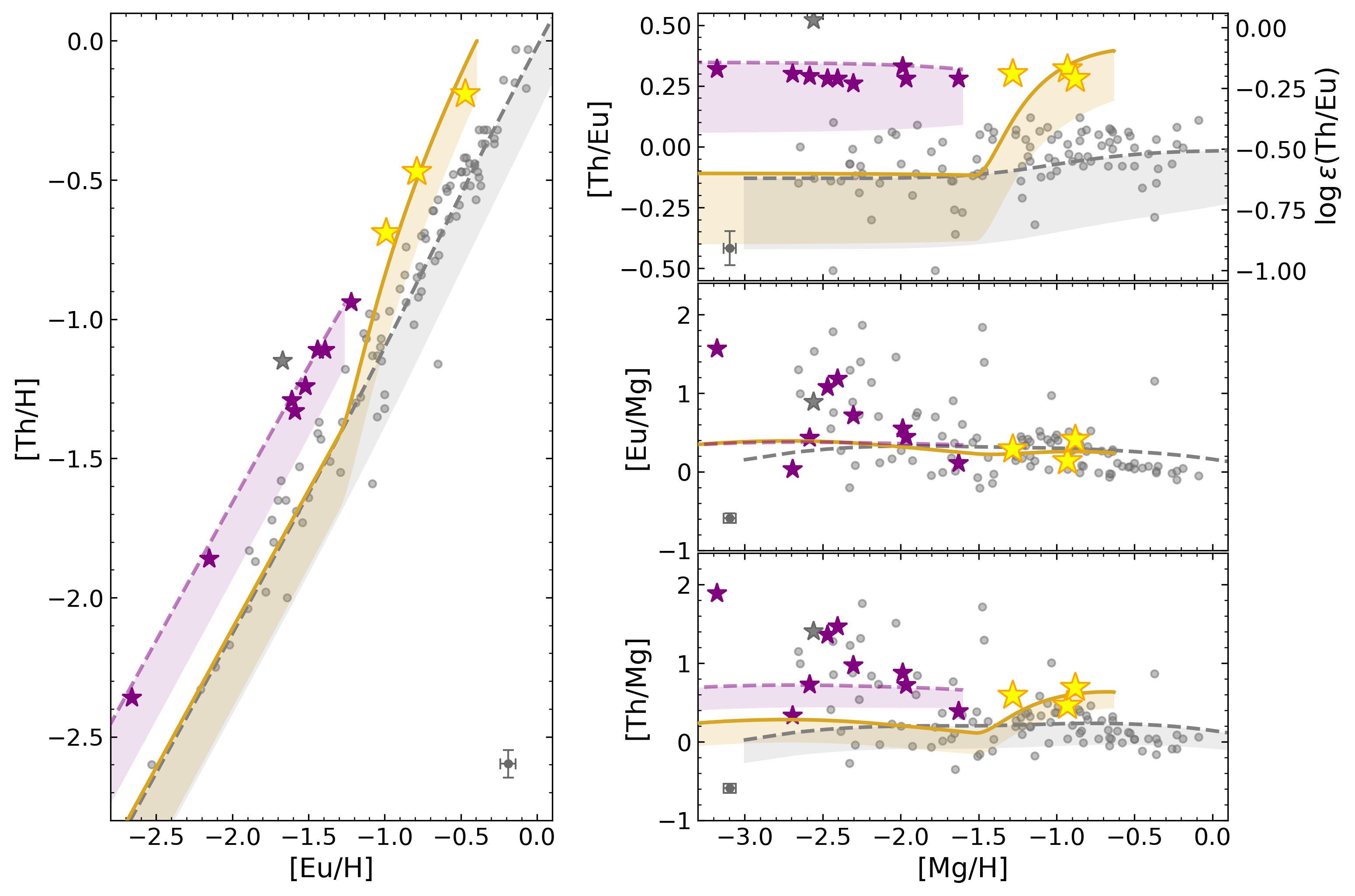}
\caption{Chemical evolution model of the formation of actinide-boost stars (star symbols). Actinide-normal stars are shown as gray dots.
At low metallicity ($\rm[Fe/H]<-1.5$, purple), such stars are formed in small gas clouds enriched by only one $r$-process event - a NSM boosted in actinides. Actinide-boost stars at $\rm[Fe/H]>-1.5$ (yellow, this work) are formed when a NSM occurs in regions that have experienced gas depletion. Lines show the model predictions at the time of star formation; shaded areas extend these down to the present-day values, accounting for the radioactive decay of Th in the stars.} 
\label{figure_model_fit}
\end{figure*}

\subsection{Statistical confirmation} \label{section_staistical_confirmation}

To verify this observed bimodality, we performed a two-stage statistical analysis, as described in detail in Appendix~\ref{section_statistical_analysis}\footnote{The star J1424 ($ \rm \log \epsilon (Th/Eu)_{\rm NLTE} > 0$) is treated as an outlier and is excluded from the following analysis.}. As a first step, a Kolmogorov–Smirnov (KS) test was conducted, which rejected the null hypothesis of unimodality ($p<0.05$). 
Gaussian Mixture Models (GMM) were used to further analyze the distributions, indicating that the two-component model 
for $ \rm \log \epsilon (Th/Eu)_{\rm NLTE}$ is optimal, while a unimodal distribution is favored for LTE. The NLTE analysis revealed two Gaussian distributions: a standard peak at $\mu_{1}$ = $-$0.54 ($\sigma_{1}$ = 0.11) comprising 89.1\% of the sample and an actinide-boost peak at $\rm \mu_{2}$ = $-$0.20 ($\rm \sigma_{2}$ = 0.02) comprising 10.9\%.
The LTE distribution is a single Gaussian centered at $\rm \mu$ = $-$0.43 with $\rm \sigma$ = 0.18 (see solid lines in the side panel of Fig.~\ref{figure_bimodality_GMM}).

A Bayesian Markov Chain Monte Carlo (MCMC) analysis provided intrinsic parameters for these peaks and confirmed a distinct observational gap, 
indicating a bimodal distribution. The intrinsic parameters for NLTE peaks based on the MCMC analysis are $\rm \mu_{1}$ = $-$0.54$\pm$0.01, $\rm \sigma_{int, 1}$ = 0.08$\pm$0.01, and $\rm \mu_{2}$ = $-$0.20$\pm$0.02, $\rm \sigma_{int, 2}$ = 0.00$\pm$0.02. The analysis verified a significant observational gap ($-$0.35 $<$ $ \rm \log \epsilon (Th/Eu)$$_{\rm NLTE}$ $<$ $-$0.25) between the peaks, with a confidence level of 94.6\% (1.9$\rm \sigma$).

The actinide-boost peak is therefore tightly
constrained with a narrow observational width ($\sim$0.02\,dex) and consistent with zero intrinsic width. 
The intrinsic mean and width of the LTE distribution are $\rm \mu$ = $-$0.43$\pm$0.02 and $\rm \sigma_{int}$ = 0.16$\pm$0.01 (Fig.~\ref{figure_bimodality_GMM}, side panel dashed lines).

This statistical analysis confirms that the $ \rm \log \epsilon (Th/Eu)$ distribution in metal-poor stars is intrinsically 
bimodal, pointing to distinct astrophysical sites for the $r$-process in the early Galactic chemical 
evolution.

\section{Origin of actinide-boost stars} \label{section_origin}
The clear separation in $ \rm \log \epsilon (Th/Eu)$$_{\rm NLTE}$ (Fig.~\ref{figure_bimodality_GMM}) favors the interpretation that actinide-boost and actinide-normal stars are enriched by two different $r$-process sites (see also Sec.~\ref{sect:discussion}), rather than by a single site with continuously varying physical conditions. Furthermore, the very narrow range of the plateau at $ \rm \log \epsilon (Th/Eu) \approx-0.20$ indicates that these stars are tracing the intrinsic yields of this site, rather than a mixture of several sites. Two key questions therefore arise: {\it 1)~What is the site that produces the actinide-boost stars? 2)~What kind of chemical enrichment scenario can explain their formation over such a large metallicity range?}

Regarding the first question, theoretical yields predict that NSMs produce higher actinide-to-lanthanide ratios \citep[e.g.][]{Wanajo2014, Holmbeck2019, Ricigliano_2024}, a result that also holds for BH--NS mergers \citep{Wanajo2024}, than other proposed $r$-process sites. These include mrSNe \citep[e.g.][]{Nishimura2017, Reichert2021, Reichert2023}, collapsars
\citep[e.g.][but see \citealt{Siegel2019}]{Miller2020, Just2022, Fujibayashi2023}, and magnetar giant flares \citep{Patel2025b}.

Regarding the second question, the large scatter of [Eu/Fe] at $\rm[Fe/H]<-2$ (e.g.~Fig.~\ref{figure_Th_Eu}) points to the $r$-process being a rare event. 
At very low metallicity, it is therefore plausible that actinide-boost stars formed in an environment that had experienced only one $r$-process event, with boosted actinide production. At higher metallicity, these stars are more challenging to explain. It is hard to imagine an environment being enriched to $\rm[Fe/H]\gtrsim-1.5$ by only one $r$-process event. Indeed, no stars at this metallicity have been identified as being very $r$-process poor. The most likely scenario is thus that, although the birth environment of these stars experienced several $r$-process events, their $r$-process abundances are {\it dominated} by a single event with high actinide production. In the following, we explore this scenario, assuming that the dominant event is a NSM.

\subsection{Chemical evolution model} \label{section_models_chemical_evolution}

Fig.~\ref{figure_model_fit} shows the results of our single-zone chemical evolution models in the [Th/Eu], [Eu/Mg], and [Th/Mg] vs [Mg/H] planes, and in the [Th/H] vs [Eu/H] plane. The models represent average trends and do not include any intrinsic scatter. In all models, the radioactive decay of $\rm ^{232}Th$ is included in the evolution of the gas. The decay that takes place in the stars after their formation is shown separately as shaded areas in Fig. \ref{figure_model_fit}. Following the theoretical predictions mentioned above, we assume that NSM produce the high actinide-to-lanthanide ratio of the plateau. We note, however, that any nucleosynthetic site with the same yields would produce similar results. The [Th/Eu] ratio depends only on the yields of the source that domiates the $r$-process budget. This degeneracy makes the plateau relatively easy to fit, and motivates some caution when interpreting the nature of the $r$-process source. 

The [Eu/Mg] and [Th/Mg] ratios, by contrast, compare an $r$-process element to the $\alpha$-element budget produced by ccSNe, and therefore also depend on the absolute $r$-process mass ejected relative to the gas it enriches. A single $r$-process event diluted in a large gas reservoir leaves a too weak imprint and the model will then underproduce the [Eu/Mg] and [Th/Mg] ratios of the actinide-boost stars, even while matching their [Th/Eu]. Reproducing all three ratios simultaneously thus requires a sufficiently high $r$-process mass per unit gas mass, that is, either a large number of events or a small gas reservoir.

\subsubsection{General chemical enrichment}
\label{section: general chemical enrichment}
The actinide-normal stars are reproduced here by adopting the chemical evolution model of GSE from \cite{Vincenzo_2019} as a reference (grey dashed lines in Fig. \ref{figure_model_fit}). In this model, the galaxy forms at high redshift by infall of gas from the intergalactic medium into its potential well, at a rate 
$\rm I(t) \propto exp(-t/\tau_{\rm inf})$, 
with an infall timescale $\rm \tau_{\rm inf}$ = 0.24 Gyr, normalized so that the total accreted gas mass is 10$\rm ^{10}$ M$\rm_{\odot}$. Stars form following 
a Schmidt–Kennicutt law, $\rm \Psi(t) \propto \nu M_{\rm gas}^{1.5}$, with a star-formation efficiency $\rm \nu = 0.42\ \rm Gyr^{-1}$. The galaxy suffers galactic winds, which develop when the thermal energy of the gas, heated by stellar winds and SNe, exceeds its binding energy (see \citealt{Bradamante_1998}); \cite{Vincenzo_2019} found the efficiency of these outflows to be low.

For the $r$-process, the model includes two sources with different characteristic timescales: a delayed source, NSMs, described by a delay-time distribution (adopted from \citealt{Simonetti_2019}) with yields 
from \citet{Ricigliano_2024} (their model $m^*_S$), and a prompt source, here represented by collapsars with nucleosynthesis prescriptions from \cite{Siegel2019}. Mg is produced by ccSNe, following \cite{Francois_2004}. This model reproduces the average trends of actinide-normal stars in all panels of Fig. \ref{figure_model_fit} but, as expected, underestimates the Th abundances of the actinide-boost stars.

\subsubsection{Actinide-boost enrichment}

Although NSMs are predicted to produce a higher [Th/Eu] ratio than prompt sources such as collapsars or mrSNe (see Sec. \ref{section_origin}), they do not dominate the $r$-process budget in the general model: when NSMs are parametrised as a delayed source with a relatively low rate, the budget is instead governed by the prompt contribution from rare massive stars (see e.g. \citealt{Molero_2023} for details). For this reason, we model the plateau enrichment as a single NSM event, ejecting 0.1\,M$_\odot$ of $r$-process material. This is at the upper end of current predictions, which typically range from $\sim 10^{-3}$ to a few $\times 10^{-2}\ \rm M_\odot$ (e.g., \citealp{Goriely2011, Rosswog2014, Ricigliano_2024}), with the more massive configuration approaching values comparable to the one adopted here.
We model this condition in two complementary ways, corresponding to the two plateau sub-populations.

The low-metallicity plateau stars (Fig.~\ref{figure_model_fit}, purple) are modeled as forming in a small environment (with a total accreted gas mass $\propto 10^{5}\ \rm M_{\odot}$), whose $r$-process enrichment comes from a single, early NSM that is the only $r$-process source. Because the gas reservoir is small, the NSM leaves a strong imprint without the gas having to be depleted first (see below). The resulting track lies on the plateau, matching the most metal-poor plateau stars in [Th/Eu], and is in reasonable agreement with the observations in the other abundance planes,  especially considering that low Eu and Th abundances are not detectable at this metallicity.

The model for the more metal-rich GSE plateau stars (Fig.~\ref{figure_model_fit}, gold) instead includes two $r$-process sources. A prompt source (e.g., collapsars), active throughout the evolution, enriches the gas as in the general model (Sec. \ref{section: general chemical enrichment}), and a single, delayed NSM is superimposed on this evolution. The host environment, a region within GSE, has a total accreted gas mass of $10^{7}\ \rm M_{\odot}$, assembled on a short timescale ($\rm \tau_{inf} = 0.5\ \rm Gyr$), forming stars with a low efficiency 
($\rm \nu$ = 0.04 Gyr$^{-1}$) and subject to a strong galactic wind with a mass-loading factor of $\omega \simeq10$.

With this gas mass, a single NSM would on its own be too diluted to leave a clear imprint. However, the strong wind progressively removes the gas, so that by the time the merger occurs the reservoir is sufficiently depleted for the actinide-rich imprint to become visible. This timing results from the combination of a short infall timescale and the delay of the NSM: the star formation rate peaks early, and the single NSM occurs about 1\,Gyr later, by the time the galactic wind has already reduced the gas reservoir by more than half. The NSM then enriches the residual gas at [Mg/H] $\approx$ $-$1.5, lifting [Th/Eu] to the plateau value and reproducing the actinide-boost stars, identified by this work. At the same time, the star formation is still high enough to form a new stellar generation. The NSM therefore occurs in a time window in which the gas reservoir is sufficiently depleted for its imprint to be visible, but still gas-rich enough to form the observed population of metal-rich actinide-boost stars.

Finally, we note that although our model reproduces the observed trends, our adopted production value of $\rm \log \epsilon (Th/Eu)$ for the actinide-boost $r$-process seems to be too low compared to the observations. However, the uncertainty of the theoretical value is very large, and the results presented here can serve as an important observational constraint, see Sec.~\ref{sec:abplateau}.

\section{Discussion} \label{sect:discussion}

\subsection{The actinide-boost plateau} \label{sec:abplateau}
As shown in Fig.~\ref{figure_bimodality_GMM}, actinide-boost stars lie on a remarkably narrow plateau of $\rm \log \epsilon (Th/Eu) = -0.2$ over a large range in metallicity. This discovery was enabled by our new detection of the three most metal-rich actinide-boost stars known to date ($\rm[Fe/H]>-1.6$), and the application of NLTE corrections to our primary and literature samples. This is an indication that the $ \rm \log \epsilon (Th/Eu)$ ratio of events with boosted actinide production must be extremely robust, since only 1/13 actinide-boost stars falls out of this plateau.

The narrowness of the plateau is very puzzling, as the scatter around the mean value, $\sigma=0.02$, is similar to, or even smaller than what is expected from the measurement errors, especially when considering that the literature sample is not homogeneously analyzed, and we have not taken into account any possible errors of the NLTE corrections, nor the variation in stellar ages. 

The most metal-poor stars ($\rm[Fe/H]\approx-3$) are expected to have ages of $\sim$13\,Gyr \citep[e.g.][]{Koutsouridou_2025}. Since the GSE merger is estimated to have happened around 10\,Gyr ago \citep[e.g.][]{Helmi_18,Feuillet_2021}, this sets a lower limit of about $\sim10\,\rm Gyr$ on the ages of the most metal-rich actinide-boost stars, which are known to belong to GSE \citep[e.g.][]{Ernandes_2025}. Recalling that $ \rm \log \epsilon (Th/Eu)$ ratio is a cosmo-chronometer\footnote{$ t = 46.67 \rm \times [\log \epsilon (Th/Eu)_{ini} - \log \epsilon (Th/Eu)_{obs}$] Gyr \citep{Hill2002}}, the $ \rm \log \epsilon (Th/Eu)$ should change by an amount 0.108\,dex if the age difference between stars is 5\,Gyr, 0.065\,dex in the case of 3\,Gyr, and 0.043\,dex in the case of 2\,Gyr. Based on our measurements, we can thus put an upper limit of an age difference between the oldest and youngest actinide-boost stars in our sample of $<$3\,Gyr, as this corresponds to the span of $ \rm \log \epsilon (Th/Eu)$ values in our sample. 
This is in agreement with our expectations of the stars belonging to the age range $\sim$10-13\,Gyr. 

If we assume an average age of $11.5\pm1.0$ Gyr for actinide-boost stars in our sample, the production value of the actinide-boost $r$-process is $\rm\log\epsilon(Th/Eu)=+0.046\pm0.021$. 
With larger samples and robust measurements and NLTE calculations, there is large potential for accurate differential ages of different populations of actinide-boost stars.

Finally, we note that the actinide-boost stars identified so far are all older than $\sim$10 Gyr irrespective of their metallicities. Observational scarcity of young actinide-boost stars is likely caused by their signature being diluted and washed out by the homogenization in more evolved galactic environments. In other words, their window of visibility is restricted to the highly structured, poorly mixed epoch of the early galactic assembly. However, systematic searches of actinide-boost stars among more metal-rich younger stars have yet to be carried out.

\subsection{The actinide-boost source} \label{sec:absource}
Perhaps the most fundamental question is: {\it What does this plateau tell us about the origin of the actinide-boost $r$-process?} In Sec.~\ref{section_origin}, we explored one plausible scenario of the formation of such stars, but that does not necessarily mean that all other options can be excluded.

Firstly, the observational gap between actinide-boost and actinide-normal stars indicates that they are most likely formed from gas enriched by two separate sites. Alternatively, there would have to be a very sharp change in either physical parameters (such as $Y_e$) within a given site, or a sharp change in $ \rm \log \epsilon (Th/Eu)$ as a function of other parameters, but current models do not predict this \citep[e.g.][]{Eichler2019}. These new results indicate that $ \rm \log \epsilon (Th/Eu)$ does not show a continuous change from actinide-normal to actinide-boost stars, as generally assumed from LTE data \citep[e.g.][]{Holmbeck_2019}.

Based on several indirect lines of evidence, here we favor a scenario where NSM produce the actinide-boost stars, and an alternative primary site associated with massive stars (prompt) is responsible for actinide-normal stars. The key arguments are as follows: \\
{\bf 1. Yields:}  According to state-of-the-art models, NSMs are typically predicted to have higher Th/Eu ratios than other proposed $r$-process sites, such as mrSNe, collapsars, and magnetar giant flares \citep[e.g.][]{Wanajo2014, Nishimura2017, Holmbeck2019, Miller2020, Reichert2023, Ricigliano_2024, Patel2025b}.\\
\noindent {\bf 2. Galactic Archaeology:} Observations of the $r$-process in the Milky Way and its satellite galaxies can only be explained by adopting at least two sources, NSM (delayed) and a prompt source associated with massive stars \citep[e.g.][]{Skuladottir_2019,skuladottir_2020}. In the Milky Way, the dominating source should be prompt to explain the general evolution of Eu \citep[e.g.][]{cote_2019,Kobayashi_2020,Molero_2023}. Given the rarity of actinide-boost stars, it is likely that their source is associated with the less dominant site, i.e.~NSM. If NSM are not the site of actinide-boost stars, we would require at minimum 3 different sources to explain the abundance patterns in stars: NSM, an actinide-normal prompt source (main contributor of the Milky Way evolution), and an actinide-boost prompt source.\\
\noindent {\bf 3. Less efficient star formation:} Most (if not all) of the known actinide-boost stars in the Milky Way are accreted from smaller galaxies. It is well known that delayed sources are particularly prominent in such systems, as seen by strong effects of SNIa and AGB stars, due to a combination of less efficient star formation and them being prone to lose their gas \citep[e.g.][]{Tolstoy_2009,skuladottir_2020}. Thus, this result points to a delayed source of the actinide-boosted $r$-process, i.e. NSM. \\

Finally, it is worth noting that the extreme narrowness of the plateau can be used to constrain the physical conditions of any suggested actinide-boost site. However, the outlier J1424 \citep{Placco_2023} seems to indicate that although a fixed intrinsic value is the most common for actinide-boost events, it might not be universal. An outlier such as this could stem from, e.g. a black hole-neutron star merger \citep{Wanajo2024}, or a NSM with unusual conditions (e.g. in relation to the mass involved). But to draw firm conclusions about the different sources, a population study of actinide-boost stars is needed, especially in relation to their frequency in different formation environments and at different metallicity ranges.

\subsection{Future prospects} \label{sec:future}

To fully explore and understand the actinide-boost stars and which astrophysical sites are capable of reproducing the observations, we need to overcome both theoretical and observational limitations. 
Improved NLTE calculations for the heavy elements, covering a large range in stellar parameters, are needed to derive more reliable abundance ratios. Together with state-of-the-art 3D atmospheric models, this can test the robustness of the observed bimodality. In addition, accurate and robust theoretical yields are needed for all proposed $r$-production sites.

On the observational side, the sample sizes are limited in all environments, especially in dwarf galaxy populations, where only a few measurements are available \citep[e.g.][]{Aoki_2007,HansenC_2018,Ji_2018}, and high-metallicity halo stars, where measurements become exceedingly challenging. The available data are expected to improve drastically with upcoming spectroscopic surveys like WEAVE \citep{dalton2016final} and 4MOST - especially the 4MOST dedicated surveys for dwarf galaxies \citep{skuladottir20234most} and the high-resolution halo survey \citep{christlieb20194most}.

These surveys, combined with high-resolution, high-quality follow-up observations using blue sensitive spectrographs like VLT-CUBES \citep{Evans_2018, ernandes2020stellar}, and the next generation of high-resolution instruments on the extremely large telescopes such as ELT-ANDES \citep{marconi2022andes}, and in the 8m-class telescopes such as HRMOS \citep{magrini2026hrmos}, will enable us to access the Th lines for fainter targets, significantly expanding the sample of stars.

\section{Conclusions} \label{section_conclusion}

From the historical sample of \cite{Nissen_2010}, we have measured high-precision abundances of Th and Eu in 50 stars. This contains Th measurements for 23 stars in Gaia Sausage Enceladus (GSE), making it the largest sample of [Th/Eu] measurements in any external galaxy to date (Sec.~\ref{section_analysis}). Among GSE stars, we identify three actinide-boost stars, which are the most metal-rich actinide-boost stars discovered to date ($\rm [Fe/H]>-1.6$). Meanwhile, no actinide-boost star was among our in-situ Milky Way sample of 27 stars (Sec.~\ref{section_Eu_Th_abundances}). In fact, most (if not all) actinide-boost stars in our sample originate from outside the Milky Way (Sec.~\ref{section_kinematics}). 

After applying NLTE corrections to our sample and stars from the literature, a bimodal distribution emerges: actinide-boost stars lie on a narrow plateau with $\rm \log \epsilon (Th/Eu)=-0.2$, clearly separated from actinide-normal stars at $\rm \log \epsilon (Th/Eu)<-0.35$. Out of 13 actinide-boost stars in our samples, 12 fall on this plateau, which has a constant value over a large range in metallicity, $\rm -3.5\lesssim[Fe/H]\lesssim-1$ (Sec.~\ref{section_bimodality}). 

The narrowness of the plateau ($\sigma=0.02$) allows us to estimate the age range of the sample as $<3$\,Gyr. Adopting an average age of $11.5\pm1.0$\,Gyr, we estimate the intrinsic value of the actinide-boosted $r$-process as $\rm\log\epsilon(Th/Eu)=+0.046\pm0.021$ (Sec.~\ref{sec:abplateau}). This provides very strong constraints on theoretical yield predictions for any $r$-process site with boosted actinide production. Our results indicate that $\rm \log \epsilon (Th/Eu)$ is extremely promising as a cosmic clock for actinide-boost stars, which seem to have a robust production ratio. For actinide-normal stars, however, the distribution should be broadened by different mixing ratios of multiple r-process events of the two types (actinide-boost, and actinide-normal).

Our chemical evolution models (Sec.~\ref{section_models_chemical_evolution}) show that the formation of actinide-boost stars, at both low and high metallicity, can be explained in a scenario with two r-process sources: 1)~Neutron star mergers (NSM), which are delayed and have enhanced [Th/Eu] ratios; 2)~a prompt source with lower [Th/Eu] ratios. In this scenario, a flat plateau arises whenever a single NSM dominates the r-process content of the gas, either in a small system or in a gas-depleted region of a larger galaxy such as GSE.  The absolute level of the plateau is instead set by the $\rm \log\epsilon (Th/Eu)$ production ratio of the actinide-boost $r$-process source, which currently remains theoretically uncertain and can potentially be better constrained by our observations.
Our model is not sensitive to which prompt site is adopted, and we draw no conclusions on which is the most likely prompt source out of those proposed in the literature. For the production site of the actinide-boost stars, here we favour NSMs, as these are predicted to produce high actinides, and this is also supported by arguments based on Galactic chemical evolution (see Sec.~\ref{sec:absource}).

Dedicated population studies of large samples are still needed to better understand the origin of actinide-boost stars (Sec.~\ref{sec:future}). Here we have demonstrated that accurate NLTE abundances are essential to provide the full picture, as LTE abundances give a very skewed view - overestimating their fraction and hiding the bimodality of [Th/Eu] abundances. Finally, our results show that even though our understanding has improved significantly in recent years, the $r$-process still offers unexpected surprises.

\begin{acknowledgements}
We thank the anonymous referee for the feedback. This project has received funding from the European Research Council (ERC) under the European Union’s Horizon 2020 research and innovation programme (grant agreement No. 101117455).
H.E. were supported by a project grant from the Knut and
Alice Wallenberg Foundation (KAW 2020.0061 Galactic Time
Machine, PI Feltzing). H.E. is supported by foundations managed by Kungliga Vetenskapsakademien (reference AST2024-0042) and supported by The Kungl. Fysiografiska Sällskapet i Lund (reference 20240013). This project was supported by funds from the Crafoord Foundation (reference 20230890).
AA, MM and JK acknowledge support by the Deutsche Forschungsgemeinschaft
(DFG, German Research Foundation) – Project-ID 279384907 – SFB 1245.
Based on data collected using VLT/UVES under the programs: PI: \'A. Skúladóttir, 109.22 and 110.240 with 
the additional data from ESO Science Archive Facility from 8.B-0475, 076.B-0133, 090.B-0605, 109.23, 71.B0529, and 68.D-0094.
\end{acknowledgements}

\bibliography{bibliography}

\begin{thebibliography}{131}
\expandafter\ifx\csname natexlab\endcsname\relax\def\natexlab#1{#1}\fi

\bibitem[{{Anders} \& {Grevesse}(1989)}]{Anders_1989}
{Anders}, E. \& {Grevesse}, N. 1989, \gca, 53, 197

\bibitem[{{Aoki} {et~al.}(2007){Aoki}, {Honda}, {Sadakane}, \&
  {Arimoto}}]{Aoki_2007}
{Aoki}, W., {Honda}, S., {Sadakane}, K., \& {Arimoto}, N. 2007, \pasj, 59, L15

\bibitem[{{Asplund} {et~al.}(2021){Asplund}, {Amarsi}, \&
  {Grevesse}}]{Asplund_2021}
{Asplund}, M., {Amarsi}, A.~M., \& {Grevesse}, N. 2021, \aap, 653, A141

\bibitem[{{Astropy Collaboration} {et~al.}(2022){Astropy Collaboration},
  {Price-Whelan}, {Lim}, {Earl}, {Starkman}, {Bradley}, {Shupe}, {Patil},
  {Corrales}, {Brasseur}, {N{"o}the}, {Donath}, {Tollerud}, {Morris},
  {Ginsburg}, {Vaher}, {Weaver}, {Tocknell}, {Jamieson}, {van Kerkwijk},
  {Robitaille}, {Merry}, {Bachetti}, {G{"u}nther}, {Aldcroft},
  {Alvarado-Montes}, {Archibald}, {B{'o}di}, {Bapat}, {Barentsen}, {Baz{'a}n},
  {Biswas}, {Boquien}, {Burke}, {Cara}, {Cara}, {Conroy}, {Conseil}, {Craig},
  {Cross}, {Cruz}, {D'Eugenio}, {Dencheva}, {Devillepoix}, {Dietrich},
  {Eigenbrot}, {Erben}, {Ferreira}, {Foreman-Mackey}, {Fox}, {Freij}, {Garg},
  {Geda}, {Glattly}, {Gondhalekar}, {Gordon}, {Grant}, {Greenfield}, {Groener},
  {Guest}, {Gurovich}, {Handberg}, {Hart}, {Hatfield-Dodds}, {Homeier},
  {Hosseinzadeh}, {Jenness}, {Jones}, {Joseph}, {Kalmbach}, {Karamehmetoglu},
  {Ka{l}uszy{'n}ski}, {Kelley}, {Kern}, {Kerzendorf}, {Koch}, {Kulumani},
  {Lee}, {Ly}, {Ma}, {MacBride}, {Maljaars}, {Muna}, {Murphy}, {Norman},
  {O'Steen}, {Oman}, {Pacifici}, {Pascual}, {Pascual-Granado}, {Patil},
  {Perren}, {Pickering}, {Rastogi}, {Roulston}, {Ryan}, {Rykoff}, {Sabater},
  {Sakurikar}, {Salgado}, {Sanghi}, {Saunders}, {Savchenko}, {Schwardt},
  {Seifert-Eckert}, {Shih}, {Jain}, {Shukla}, {Sick}, {Simpson},
  {Singanamalla}, {Singer}, {Singhal}, {Sinha}, {Sip{H{o}}cz}, {Spitler},
  {Stansby}, {Streicher}, {{{S}}umak}, {Swinbank}, {Taranu}, {Tewary},
  {Tremblay}, {Val-Borro}, {Van Kooten}, {Vasovi{'c}}, {Verma}, {de Miranda
  Cardoso}, {Williams}, {Wilson}, {Winkel}, {Wood-Vasey}, {Xue}, {Yoachim},
  {Zhang}, {Zonca}, \& {Astropy Project Contributors}}]{astropy22}
{Astropy Collaboration}, {Price-Whelan}, A.~M., {Lim}, P.~L., {et~al.} 2022,
  \apj, 935, 167

\bibitem[{{Astropy Collaboration} {et~al.}(2018){Astropy Collaboration},
  {Price-Whelan}, {Sip{\H{o}}cz}, {G{\"u}nther}, {Lim}, {Crawford}, {Conseil},
  {Shupe}, {Craig}, {Dencheva}, {Ginsburg}, {Vand erPlas}, {Bradley},
  {P{\'e}rez-Su{\'a}rez}, {de Val-Borro}, {Aldcroft}, {Cruz}, {Robitaille},
  {Tollerud}, {Ardelean}, {Babej}, {Bach}, {Bachetti}, {Bakanov}, {Bamford},
  {Barentsen}, {Barmby}, {Baumbach}, {Berry}, {Biscani}, {Boquien}, {Bostroem},
  {Bouma}, {Brammer}, {Bray}, {Breytenbach}, {Buddelmeijer}, {Burke},
  {Calderone}, {Cano Rodr{\'\i}guez}, {Cara}, {Cardoso}, {Cheedella}, {Copin},
  {Corrales}, {Crichton}, {D'Avella}, {Deil}, {Depagne}, {Dietrich}, {Donath},
  {Droettboom}, {Earl}, {Erben}, {Fabbro}, {Ferreira}, {Finethy}, {Fox},
  {Garrison}, {Gibbons}, {Goldstein}, {Gommers}, {Greco}, {Greenfield},
  {Groener}, {Grollier}, {Hagen}, {Hirst}, {Homeier}, {Horton}, {Hosseinzadeh},
  {Hu}, {Hunkeler}, {Ivezi{\'c}}, {Jain}, {Jenness}, {Kanarek}, {Kendrew},
  {Kern}, {Kerzendorf}, {Khvalko}, {King}, {Kirkby}, {Kulkarni}, {Kumar},
  {Lee}, {Lenz}, {Littlefair}, {Ma}, {Macleod}, {Mastropietro}, {McCully},
  {Montagnac}, {Morris}, {Mueller}, {Mumford}, {Muna}, {Murphy}, {Nelson},
  {Nguyen}, {Ninan}, {N{\"o}the}, {Ogaz}, {Oh}, {Parejko}, {Parley}, {Pascual},
  {Patil}, {Patil}, {Plunkett}, {Prochaska}, {Rastogi}, {Reddy Janga},
  {Sabater}, {Sakurikar}, {Seifert}, {Sherbert}, {Sherwood-Taylor}, {Shih},
  {Sick}, {Silbiger}, {Singanamalla}, {Singer}, {Sladen}, {Sooley},
  {Sornarajah}, {Streicher}, {Teuben}, {Thomas}, {Tremblay}, {Turner},
  {Terr{\'o}n}, {van Kerkwijk}, {de la Vega}, {Watkins}, {Weaver}, {Whitmore},
  {Woillez}, {Zabalza}, \& {Astropy Contributors}}]{astropy18}
{Astropy Collaboration}, {Price-Whelan}, A.~M., {Sip{\H{o}}cz}, B.~M., {et~al.}
  2018, \aj, 156, 123

\bibitem[{{Astropy Collaboration} {et~al.}(2013){Astropy Collaboration},
  {Robitaille}, {Tollerud}, {Greenfield}, {Droettboom}, {Bray}, {Aldcroft},
  {Davis}, {Ginsburg}, {Price-Whelan}, {Kerzendorf}, {Conley}, {Crighton},
  {Barbary}, {Muna}, {Ferguson}, {Grollier}, {Parikh}, {Nair}, {Unther},
  {Deil}, {Woillez}, {Conseil}, {Kramer}, {Turner}, {Singer}, {Fox}, {Weaver},
  {Zabalza}, {Edwards}, {Azalee Bostroem}, {Burke}, {Casey}, {Crawford},
  {Dencheva}, {Ely}, {Jenness}, {Labrie}, {Lim}, {Pierfederici}, {Pontzen},
  {Ptak}, {Refsdal}, {Servillat}, \& {Streicher}}]{astropy13}
{Astropy Collaboration}, {Robitaille}, T.~P., {Tollerud}, E.~J., {et~al.} 2013,
  \aap, 558, A33

\bibitem[{{Beers} \& {Christlieb}(2005)}]{Beers_2005}
{Beers}, T.~C. \& {Christlieb}, N. 2005, \araa, 43, 531

\bibitem[{{Bovy}(2015)}]{Bovy_2015}
{Bovy}, J. 2015, \apjs, 216, 29

\bibitem[{{Bradamante} {et~al.}(1998){Bradamante}, {Matteucci}, \&
  {D'Ercole}}]{Bradamante_1998}
{Bradamante}, F., {Matteucci}, F., \& {D'Ercole}, A. 1998, \aap, 337, 338

\bibitem[{{Burbidge} {et~al.}(1957){Burbidge}, {Burbidge}, {Fowler}, \&
  {Hoyle}}]{B2FH}
{Burbidge}, E.~M., {Burbidge}, G.~R., {Fowler}, W.~A., \& {Hoyle}, F. 1957,
  Reviews of Modern Physics, 29, 547

\bibitem[{{Burris} {et~al.}(2000){Burris}, {Pilachowski}, {Armandroff},
  {Sneden}, {Cowan}, \& {Roe}}]{Burris2000}
{Burris}, D.~L., {Pilachowski}, C.~A., {Armandroff}, T.~E., {et~al.} 2000,
  \apj, 544, 302

\bibitem[{{Cain} {et~al.}(2018){Cain}, {Frebel}, {Gull}, {Ji}, {Placco},
  {Beers}, {Mel{\'e}ndez}, {Ezzeddine}, {Casey}, {Hansen}, {Roederer}, \&
  {Sakari}}]{Cain_2018}
{Cain}, M., {Frebel}, A., {Gull}, M., {et~al.} 2018, \apj, 864, 43

\bibitem[{{Cain} {et~al.}(2020){Cain}, {Frebel}, {Ji}, {Placco}, {Ezzeddine},
  {Roederer}, {Hattori}, {Beers}, {Mel{\'e}ndez}, {Hansen}, \&
  {Sakari}}]{Cain_2020}
{Cain}, M., {Frebel}, A., {Ji}, A.~P., {et~al.} 2020, \apj, 898, 40

\bibitem[{{Castelli} \& {Kurucz}(2003)}]{Castelli_2003}
{Castelli}, F. \& {Kurucz}, R.~L. 2003, in Proceedings of the IAU Symp. No 210,
  Vol. 210, Modelling of Stellar Atmospheres, ed. N.~{Piskunov}, W.~W. {Weiss},
  \& D.~F. {Gray}, A20

\bibitem[{{Cehula} {et~al.}(2024){Cehula}, {Thompson}, \&
  {Metzger}}]{Cehula2024}
{Cehula}, J., {Thompson}, T.~A., \& {Metzger}, B.~D. 2024, \mnras, 528, 5323

\bibitem[{{Cescutti} {et~al.}(2015){Cescutti}, {Romano}, {Matteucci},
  {Chiappini}, \& {Hirschi}}]{Cescutti_2015}
{Cescutti}, G., {Romano}, D., {Matteucci}, F., {Chiappini}, C., \& {Hirschi},
  R. 2015, \aap, 577, A139

\bibitem[{Christlieb {et~al.}(2019)Christlieb, Battistini, Bonifacio, Caffau,
  Ludwig, Asplund, Barklem, Bergemann, Church, Feltzing,
  {et~al.}}]{christlieb20194most}
Christlieb, N., Battistini, C., Bonifacio, P., {et~al.} 2019, arXiv preprint
  arXiv:1903.02468

\bibitem[{{Christlieb} {et~al.}(2004){Christlieb}, {Beers}, {Barklem},
  {Bessell}, {Hill}, {Holmberg}, {Korn}, {Marsteller}, {Mashonkina}, {Qian},
  {Rossi}, {Wasserburg}, {Zickgraf}, {Kratz}, {Nordstr{\"o}m}, {Pfeiffer},
  {Rhee}, \& {Ryan}}]{Christlieb_2004}
{Christlieb}, N., {Beers}, T.~C., {Barklem}, P.~S., {et~al.} 2004, \aap, 428,
  1027

\bibitem[{{C{\^o}t{\'e}} {et~al.}(2019){C{\^o}t{\'e}}, {Eichler}, {Arcones},
  {Hansen}, {Simonetti}, {Frebel}, {Fryer}, {Pignatari}, {Reichert},
  {Belczynski}, \& {Matteucci}}]{cote_2019}
{C{\^o}t{\'e}}, B., {Eichler}, M., {Arcones}, A., {et~al.} 2019, \apj, 875, 106

\bibitem[{{Cowan} {et~al.}(2002){Cowan}, {Sneden}, {Burles}, {Ivans}, {Beers},
  {Truran}, {Lawler}, {Primas}, {Fuller}, {Pfeiffer}, \& {Kratz}}]{Cowan_2002}
{Cowan}, J.~J., {Sneden}, C., {Burles}, S., {et~al.} 2002, \apj, 572, 861

\bibitem[{{da Silva} \& {Smiljanic}(2025)}]{dasilva_2025}
{da Silva}, A.~R. \& {Smiljanic}, R. 2025, \aap, 696, A122

\bibitem[{Dalton {et~al.}(2016)Dalton, Trager, Abrams, Bonifacio, Aguerri,
  Middleton, Benn, Dee, Say{\`e}de, Lewis, {et~al.}}]{dalton2016final}
Dalton, G., Trager, S., Abrams, D.~C., {et~al.} 2016, in Ground-based and
  Airborne Instrumentation for Astronomy VI, Vol. 9908, SPIE, 394--403

\bibitem[{{Eichler} {et~al.}(2019){Eichler}, {Sayar}, {Arcones}, \&
  {Rauscher}}]{Eichler2019}
{Eichler}, M., {Sayar}, W., {Arcones}, A., \& {Rauscher}, T. 2019, \apj, 879,
  47

\bibitem[{Ernandes {et~al.}(2020)Ernandes, Evans, Barbuy, Castilho, Cescutti,
  Christlieb, Cristiani, Di~Marcantonio, Hansen, Quirrenbach,
  {et~al.}}]{ernandes2020stellar}
Ernandes, H., Evans, C., Barbuy, B., {et~al.} 2020, in Ground-based and
  Airborne Instrumentation for Astronomy VIII, Vol. 11447, SPIE, 1327--1337

\bibitem[{{Ernandes} {et~al.}(2024){Ernandes}, {Feuillet}, {Feltzing}, \&
  {Sk{\'u}lad{\'o}ttir}}]{Ernandes_2024}
{Ernandes}, H., {Feuillet}, D., {Feltzing}, S., \& {Sk{\'u}lad{\'o}ttir},
  {\'A}. 2024, \aap, 691, A333

\bibitem[{{Ernandes} {et~al.}(2025){Ernandes}, {Sk{\'u}lad{\'o}ttir},
  {Feltzing}, \& {Feuillet}}]{Ernandes_2025}
{Ernandes}, H., {Sk{\'u}lad{\'o}ttir}, {\'A}., {Feltzing}, S., \& {Feuillet},
  D. 2025, \aap, 703, A256

\bibitem[{Evans {et~al.}(2018)Evans, Barbuy, Castilho, Smiljanic, Melendez,
  Japelj, Cristiani, Snodgrass, Bonifacio, Puech, \& Quirrenbach}]{Evans_2018}
Evans, C.~J., Barbuy, B., Castilho, B., {et~al.} 2018, in Ground-based and
  Airborne Instrumentation for Astronomy VII, ed. C.~J. Evans, L.~Simard, \&
  H.~Takami, Vol. 10702, International Society for Optics and Photonics (SPIE),
  107022E

\bibitem[{{Feuillet} {et~al.}(2021){Feuillet}, {Sahlholdt}, {Feltzing}, \&
  {Casagrande}}]{Feuillet_2021}
{Feuillet}, D.~K., {Sahlholdt}, C.~L., {Feltzing}, S., \& {Casagrande}, L.
  2021, \mnras, 508, 1489

\bibitem[{{Fishlock} {et~al.}(2017){Fishlock}, {Yong}, {Karakas},
  {Alves-Brito}, {Mel{\'e}ndez}, {Nissen}, {Kobayashi}, \&
  {Casey}}]{Fishlock_2017}
{Fishlock}, C.~K., {Yong}, D., {Karakas}, A.~I., {et~al.} 2017, \mnras, 466,
  4672

\bibitem[{{Fran{\c{c}}ois} {et~al.}(2007){Fran{\c{c}}ois}, {Depagne}, {Hill},
  {Spite}, {Spite}, {Plez}, {Beers}, {Andersen}, {James}, {Barbuy}, {Cayrel},
  {Bonifacio}, {Molaro}, {Nordstr{\"o}m}, \& {Primas}}]{Francois_2007}
{Fran{\c{c}}ois}, P., {Depagne}, E., {Hill}, V., {et~al.} 2007, \aap, 476, 935

\bibitem[{{Fran{\c{c}}ois} {et~al.}(2004){Fran{\c{c}}ois}, {Matteucci},
  {Cayrel}, {Spite}, {Spite}, \& {Chiappini}}]{Francois_2004}
{Fran{\c{c}}ois}, P., {Matteucci}, F., {Cayrel}, R., {et~al.} 2004, \aap, 421,
  613

\bibitem[{{Frebel} {et~al.}(2007){Frebel}, {Christlieb}, {Norris}, {Thom},
  {Beers}, \& {Rhee}}]{Frebel_2007}
{Frebel}, A., {Christlieb}, N., {Norris}, J.~E., {et~al.} 2007, \apjl, 660,
  L117

\bibitem[{{Freudling} {et~al.}(2013){Freudling}, {Romaniello}, {Bramich},
  {Ballester}, {Forchi}, {Garc{\'\i}a-Dabl{\'o}}, {Moehler}, \&
  {Neeser}}]{Freudling_2013}
{Freudling}, W., {Romaniello}, M., {Bramich}, D.~M., {et~al.} 2013, \aap, 559,
  A96

\bibitem[{{Fujibayashi} {et~al.}(2023){Fujibayashi}, {Sekiguchi}, {Shibata}, \&
  {Wanajo}}]{Fujibayashi2023}
{Fujibayashi}, S., {Sekiguchi}, Y., {Shibata}, M., \& {Wanajo}, S. 2023, \apj,
  956, 100

\bibitem[{{\relax Gaia Collaboration} {et~al.}(2016){\relax Gaia
  Collaboration}, {Prusti}, {de Bruijne}, {Brown}, {Vallenari}, {Babusiaux},
  {Bailer-Jones}, {Bastian}, {Biermann}, {Evans}, {Eyer}, {Jansen}, {Jordi},
  {Klioner}, {Lammers}, {Lindegren}, {Luri}, {Mignard}, {Milligan}, {Panem},
  {Poinsignon}, {Pourbaix}, {Randich}, {Sarri}, {Sartoretti}, {Siddiqui},
  {Soubiran}, {Valette}, {van Leeuwen}, {Walton}, {Aerts}, {Arenou}, {Cropper},
  {Drimmel}, {H{\o}g}, {Katz}, {Lattanzi}, {O'Mullane}, {Grebel}, {Holland},
  {Huc}, {Passot}, {Bramante}, {Cacciari}, {Casta{\~n}eda}, {Chaoul}, {Cheek},
  {De Angeli}, {Fabricius}, {Guerra}, {Hern{\'a}ndez}, {Jean-Antoine-Piccolo},
  {Masana}, {Messineo}, {Mowlavi}, {Nienartowicz}, {Ord{\'o}{\~n}ez-Blanco},
  {Panuzzo}, {Portell}, {Richards}, {Riello}, {Seabroke}, {Tanga},
  {Th{\'e}venin}, {Torra}, {Els}, {Gracia-Abril}, {Comoretto},
  {Garcia-Reinaldos}, {Lock}, {Mercier}, {Altmann}, {Andrae}, {Astraatmadja},
  {Bellas-Velidis}, {Benson}, {Berthier}, {Blomme}, {Busso}, {Carry},
  {Cellino}, {Clementini}, {Cowell}, {Creevey}, {Cuypers}, {Davidson}, {De
  Ridder}, {de Torres}, {Delchambre}, {Dell'Oro}, {Ducourant}, {Fr{\'e}mat},
  {Garc{\'\i}a-Torres}, {Gosset}, {Halbwachs}, {Hambly}, {Harrison}, {Hauser},
  {Hestroffer}, {Hodgkin}, {Huckle}, {Hutton}, {Jasniewicz}, {Jordan},
  {Kontizas}, {Korn}, {Lanzafame}, {Manteiga}, {Moitinho}, {Muinonen},
  {Osinde}, {Pancino}, {Pauwels}, {Petit}, {Recio-Blanco}, {Robin}, {Sarro},
  {Siopis}, {Smith}, {Smith}, {Sozzetti}, {Thuillot}, {van Reeven}, {Viala},
  {Abbas}, {Abreu Aramburu}, {Accart}, {Aguado}, {Allan}, {Allasia},
  {Altavilla}, {{\'A}lvarez}, {Alves}, {Anderson}, {Andrei}, {Anglada Varela},
  {Antiche}, {Antoja}, {Ant{\'o}n}, {Arcay}, {Atzei}, {Ayache}, {Bach},
  {Baker}, {Balaguer-N{\'u}{\~n}ez}, {Barache}, {Barata}, {Barbier}, {Barblan},
  {Baroni}, {Barrado y Navascu{\'e}s}, {Barros}, {Barstow}, {Becciani},
  {Bellazzini}, {Bellei}, {Bello Garc{\'\i}a}, {Belokurov}, {Bendjoya},
  {Berihuete}, {Bianchi}, {Bienaym{\'e}}, {Billebaud}, {Blagorodnova},
  {Blanco-Cuaresma}, {Boch}, {Bombrun}, {Borrachero}, {Bouquillon}, {Bourda},
  {Bouy}, {Bragaglia}, {Breddels}, {Brouillet}, {Br{\"u}semeister},
  {Bucciarelli}, {Budnik}, {Burgess}, {Burgon}, {Burlacu}, {Busonero}, {Buzzi},
  {Caffau}, {Cambras}, {Campbell}, {Cancelliere}, {Cantat-Gaudin}, {Carlucci},
  {Carrasco}, {Castellani}, {Charlot}, {Charnas}, {Charvet}, {Chassat},
  {Chiavassa}, {Clotet}, {Cocozza}, {Collins}, {Collins}, {Costigan}, {Crifo},
  {Cross}, {Crosta}, {Crowley}, {Dafonte}, {Damerdji}, {Dapergolas}, {David},
  {David}, {De Cat}, {de Felice}, {de Laverny}, {De Luise}, {De March}, {de
  Martino}, {de Souza}, {Debosscher}, {del Pozo}, {Delbo}, {Delgado},
  {Delgado}, {di Marco}, {Di Matteo}, {Diakite}, {Distefano}, {Dolding}, {Dos
  Anjos}, {Drazinos}, {Dur{\'a}n}, {Dzigan}, {Ecale}, {Edvardsson}, {Enke},
  {Erdmann}, {Escolar}, {Espina}, {Evans}, {Eynard Bontemps}, {Fabre},
  {Fabrizio}, {Faigler}, {Falc{\~a}o}, {Farr{\`a}s Casas}, {Faye}, {Federici},
  {Fedorets}, {Fern{\'a}ndez-Hern{\'a}ndez}, {Fernique}, {Fienga}, {Figueras},
  {Filippi}, {Findeisen}, {Fonti}, {Fouesneau}, {Fraile}, {Fraser}, {Fuchs},
  {Furnell}, {Gai}, {Galleti}, {Galluccio}, {Garabato}, {Garc{\'\i}a-Sedano},
  {Gar{\'e}}, {Garofalo}, {Garralda}, {Gavras}, {Gerssen}, {Geyer}, {Gilmore},
  {Girona}, {Giuffrida}, {Gomes}, {Gonz{\'a}lez-Marcos},
  {Gonz{\'a}lez-N{\'u}{\~n}ez}, {Gonz{\'a}lez-Vidal}, {Granvik}, {Guerrier},
  {Guillout}, {Guiraud}, {G{\'u}rpide}, {Guti{\'e}rrez-S{\'a}nchez}, {Guy},
  {Haigron}, {Hatzidimitriou}, {Haywood}, {Heiter}, {Helmi}, {Hobbs},
  {Hofmann}, {Holl}, {Holland}, {Hunt}, {Hypki}, {Icardi}, {Irwin}, {Jevardat
  de Fombelle}, {Jofr{\'e}}, {Jonker}, {Jorissen}, {Julbe}, {Karampelas},
  {Kochoska}, {Kohley}, {Kolenberg}, {Kontizas}, {Koposov}, {Kordopatis},
  {Koubsky}, {Kowalczyk}, {Krone-Martins}, {Kudryashova}, {Kull}, {Bachchan},
  {Lacoste-Seris}, {Lanza}, {Lavigne}, {Le Poncin-Lafitte}, {Lebreton},
  {Lebzelter}, {Leccia}, {Leclerc}, {Lecoeur-Taibi}, {Lemaitre}, {Lenhardt},
  {Leroux}, {Liao}, {Licata}, {Lindstr{\o}m}, {Lister}, {Livanou}, {Lobel},
  {L{\"o}ffler}, {L{\'o}pez}, {Lopez-Lozano}, {Lorenz}, {Loureiro},
  {MacDonald}, {Magalh{\~a}es Fernandes}, {Managau}, {Mann}, {Mantelet},
  {Marchal}, {Marchant}, {Marconi}, {Marie}, {Marinoni}, {Marrese},
  {Marschalk{\'o}}, {Marshall}, {Mart{\'\i}n-Fleitas}, {Martino}, {Mary},
  {Matijevi{\v{c}}}, {Mazeh}, {McMillan}, {Messina}, {Mestre}, {Michalik},
  {Millar}, {Miranda}, {Molina}, {Molinaro}, {Molinaro}, {Moln{\'a}r},
  {Moniez}, {Montegriffo}, {Monteiro}, {Mor}, {Mora}, {Morbidelli}, {Morel},
  {Morgenthaler}, {Morley}, {Morris}, {Mulone}, {Muraveva}, {Musella},
  {Narbonne}, {Nelemans}, {Nicastro}, {Noval}, {Ord{\'e}novic},
  {Ordieres-Mer{\'e}}, {Osborne}, {Pagani}, {Pagano}, {Pailler}, {Palacin},
  {Palaversa}, {Parsons}, {Paulsen}, {Pecoraro}, {Pedrosa}, {Pentik{\"a}inen},
  {Pereira}, {Pichon}, {Piersimoni}, {Pineau}, {Plachy}, {Plum}, {Poujoulet},
  {Pr{\v{s}}a}, {Pulone}, {Ragaini}, {Rago}, {Rambaux}, {Ramos-Lerate},
  {Ranalli}, {Rauw}, {Read}, {Regibo}, {Renk}, {Reyl{\'e}}, {Ribeiro},
  {Rimoldini}, {Ripepi}, {Riva}, {Rixon}, {Roelens}, {Romero-G{\'o}mez},
  {Rowell}, {Royer}, {Rudolph}, {Ruiz-Dern}, {Sadowski}, {Sagrist{\`a}
  Sell{\'e}s}, {Sahlmann}, {Salgado}, {Salguero}, {Sarasso}, {Savietto},
  {Schnorhk}, {Schultheis}, {Sciacca}, {Segol}, {Segovia}, {Segransan},
  {Serpell}, {Shih}, {Smareglia}, {Smart}, {Smith}, {Solano}, {Solitro},
  {Sordo}, {Soria Nieto}, {Souchay}, {Spagna}, {Spoto}, {Stampa}, {Steele},
  {Steidelm{\"u}ller}, {Stephenson}, {Stoev}, {Suess}, {S{\"u}veges}, {Surdej},
  {Szabados}, {Szegedi-Elek}, {Tapiador}, {Taris}, {Tauran}, {Taylor},
  {Teixeira}, {Terrett}, {Tingley}, {Trager}, {Turon}, {Ulla}, {Utrilla},
  {Valentini}, {van Elteren}, {Van Hemelryck}, {van Leeuwen}, {Varadi},
  {Vecchiato}, {Veljanoski}, {Via}, {Vicente}, {Vogt}, {Voss}, {Votruba},
  {Voutsinas}, {Walmsley}, {Weiler}, {Weingrill}, {Werner}, {Wevers},
  {Whitehead}, {Wyrzykowski}, {Yoldas}, {{\v{Z}}erjal}, {Zucker}, {Zurbach},
  {Zwitter}, {Alecu}, {Allen}, {Allende Prieto}, {Amorim},
  {Anglada-Escud{\'e}}, {Arsenijevic}, {Azaz}, {Balm}, {Beck}, {Bernstein},
  {Bigot}, {Bijaoui}, {Blasco}, {Bonfigli}, {Bono}, {Boudreault}, {Bressan},
  {Brown}, {Brunet}, {Bunclark}, {Buonanno}, {Butkevich}, {Carret}, {Carrion},
  {Chemin}, {Ch{\'e}reau}, {Corcione}, {Darmigny}, {de Boer}, {de Teodoro}, {de
  Zeeuw}, {Delle Luche}, {Domingues}, {Dubath}, {Fodor}, {Fr{\'e}zouls},
  {Fries}, {Fustes}, {Fyfe}, {Gallardo}, {Gallegos}, {Gardiol}, {Gebran},
  {Gomboc}, {G{\'o}mez}, {Grux}, {Gueguen}, {Heyrovsky}, {Hoar}, {Iannicola},
  {Isasi Parache}, {Janotto}, {Joliet}, {Jonckheere}, {Keil}, {Kim},
  {Klagyivik}, {Klar}, {Knude}, {Kochukhov}, {Kolka}, {Kos}, {Kutka}, {Lainey},
  {LeBouquin}, {Liu}, {Loreggia}, {Makarov}, {Marseille}, {Martayan},
  {Martinez-Rubi}, {Massart}, {Meynadier}, {Mignot}, {Munari}, {Nguyen},
  {Nordlander}, {Ocvirk}, {O'Flaherty}, {Olias Sanz}, {Ortiz}, {Osorio},
  {Oszkiewicz}, {Ouzounis}, {Palmer}, {Park}, {Pasquato}, {Peltzer}, {Peralta},
  {P{\'e}turaud}, {Pieniluoma}, {Pigozzi}, {Poels}, {Prat}, {Prod'homme},
  {Raison}, {Rebordao}, {Risquez}, {Rocca-Volmerange}, {Rosen}, {Ruiz-Fuertes},
  {Russo}, {Sembay}, {Serraller Vizcaino}, {Short}, {Siebert}, {Silva},
  {Sinachopoulos}, {Slezak}, {Soffel}, {Sosnowska}, {Strai{\v{z}}ys}, {ter
  Linden}, {Terrell}, {Theil}, {Tiede}, {Troisi}, {Tsalmantza}, {Tur},
  {Vaccari}, {Vachier}, {Valles}, {Van Hamme}, {Veltz}, {Virtanen}, {Wallut},
  {Wichmann}, {Wilkinson}, {Ziaeepour}, \& {Zschocke}}]{Gaia_2016b}
{\relax Gaia Collaboration}, {Prusti}, T., {de Bruijne}, J.~H.~J., {et~al.}
  2016, \aap, 595, A1

\bibitem[{{\relax Gaia Collaboration} {et~al.}(2023){\relax Gaia
  Collaboration}, {Vallenari}, {Brown}, {Prusti}, {de Bruijne}, {Arenou},
  {Babusiaux}, {Biermann}, {Creevey}, {Ducourant}, {Evans}, {Eyer}, {Guerra},
  {Hutton}, {Jordi}, {Klioner}, {Lammers}, {Lindegren}, {Luri}, {Mignard},
  {Panem}, {Pourbaix}, {Randich}, {Sartoretti}, {Soubiran}, {Tanga}, {Walton},
  {Bailer-Jones}, {Bastian}, {Drimmel}, {Jansen}, {Katz}, {Lattanzi}, {van
  Leeuwen}, {Bakker}, {Cacciari}, {Casta{\~n}eda}, {De Angeli}, {Fabricius},
  {Fouesneau}, {Fr{\'e}mat}, {Galluccio}, {Guerrier}, {Heiter}, {Masana},
  {Messineo}, {Mowlavi}, {Nicolas}, {Nienartowicz}, {Pailler}, {Panuzzo},
  {Riclet}, {Roux}, {Seabroke}, {Sordo}, {Th{\'e}venin}, {Gracia-Abril},
  {Portell}, {Teyssier}, {Altmann}, {Andrae}, {Audard}, {Bellas-Velidis},
  {Benson}, {Berthier}, {Blomme}, {Burgess}, {Busonero}, {Busso},
  {C{\'a}novas}, {Carry}, {Cellino}, {Cheek}, {Clementini}, {Damerdji},
  {Davidson}, {de Teodoro}, {Nu{\~n}ez Campos}, {Delchambre}, {Dell'Oro},
  {Esquej}, {Fern{\'a}ndez-Hern{\'a}ndez}, {Fraile}, {Garabato},
  {Garc{\'\i}a-Lario}, {Gosset}, {Haigron}, {Halbwachs}, {Hambly}, {Harrison},
  {Hern{\'a}ndez}, {Hestroffer}, {Hodgkin}, {Holl}, {Jan{\ss}en}, {Jevardat de
  Fombelle}, {Jordan}, {Krone-Martins}, {Lanzafame}, {L{\"o}ffler}, {Marchal},
  {Marrese}, {Moitinho}, {Muinonen}, {Osborne}, {Pancino}, {Pauwels},
  {Recio-Blanco}, {Reyl{\'e}}, {Riello}, {Rimoldini}, {Roegiers}, {Rybizki},
  {Sarro}, {Siopis}, {Smith}, {Sozzetti}, {Utrilla}, {van Leeuwen}, {Abbas},
  {{\'A}brah{\'a}m}, {Abreu Aramburu}, {Aerts}, {Aguado}, {Ajaj},
  {Aldea-Montero}, {Altavilla}, {{\'A}lvarez}, {Alves}, {Anders}, {Anderson},
  {Anglada Varela}, {Antoja}, {Baines}, {Baker}, {Balaguer-N{\'u}{\~n}ez},
  {Balbinot}, {Balog}, {Barache}, {Barbato}, {Barros}, {Barstow},
  {Bartolom{\'e}}, {Bassilana}, {Bauchet}, {Becciani}, {Bellazzini},
  {Berihuete}, {Bernet}, {Bertone}, {Bianchi}, {Binnenfeld}, {Blanco-Cuaresma},
  {Blazere}, {Boch}, {Bombrun}, {Bossini}, {Bouquillon}, {Bragaglia},
  {Bramante}, {Breedt}, {Bressan}, {Brouillet}, {Brugaletta}, {Bucciarelli},
  {Burlacu}, {Butkevich}, {Buzzi}, {Caffau}, {Cancelliere}, {Cantat-Gaudin},
  {Carballo}, {Carlucci}, {Carnerero}, {Carrasco}, {Casamiquela}, {Castellani},
  {Castro-Ginard}, {Chaoul}, {Charlot}, {Chemin}, {Chiaramida}, {Chiavassa},
  {Chornay}, {Comoretto}, {Contursi}, {Cooper}, {Cornez}, {Cowell}, {Crifo},
  {Cropper}, {Crosta}, {Crowley}, {Dafonte}, {Dapergolas}, {David}, {David},
  {de Laverny}, {De Luise}, \& {De March}}]{Gaia_2023}
{\relax Gaia Collaboration}, {Vallenari}, A., {Brown}, A.~G.~A., {et~al.} 2023,
  \aap, 674, A1

\bibitem[{{Gehrels}(1986)}]{Gehrels_1986}
{Gehrels}, N. 1986, \apj, 303, 336

\bibitem[{{Ginsburg} {et~al.}(2019){Ginsburg}, {Sip{\H{o}}cz}, {Brasseur},
  {Cowperthwaite}, {Craig}, {Deil}, {Guillochon}, {Guzman}, {Liedtke}, {Lian
  Lim}, {Lockhart}, {Mommert}, {Morris}, {Norman}, {Parikh}, {Persson},
  {Robitaille}, {Segovia}, {Singer}, {Tollerud}, {de Val-Borro}, {Valtchanov},
  {Woillez}, {Astroquery Collaboration}, \& {a subset of astropy
  Collaboration}}]{Ginsburg_2019}
{Ginsburg}, A., {Sip{\H{o}}cz}, B.~M., {Brasseur}, C.~E., {et~al.} 2019, \aj,
  157, 98

\bibitem[{{Goriely} {et~al.}(2011){Goriely}, {Bauswein}, \&
  {Janka}}]{Goriely2011}
{Goriely}, S., {Bauswein}, A., \& {Janka}, H.-T. 2011, \apjl, 738, L32

\bibitem[{{Gudin} {et~al.}(2021){Gudin}, {Shank}, {Beers}, {Yuan}, {Limberg},
  {Roederer}, {Placco}, {Holmbeck}, {Dietz}, {Rasmussen}, {Hansen}, {Sakari},
  {Ezzeddine}, \& {Frebel}}]{Gudin_2021}
{Gudin}, D., {Shank}, D., {Beers}, T.~C., {et~al.} 2021, \apj, 908, 79

\bibitem[{{Gull} {et~al.}(2018){Gull}, {Frebel}, {Cain}, {Placco}, {Ji},
  {Abate}, {Ezzeddine}, {Karakas}, {Hansen}, {Sakari}, {Holmbeck}, {Santucci},
  {Casey}, \& {Beers}}]{Gull_2018}
{Gull}, M., {Frebel}, A., {Cain}, M.~G., {et~al.} 2018, \apj, 862, 174

\bibitem[{{Gull} {et~al.}(2021){Gull}, {Frebel}, {Hinojosa}, {Roederer}, {Ji},
  \& {Brauer}}]{Gull_2021}
{Gull}, M., {Frebel}, A., {Hinojosa}, K., {et~al.} 2021, \apj, 912, 52

\bibitem[{{Guo} {et~al.}(2025){Guo}, {Storm}, {Bergemann}, {Lian}, {Alexeeva},
  {Li}, {Ezzeddine}, {Jeffrey}, \& {Chen}}]{Guo_2025}
{Guo}, Y., {Storm}, N., {Bergemann}, M., {et~al.} 2025, \aap, 693, A211

\bibitem[{{Hanke} {et~al.}(2020){Hanke}, {Hansen}, {Ludwig}, {Cristallo},
  {McWilliam}, {Grebel}, \& {Piersanti}}]{Hanke_2020}
{Hanke}, M., {Hansen}, C.~J., {Ludwig}, H.-G., {et~al.} 2020, \aap, 635, A104

\bibitem[{{Hansen} {et~al.}(2018){Hansen}, {El-Souri}, {Monaco}, {Villanova},
  {Bonifacio}, {Caffau}, \& {Sbordone}}]{HansenC_2018}
{Hansen}, C.~J., {El-Souri}, M., {Monaco}, L., {et~al.} 2018, \apj, 855, 83

\bibitem[{{Hansen} {et~al.}(2025){Hansen}, {Roederer}, {Shah}, {Ezzeddine},
  {Beers}, {Frebel}, {Holmbeck}, {Placco}, {Sakari}, {Ji}, {Marshall},
  {Mardini}, \& {Chiti}}]{Hansen_2025}
{Hansen}, T.~T., {Roederer}, I.~U., {Shah}, S.~P., {et~al.} 2025, \aap, 697,
  A127

\bibitem[{{Hansen} {et~al.}(2017){Hansen}, {Simon}, {Marshall}, {Li},
  {Carollo}, {DePoy}, {Nagasawa}, {Bernstein}, {Drlica-Wagner}, {Abdalla},
  {Allam}, {Annis}, {Bechtol}, {Benoit-L{\'e}vy}, {Brooks}, {Buckley-Geer},
  {Carnero Rosell}, {Carrasco Kind}, {Carretero}, {Cunha}, {da Costa}, {Desai},
  {Eifler}, {Fausti Neto}, {Flaugher}, {Frieman}, {Garc{\'\i}a-Bellido},
  {Gaztanaga}, {Gerdes}, {Gruen}, {Gruendl}, {Gschwend}, {Gutierrez}, {James},
  {Krause}, {Kuehn}, {Kuropatkin}, {Lahav}, {Miquel}, {Plazas}, {Romer},
  {Sanchez}, {Santiago}, {Scarpine}, {Smith}, {Soares-Santos}, {Sobreira},
  {Suchyta}, {Swanson}, {Tarle}, {Walker}, \& {DES
  Collaboration}}]{HansenT_2017}
{Hansen}, T.~T., {Simon}, J.~D., {Marshall}, J.~L., {et~al.} 2017, \apj, 838,
  44

\bibitem[{{Hattori} {et~al.}(2023){Hattori}, {Okuno}, \&
  {Roederer}}]{Hattori_2023}
{Hattori}, K., {Okuno}, A., \& {Roederer}, I.~U. 2023, \apj, 946, 48

\bibitem[{{Hayek} {et~al.}(2009){Hayek}, {Wiesendahl}, {Christlieb},
  {Eriksson}, {Korn}, {Barklem}, {Hill}, {Beers}, {Farouqi}, {Pfeiffer}, \&
  {Kratz}}]{Hayek_2009}
{Hayek}, W., {Wiesendahl}, U., {Christlieb}, N., {et~al.} 2009, \aap, 504, 511

\bibitem[{{Helmi} {et~al.}(2018){Helmi}, {Babusiaux}, {Koppelman}, {Massari},
  {Veljanoski}, \& {Brown}}]{Helmi_18}
{Helmi}, A., {Babusiaux}, C., {Koppelman}, H.~H., {et~al.} 2018, \nat, 563, 85

\bibitem[{{Hill} {et~al.}(2002){Hill}, {Plez}, {Cayrel}, {Beers},
  {Nordstr{\"o}m}, {Andersen}, {Spite}, {Spite}, {Barbuy}, {Bonifacio},
  {Depagne}, {Fran{\c{c}}ois}, \& {Primas}}]{Hill2002}
{Hill}, V., {Plez}, B., {Cayrel}, R., {et~al.} 2002, \aap, 387, 560

\bibitem[{{Hirai} {et~al.}(2022){Hirai}, {Beers}, {Chiba}, {Aoki}, {Shank},
  {Saitoh}, {Okamoto}, \& {Makino}}]{Hirai_2022}
{Hirai}, Y., {Beers}, T.~C., {Chiba}, M., {et~al.} 2022, \mnras, 517, 4856

\bibitem[{{Hirai} {et~al.}(2025){Hirai}, {Beers}, {Lee}, {Wanajo}, {Roederer},
  {Tanaka}, {Chiba}, {Saitoh}, {Placco}, {Hansen}, {Ezzeddine}, {Frebel},
  {Holmbeck}, \& {Sakari}}]{Hirai_2025}
{Hirai}, Y., {Beers}, T.~C., {Lee}, Y.~S., {et~al.} 2025, \apj, 990, 125

\bibitem[{{Holmbeck} {et~al.}(2018){Holmbeck}, {Beers}, {Roederer}, {Placco},
  {Hansen}, {Sakari}, {Sneden}, {Liu}, {Lee}, {Cowan}, \&
  {Frebel}}]{Holmbeck2018}
{Holmbeck}, E.~M., {Beers}, T.~C., {Roederer}, I.~U., {et~al.} 2018, \apjl,
  859, L24

\bibitem[{{Holmbeck} {et~al.}(2019{\natexlab{a}}){Holmbeck}, {Frebel},
  {McLaughlin}, {Mumpower}, {Sprouse}, \& {Surman}}]{Holmbeck_2019}
{Holmbeck}, E.~M., {Frebel}, A., {McLaughlin}, G.~C., {et~al.}
  2019{\natexlab{a}}, \apj, 881, 5

\bibitem[{{Holmbeck} {et~al.}(2019{\natexlab{b}}){Holmbeck}, {Sprouse},
  {Mumpower}, {Vassh}, {Surman}, {Beers}, \& {Kawano}}]{Holmbeck2019}
{Holmbeck}, E.~M., {Sprouse}, T.~M., {Mumpower}, M.~R., {et~al.}
  2019{\natexlab{b}}, \apj, 870, 23

\bibitem[{Holmbeck {et~al.}(2020)Holmbeck, Surman, Frebel, McLaughlin,
  Mumpower, Sprouse, Kawano, Vassh, \& Beers}]{Holmbeck_2020}
Holmbeck, E.~M., Surman, R., Frebel, A., {et~al.} 2020, Journal of Physics:
  Conference Series, 1668, 012020

\bibitem[{{Honda} {et~al.}(2004){Honda}, {Aoki}, {Kajino}, {Ando}, {Beers},
  {Izumiura}, {Sadakane}, \& {Takada-Hidai}}]{Honda_2004}
{Honda}, S., {Aoki}, W., {Kajino}, T., {et~al.} 2004, \apj, 607, 474

\bibitem[{{Ivans} {et~al.}(2006){Ivans}, {Simmerer}, {Sneden}, {Lawler},
  {Cowan}, {Gallino}, \& {Bisterzo}}]{Ivans_2006}
{Ivans}, I.~I., {Simmerer}, J., {Sneden}, C., {et~al.} 2006, \apj, 645, 613

\bibitem[{{Ji} \& {Frebel}(2018)}]{Ji_2018}
{Ji}, A.~P. \& {Frebel}, A. 2018, \apj, 856, 138

\bibitem[{{Ji} {et~al.}(2016){Ji}, {Frebel}, {Chiti}, \& {Simon}}]{Ji_2016}
{Ji}, A.~P., {Frebel}, A., {Chiti}, A., \& {Simon}, J.~D. 2016, \nat, 531, 610

\bibitem[{{Johnson} \& {Bolte}(2001)}]{Johnson_2001}
{Johnson}, J.~A. \& {Bolte}, M. 2001, \apj, 554, 888

\bibitem[{{Just} {et~al.}(2022){Just}, {Aloy}, {Obergaulinger}, \&
  {Nagataki}}]{Just2022}
{Just}, O., {Aloy}, M.~A., {Obergaulinger}, M., \& {Nagataki}, S. 2022, \apjl,
  934, L30

\bibitem[{Kass \& Raftery(1995)}]{Kass_1995}
Kass, R.~E. \& Raftery, A.~E. 1995, Journal of the American Statistical
  Association, 90, 773

\bibitem[{{Kobayashi} {et~al.}(2020){Kobayashi}, {Karakas}, \&
  {Lugaro}}]{Kobayashi_2020}
{Kobayashi}, C., {Karakas}, A.~I., \& {Lugaro}, M. 2020, \apj, 900, 179

\bibitem[{{Kobayashi} {et~al.}(2023){Kobayashi}, {Mandel}, {Belczynski},
  {Goriely}, {Janka}, {Just}, {Ruiter}, {Vanbeveren}, {Kruckow}, {Briel},
  {Eldridge}, \& {Stanway}}]{Kobayashi_2023}
{Kobayashi}, C., {Mandel}, I., {Belczynski}, K., {et~al.} 2023, \apjl, 943, L12

\bibitem[{{Koutsouridou} {et~al.}(2026){Koutsouridou}, {Salvadori},
  {Sk{\'u}lad{\'o}ttir}, {Gelli}, {Rusta}, {Querci}, {Aguado}, \&
  {Mori}}]{Koutsouridou_2026}
{Koutsouridou}, I., {Salvadori}, S., {Sk{\'u}lad{\'o}ttir}, {\'A}., {et~al.}
  2026, arXiv e-prints, arXiv:2605.00990

\bibitem[{{Koutsouridou} {et~al.}(2025){Koutsouridou}, {Sk{\'u}lad{\'o}ttir},
  \& {Salvadori}}]{Koutsouridou_2025}
{Koutsouridou}, I., {Sk{\'u}lad{\'o}ttir}, {\'A}., \& {Salvadori}, S. 2025,
  \aap, 699, A32

\bibitem[{{Lai} {et~al.}(2008){Lai}, {Bolte}, {Johnson}, {Lucatello}, {Heger},
  \& {Woosley}}]{Lai_2008}
{Lai}, D.~K., {Bolte}, M., {Johnson}, J.~A., {et~al.} 2008, \apj, 681, 1524

\bibitem[{{Lattimer} \& {Schramm}(1974)}]{Lattimer1974}
{Lattimer}, J.~M. \& {Schramm}, D.~N. 1974, \apjl, 192, L145

\bibitem[{{Lin} {et~al.}(2025){Lin}, {Li}, {Jiang}, {Aoki}, {Honda}, {He},
  {Zhang}, {Li}, \& {Zhao}}]{Lin2025}
{Lin}, Y., {Li}, H., {Jiang}, R., {et~al.} 2025, \apjl, 984, L43

\bibitem[{{Lind} {et~al.}(2022){Lind}, {Nordlander}, {Wehrhahn}, {Montelius},
  {Osorio}, {Barklem}, {Af{\textcommabelow s}ar}, {Sneden}, \&
  {Kobayashi}}]{Lind_2022}
{Lind}, K., {Nordlander}, T., {Wehrhahn}, A., {et~al.} 2022, \aap, 665, A33

\bibitem[{{Lindegren} {et~al.}(2021){Lindegren}, {Bastian}, {Biermann},
  {Bombrun}, {de Torres}, {Gerlach}, {Geyer}, {Hern{\'a}ndez}, {Hilger},
  {Hobbs}, {Klioner}, {Lammers}, {McMillan}, {Ramos-Lerate},
  {Steidelm{\"u}ller}, {Stephenson}, \& {van Leeuwen}}]{Lindegren_2021}
{Lindegren}, L., {Bastian}, U., {Biermann}, M., {et~al.} 2021, \aap, 649, A4

\bibitem[{{Lucchesi} {et~al.}(2024){Lucchesi}, {Jablonka},
  {Sk{\'u}lad{\'o}ttir}, {Lardo}, {Mashonkina}, {Primas}, {Venn}, {Hill}, \&
  {Minniti}}]{Lucchesi_2024}
{Lucchesi}, R., {Jablonka}, P., {Sk{\'u}lad{\'o}ttir}, {\'A}., {et~al.} 2024,
  \aap, 686, A266

\bibitem[{Magrini {et~al.}(2026)Magrini, Bensby, Randich, Bianco, Gonzalez,
  Fernandez-Alvar, Sousa, Caito, Riva, Adibekyan, {et~al.}}]{magrini2026hrmos}
Magrini, L., Bensby, T., Randich, S., {et~al.} 2026, arXiv preprint
  arXiv:2607.04916

\bibitem[{Marconi {et~al.}(2022)Marconi, Abreu, Adibekyan, Alberti, Albrecht,
  Alcaniz, Aliverti, Prieto, G{\'o}mez, Amado, {et~al.}}]{marconi2022andes}
Marconi, A., Abreu, M., Adibekyan, V., {et~al.} 2022, in Ground-based and
  Airborne Instrumentation for Astronomy IX, Vol. 12184, SPIE, 720--735

\bibitem[{{Mashonkina} {et~al.}(2010){Mashonkina}, {Christlieb}, {Barklem},
  {Hill}, {Beers}, \& {Velichko}}]{Mashonkina_2010}
{Mashonkina}, L., {Christlieb}, N., {Barklem}, P.~S., {et~al.} 2010, \aap, 516,
  A46

\bibitem[{{Mashonkina} {et~al.}(2014){Mashonkina}, {Christlieb}, \&
  {Eriksson}}]{Mashonkina_2014}
{Mashonkina}, L., {Christlieb}, N., \& {Eriksson}, K. 2014, \aap, 569, A43

\bibitem[{{Mashonkina} \& {Gehren}(2000)}]{Mashonkina_2000}
{Mashonkina}, L. \& {Gehren}, T. 2000, \aap, 364, 249

\bibitem[{{Mashonkina} {et~al.}(2012){Mashonkina}, {Ryabtsev}, \&
  {Frebel}}]{Mashonkina_2012}
{Mashonkina}, L., {Ryabtsev}, A., \& {Frebel}, A. 2012, \aap, 540, A98

\bibitem[{{Miller} {et~al.}(2020){Miller}, {Sprouse}, {Fryer}, {Ryan},
  {Dolence}, {Mumpower}, \& {Surman}}]{Miller2020}
{Miller}, J.~M., {Sprouse}, T.~M., {Fryer}, C.~L., {et~al.} 2020, \apj, 902, 66

\bibitem[{{Molero} {et~al.}(2023){Molero}, {Magrini}, {Matteucci}, {Romano},
  {Palla}, {Cescutti}, {Viscasillas V{\'a}zquez}, \& {Spitoni}}]{Molero_2023}
{Molero}, M., {Magrini}, L., {Matteucci}, F., {et~al.} 2023, \mnras, 523, 2974

\bibitem[{{M{\"o}sta} {et~al.}(2018){M{\"o}sta}, {Roberts}, {Halevi}, {Ott},
  {Lippuner}, {Haas}, \& {Schnetter}}]{Mosta2018}
{M{\"o}sta}, P., {Roberts}, L.~F., {Halevi}, G., {et~al.} 2018, \apj, 864, 171

\bibitem[{{Nishimura} {et~al.}(2017){Nishimura}, {Sawai}, {Takiwaki}, {Yamada},
  \& {Thielemann}}]{Nishimura2017}
{Nishimura}, N., {Sawai}, H., {Takiwaki}, T., {Yamada}, S., \& {Thielemann},
  F.-K. 2017, \apjl, 836, L21

\bibitem[{{Nissen} {et~al.}(2024){Nissen}, {Amarsi}, {Sk{\'u}lad{\'o}ttir}, \&
  {Schuster}}]{Nissen_2024}
{Nissen}, P.~E., {Amarsi}, A.~M., {Sk{\'u}lad{\'o}ttir}, {\'A}., \& {Schuster},
  W.~J. 2024, \aap, 682, A116

\bibitem[{{Nissen} \& {Schuster}(2010)}]{Nissen_2010}
{Nissen}, P.~E. \& {Schuster}, W.~J. 2010, \aap, 511, L10

\bibitem[{{Nissen} \& {Schuster}(2011)}]{Nissen_2011}
{Nissen}, P.~E. \& {Schuster}, W.~J. 2011, \aap, 530, A15

\bibitem[{{Patel} {et~al.}(2025{\natexlab{a}}){Patel}, {Metzger}, {Cehula},
  {Burns}, {Goldberg}, \& {Thompson}}]{Patel2025a}
{Patel}, A., {Metzger}, B.~D., {Cehula}, J., {et~al.} 2025{\natexlab{a}},
  \apjl, 984, L29

\bibitem[{{Patel} {et~al.}(2025{\natexlab{b}}){Patel}, {Metzger}, {Goldberg},
  {Cehula}, {Thompson}, \& {Renzo}}]{Patel2025b}
{Patel}, A., {Metzger}, B.~D., {Goldberg}, J.~A., {et~al.} 2025{\natexlab{b}},
  \apj, 985, 234

\bibitem[{{Pian} {et~al.}(2017){Pian}, {D'Avanzo}, {Benetti}, {Branchesi},
  {Brocato}, {Campana}, {Cappellaro}, {Covino}, {D'Elia}, {Fynbo}, {Getman},
  {Ghirlanda}, {Ghisellini}, {Grado}, {Greco}, {Hjorth}, {Kouveliotou},
  {Levan}, {Limatola}, {Malesani}, {Mazzali}, {Melandri}, {M{\o}ller},
  {Nicastro}, {Palazzi}, {Piranomonte}, {Rossi}, {Salafia}, {Selsing},
  {Stratta}, {Tanaka}, {Tanvir}, {Tomasella}, {Watson}, {Yang}, {Amati},
  {Antonelli}, {Ascenzi}, {Bernardini}, {Bo{\"e}r}, {Bufano}, {Bulgarelli},
  {Capaccioli}, {Casella}, {Castro-Tirado}, {Chassande-Mottin}, {Ciolfi},
  {Copperwheat}, {Dadina}, {De Cesare}, {di Paola}, {Fan}, {Gendre},
  {Giuffrida}, {Giunta}, {Hunt}, {Israel}, {Jin}, {Kasliwal}, {Klose}, {Lisi},
  {Longo}, {Maiorano}, {Mapelli}, {Masetti}, {Nava}, {Patricelli}, {Perley},
  {Pescalli}, {Piran}, {Possenti}, {Pulone}, {Razzano}, {Salvaterra},
  {Schipani}, {Spera}, {Stamerra}, {Stella}, {Tagliaferri}, {Testa}, {Troja},
  {Turatto}, {Vergani}, \& {Vergani}}]{Pian2017}
{Pian}, E., {D'Avanzo}, P., {Benetti}, S., {et~al.} 2017, \nat, 551, 67

\bibitem[{{Placco} {et~al.}(2023){Placco}, {Almeida-Fernandes}, {Holmbeck},
  {Roederer}, {Mardini}, {Hayes}, {Venn}, {Chiboucas}, {Deibert}, {Gamen},
  {Heo}, {Jeong}, {Kalari}, {Martioli}, {Xu}, {Diaz}, {Gomez-Jimenez},
  {Henderson}, {Prado}, {Quiroz}, {Ruiz-Carmona}, {Simpson}, {Urrutia},
  {McConnachie}, {Pazder}, {Burley}, {Ireland}, {Waller}, {Berg}, {Robertson},
  {Hartman}, {Jones}, {Labrie}, {Perez}, {Ridgway}, \&
  {Thomas-Osip}}]{Placco_2023}
{Placco}, V.~M., {Almeida-Fernandes}, F., {Holmbeck}, E.~M., {et~al.} 2023,
  \apj, 959, 60

\bibitem[{{Placco} {et~al.}(2017){Placco}, {Holmbeck}, {Frebel}, {Beers},
  {Surman}, {Ji}, {Ezzeddine}, {Points}, {Kaleida}, {Hansen}, {Sakari}, \&
  {Casey}}]{Placco_2017}
{Placco}, V.~M., {Holmbeck}, E.~M., {Frebel}, A., {et~al.} 2017, \apj, 844, 18

\bibitem[{{Placco} {et~al.}(2021){Placco}, {Sneden}, {Roederer}, {Lawler}, {Den
  Hartog}, {Hejazi}, {Maas}, \& {Bernath}}]{Placco_2021}
{Placco}, V.~M., {Sneden}, C., {Roederer}, I.~U., {et~al.} 2021, Research Notes
  of the American Astronomical Society, 5, 92

\bibitem[{{Reichert} {et~al.}(2021{\natexlab{a}}){Reichert}, {Hansen}, \&
  {Arcones}}]{Reichert_2021b}
{Reichert}, M., {Hansen}, C.~J., \& {Arcones}, A. 2021{\natexlab{a}}, \apj,
  912, 157

\bibitem[{{Reichert} {et~al.}(2023){Reichert}, {Obergaulinger}, {Aloy},
  {Gabler}, {Arcones}, \& {Thielemann}}]{Reichert2023}
{Reichert}, M., {Obergaulinger}, M., {Aloy}, M.~{\'A}., {et~al.} 2023, \mnras,
  518, 1557

\bibitem[{{Reichert} {et~al.}(2021{\natexlab{b}}){Reichert}, {Obergaulinger},
  {Eichler}, {Aloy}, \& {Arcones}}]{Reichert2021}
{Reichert}, M., {Obergaulinger}, M., {Eichler}, M., {Aloy}, M.~{\'A}., \&
  {Arcones}, A. 2021{\natexlab{b}}, \mnras, 501, 5733

\bibitem[{{Ricigliano} {et~al.}(2024){Ricigliano}, {Jacobi}, \&
  {Arcones}}]{Ricigliano_2024}
{Ricigliano}, G., {Jacobi}, M., \& {Arcones}, A. 2024, \mnras, 533, 2096

\bibitem[{{Roederer} {et~al.}(2024){Roederer}, {Beers}, {Hattori}, {Placco},
  {Hansen}, {Ezzeddine}, {Frebel}, {Holmbeck}, \& {Sakari}}]{Roederer_2024}
{Roederer}, I.~U., {Beers}, T.~C., {Hattori}, K., {et~al.} 2024, \apj, 971, 158

\bibitem[{{Roederer} {et~al.}(2018{\natexlab{a}}){Roederer}, {Hattori}, \&
  {Valluri}}]{Roederer_2018a}
{Roederer}, I.~U., {Hattori}, K., \& {Valluri}, M. 2018{\natexlab{a}}, \aj,
  156, 179

\bibitem[{{Roederer} {et~al.}(2009){Roederer}, {Kratz}, {Frebel}, {Christlieb},
  {Pfeiffer}, {Cowan}, \& {Sneden}}]{Roederer_2009}
{Roederer}, I.~U., {Kratz}, K.-L., {Frebel}, A., {et~al.} 2009, \apj, 698, 1963

\bibitem[{{Roederer} {et~al.}(2012){Roederer}, {Lawler}, {Sobeck}, {Beers},
  {Cowan}, {Frebel}, {Ivans}, {Schatz}, {Sneden}, \&
  {Thompson}}]{Roederer_2012}
{Roederer}, I.~U., {Lawler}, J.~E., {Sobeck}, J.~S., {et~al.} 2012, \apjs, 203,
  27

\bibitem[{{Roederer} {et~al.}(2014){Roederer}, {Preston}, {Thompson},
  {Shectman}, {Sneden}, {Burley}, \& {Kelson}}]{Roederer_2014}
{Roederer}, I.~U., {Preston}, G.~W., {Thompson}, I.~B., {et~al.} 2014, \aj,
  147, 136

\bibitem[{{Roederer} {et~al.}(2018{\natexlab{b}}){Roederer}, {Sakari},
  {Placco}, {Beers}, {Ezzeddine}, {Frebel}, \& {Hansen}}]{Roederer_2018}
{Roederer}, I.~U., {Sakari}, C.~M., {Placco}, V.~M., {et~al.}
  2018{\natexlab{b}}, \apj, 865, 129

\bibitem[{{Roederer} {et~al.}(2010){Roederer}, {Sneden}, {Thompson}, {Preston},
  \& {Shectman}}]{Roederer_2010}
{Roederer}, I.~U., {Sneden}, C., {Thompson}, I.~B., {Preston}, G.~W., \&
  {Shectman}, S.~A. 2010, \apj, 711, 573

\bibitem[{{Rosswog} {et~al.}(2014){Rosswog}, {Korobkin}, {Arcones},
  {Thielemann}, \& {Piran}}]{Rosswog2014}
{Rosswog}, S., {Korobkin}, O., {Arcones}, A., {Thielemann}, F.-K., \& {Piran},
  T. 2014, \mnras, 439, 744

\bibitem[{{Sakari} {et~al.}(2018){Sakari}, {Placco}, {Farrell}, {Roederer},
  {Wallerstein}, {Beers}, {Ezzeddine}, {Frebel}, {Hansen}, {Holmbeck},
  {Sneden}, {Cowan}, {Venn}, {Davis}, {Matijevi{\v{c}}}, {Wyse},
  {Bland-Hawthorn}, {Chiappini}, {Freeman}, {Gibson}, {Grebel}, {Helmi},
  {Kordopatis}, {Kunder}, {Navarro}, {Reid}, {Seabroke}, {Steinmetz}, \&
  {Watson}}]{Sakari_2018}
{Sakari}, C.~M., {Placco}, V.~M., {Farrell}, E.~M., {et~al.} 2018, \apj, 868,
  110

\bibitem[{{Saraf} {et~al.}(2023){Saraf}, {Allende Prieto}, {Sivarani},
  {Bandyopadhyay}, {Beers}, \& {Susmitha}}]{Saraf_2023}
{Saraf}, P., {Allende Prieto}, C., {Sivarani}, T., {et~al.} 2023, \mnras, 524,
  5607

\bibitem[{{Schatz} {et~al.}(2022){Schatz}, {Becerril Reyes}, {Best}, {Brown},
  {Chatziioannou}, {Chipps}, {Deibel}, {Ezzeddine}, {Galloway}, {Hansen},
  {Herwig}, {Ji}, {Lugaro}, {Meisel}, {Norman}, {Read}, {Roberts}, {Spyrou},
  {Tews}, {Timmes}, {Travaglio}, {Vassh}, {Abia}, {Adsley}, {Agarwal},
  {Aliotta}, {Aoki}, {Arcones}, {Aryan}, {Bandyopadhyay}, {Banu}, {Bardayan},
  {Barnes}, {Bauswein}, {Beers}, {Bishop}, {Boztepe}, {C{\^o}t{\'e}}, {Caplan},
  {Champagne}, {Clark}, {Couder}, {Couture}, {de Mink}, {Debnath}, {deBoer},
  {den Hartogh}, {Denissenkov}, {Dexheimer}, {Dillmann}, {Escher}, {Famiano},
  {Farmer}, {Fisher}, {Fr{\"o}hlich}, {Frebel}, {Fryer}, {Fuller}, {Ganguly},
  {Ghosh}, {Gibson}, {Gorda}, {Gourgouliatos}, {Graber}, {Gupta}, {Haxton},
  {Heger}, {Hix}, {Ho}, {Holmbeck}, {Hood}, {Huth}, {Imbriani}, {Izzard},
  {Jain}, {Jayatissa}, {Johnston}, {Kajino}, {Kankainen}, {Kiss},
  {Kwiatkowski}, {La Cognata}, {Laird}, {Lamia}, {Landry}, {Laplace}, {Launey},
  {Leahy}, {Leckenby}, {Lennarz}, {Longfellow}, {Lovell}, {Lynch}, {Lyons},
  {Maeda}, {Masha}, {Matei}, {Merc}, {Messer}, {Montes}, {Mukherjee},
  {Mumpower}, {Neto}, {Nevins}, {Newton}, {Nguyen}, {Nishikawa}, {Nishimura},
  {Nunes}, {O'Connor}, {O'Shea}, {Ong}, {Pain}, {Pajkos}, {Pignatari},
  {Pizzone}, {Placco}, {Plewa}, {Pritychenko}, {Psaltis}, {Puentes}, {Qian},
  {Radice}, {Rapagnani}, {Rebeiro}, {Reifarth}, {Richard}, {Rijal}, {Roederer},
  {Rojo}, {J S K}, {Saito}, {Schwenk}, {Sergi}, {Sidhu}, {Simon}, {Sivarani},
  {Sk{\'u}lad{\'o}ttir}, {Smith}, {Spiridon}, {Sprouse}, {Starrfield},
  {Steiner}, {Strieder}, {Sultana}, {Surman}, {Sz{\"u}cs}, {Tawfik},
  {Thielemann}, {Trache}, {Trappitsch}, {Tsang}, {Tumino}, {Upadhyayula},
  {Valle Mart{\'\i}nez}, {Van der Swaelmen}, {Viscasillas V{\'a}zquez},
  {Watts}, {Wehmeyer}, {Wiescher}, {Wrede}, {Yoon}, {Zegers}, {Zermane}, \&
  {Zingale}}]{Schatz2022}
{Schatz}, H., {Becerril Reyes}, A.~D., {Best}, A., {et~al.} 2022, Journal of
  Physics G Nuclear Physics, 49, 110502

\bibitem[{{Shah} {et~al.}(2023){Shah}, {Ezzeddine}, {Ji}, {Hansen}, {Roederer},
  {Catelan}, {Hackshaw}, {Holmbeck}, {Beers}, \& {Surman}}]{Shah_2023}
{Shah}, S.~P., {Ezzeddine}, R., {Ji}, A.~P., {et~al.} 2023, \apj, 948, 122

\bibitem[{{Shejeelammal} {et~al.}(2020){Shejeelammal}, {Goswami}, {Goswami},
  {Rathour}, \& {Masseron}}]{Shejeelammal_2020}
{Shejeelammal}, J., {Goswami}, A., {Goswami}, P.~P., {Rathour}, R.~S., \&
  {Masseron}, T. 2020, \mnras, 492, 3708

\bibitem[{{Siegel} {et~al.}(2019){Siegel}, {Barnes}, \& {Metzger}}]{Siegel2019}
{Siegel}, D.~M., {Barnes}, J., \& {Metzger}, B.~D. 2019, \nat, 569, 241

\bibitem[{{Simonetti} {et~al.}(2019){Simonetti}, {Matteucci}, {Greggio}, \&
  {Cescutti}}]{Simonetti_2019}
{Simonetti}, P., {Matteucci}, F., {Greggio}, L., \& {Cescutti}, G. 2019,
  \mnras, 486, 2896

\bibitem[{{Siqueira Mello} {et~al.}(2014){Siqueira Mello}, {Hill}, {Barbuy},
  {Spite}, {Spite}, {Beers}, {Caffau}, {Bonifacio}, {Cayrel}, {Fran{\c{c}}ois},
  {Schatz}, \& {Wanajo}}]{SiqueiraMello_2014}
{Siqueira Mello}, C., {Hill}, V., {Barbuy}, B., {et~al.} 2014, \aap, 565, A93

\bibitem[{{Sk{\'u}lad{\'o}ttir} {et~al.}(2025){Sk{\'u}lad{\'o}ttir},
  {Ernandes}, {Feuillet}, {Mori}, {Feltzing}, {Lucchesi}, \& {Di
  Matteo}}]{Skuladottir_2025}
{Sk{\'u}lad{\'o}ttir}, {\'A}., {Ernandes}, H., {Feuillet}, D.~K., {et~al.}
  2025, \apjl, 986, L21

\bibitem[{{Sk{\'u}lad{\'o}ttir} {et~al.}(2019){Sk{\'u}lad{\'o}ttir}, {Hansen},
  {Salvadori}, \& {Choplin}}]{Skuladottir_2019}
{Sk{\'u}lad{\'o}ttir}, {\'A}., {Hansen}, C.~J., {Salvadori}, S., \& {Choplin},
  A. 2019, \aap, 631, A171

\bibitem[{Sk{\'u}lad{\'o}ttir {et~al.}(2023)Sk{\'u}lad{\'o}ttir, Puls, Amarsi,
  Battaglia, Buder, Campbell, Cardona-Barrero, Christlieb, Feuillet, Gelli,
  {et~al.}}]{skuladottir20234most}
Sk{\'u}lad{\'o}ttir, {\'A}., Puls, A.~A., Amarsi, A.~M., {et~al.} 2023,
  Messenger, 190, 19

\bibitem[{{Sk{\'u}lad{\'o}ttir} \& {Salvadori}(2020)}]{skuladottir_2020}
{Sk{\'u}lad{\'o}ttir}, {\'A}. \& {Salvadori}, S. 2020, \aap, 634, L2

\bibitem[{{Sneden} {et~al.}(2012){Sneden}, {Bean}, {Ivans}, {Lucatello}, \&
  {Sobeck}}]{Sneden_2012}
{Sneden}, C., {Bean}, J., {Ivans}, I., {Lucatello}, S., \& {Sobeck}, J. 2012,
  {MOOG: LTE line analysis and spectrum synthesis}, Astrophysics Source Code
  Library, record ascl:1202.009

\bibitem[{{Sneden} {et~al.}(2003){Sneden}, {Cowan}, {Lawler}, {Ivans},
  {Burles}, {Beers}, {Primas}, {Hill}, {Truran}, {Fuller}, {Pfeiffer}, \&
  {Kratz}}]{Sneden2003}
{Sneden}, C., {Cowan}, J.~J., {Lawler}, J.~E., {et~al.} 2003, \apj, 591, 936

\bibitem[{{Sneden}(1973)}]{Sneden_1973}
{Sneden}, C.~A. 1973, PhD thesis, THE UNIVERSITY OF TEXAS AT AUSTIN.

\bibitem[{{Storm} {et~al.}(2024){Storm}, {Barklem}, {Yakovleva}, {Belyaev},
  {Palmeri}, {Quinet}, {Lodders}, {Bergemann}, \& {Hoppe}}]{Storm_2024}
{Storm}, N., {Barklem}, P.~S., {Yakovleva}, S.~A., {et~al.} 2024, \aap, 683,
  A200

\bibitem[{{Storm} {et~al.}(2025){Storm}, {Bergemann}, {Eitner}, {Hoppe},
  {Kemp}, {Ruiter}, {Janka}, {Sieverding}, {de Mink}, {Seitenzahl}, \&
  {Owusu}}]{Storm_2025}
{Storm}, N., {Bergemann}, M., {Eitner}, P., {et~al.} 2025, \mnras, 538, 3284

\bibitem[{{Tolstoy} {et~al.}(2009){Tolstoy}, {Hill}, \& {Tosi}}]{Tolstoy_2009}
{Tolstoy}, E., {Hill}, V., \& {Tosi}, M. 2009, \araa, 47, 371

\bibitem[{{Vincenzo} {et~al.}(2019){Vincenzo}, {Spitoni}, {Calura},
  {Matteucci}, {Silva Aguirre}, {Miglio}, \& {Cescutti}}]{Vincenzo_2019}
{Vincenzo}, F., {Spitoni}, E., {Calura}, F., {et~al.} 2019, \mnras, 487, L47

\bibitem[{{Wanajo} {et~al.}(2024){Wanajo}, {Fujibayashi}, {Hayashi}, {Kiuchi},
  {Sekiguchi}, \& {Shibata}}]{Wanajo2024}
{Wanajo}, S., {Fujibayashi}, S., {Hayashi}, K., {et~al.} 2024, \prl, 133,
  241201

\bibitem[{{Wanajo} {et~al.}(2014){Wanajo}, {Sekiguchi}, {Nishimura}, {Kiuchi},
  {Kyutoku}, \& {Shibata}}]{Wanajo2014}
{Wanajo}, S., {Sekiguchi}, Y., {Nishimura}, N., {et~al.} 2014, \apjl, 789, L39

\bibitem[{{Watson} {et~al.}(2019){Watson}, {Hansen}, {Selsing}, {Koch},
  {Malesani}, {Andersen}, {Fynbo}, {Arcones}, {Bauswein}, {Covino}, {Grado},
  {Heintz}, {Hunt}, {Kouveliotou}, {Leloudas}, {Levan}, {Mazzali}, \&
  {Pian}}]{Watson2019}
{Watson}, D., {Hansen}, C.~J., {Selsing}, J., {et~al.} 2019, \nat, 574, 497

\bibitem[{{Xie} {et~al.}(2024){Xie}, {Shi}, {Yan}, {Chen}, {Allende Prieto},
  {Beers}, {Liu}, {Li}, {Ding}, {Tang}, {Zhang}, \& {Xie}}]{Xie_2024}
{Xie}, X.-J., {Shi}, J., {Yan}, H.-L., {et~al.} 2024, \apjl, 970, L30

\bibitem[{{Xing} {et~al.}(2024){Xing}, {Zhao}, {Aoki}, {Li}, {Zhao}, {Matsuno},
  \& {Suda}}]{Xing_2024}
{Xing}, Q., {Zhao}, G., {Aoki}, W., {et~al.} 2024, \apj, 965, 79

\bibitem[{{Yong} {et~al.}(2021){Yong}, {Kobayashi}, {Da Costa}, {Bessell},
  {Chiti}, {Frebel}, {Lind}, {Mackey}, {Nordlander}, {Asplund}, {Casey},
  {Marino}, {Murphy}, \& {Schmidt}}]{Yong_2021}
{Yong}, D., {Kobayashi}, C., {Da Costa}, G.~S., {et~al.} 2021, \nat, 595, 223

\bibitem[{{Yuan} {et~al.}(2020){Yuan}, {Myeong}, {Beers}, {Evans}, {Lee},
  {Banerjee}, {Gudin}, {Hattori}, {Li}, {Matsuno}, {Placco}, {Smith},
  {Whitten}, \& {Zhao}}]{Yuan_2020}
{Yuan}, Z., {Myeong}, G.~C., {Beers}, T.~C., {et~al.} 2020, \apj, 891, 39

\end{thebibliography}
\bibliographystyle{aa}

\begin{appendix}
\onecolumn

\section{Stellar parameters, abundances, and kinematic parameters} \label{appendix tables}
Stellar atmospheric parameters, elemental abundances, and kinematic parameters of the primary sample and the literature sample. 

\begin{table*}[htp]
\centering
\caption{\small {Stellar atmospheric parameters and elemental abundances of the stellar sample analyzed in this work}.} \label{Table_results_abundances} 
\resizebox{1.0\textwidth}{!}{ 
\begin{tabular}{lHccccccccccc} 
\hline                       
ID          & Gaia DR3 ID     & T$_{\rm eff}$  & log g  & v$_{\rm mic}$      & [Fe/H]     & [Mg/Fe]$^{\dagger}$  & $ \rm \log \epsilon (Eu)_{LTE}$  & $ \rm \log \epsilon (Th)_{LTE}$ & $ \rm \log \epsilon (Th/Eu)_{LTE}$ 
& $\rm \Delta_{NLTE}$(Eu)    & $ \rm \log \epsilon (Th/Eu)_{NLTE}^{\ast}$  & Pop    \\  
NS24        &                 & (K)       & (dex)   & (km s$^{-1}$)  & (dex)      & (dex)                 & (dex)           & (dex)           & (dex)              
& (dex)                    & (dex)                          &   \\
\hline
 G 112$-$43   	&	 3085891537839267328 	&	6209	&	4.02	&	1.17	&	$-$1.27	&	0.34	&	$-$0.46	&	$-$0.45	&	0.01	&	0.19	&	$-$0.17	&	GSE	\\
 G 13$-$38    	&	 3700341138433848832 	&	5263	&	4.54	&	0.90	&	$-$0.88	&	0.35	&	$-$0.09	&	$-$0.40	&	$-$0.32	&	0.14	&	$-$0.45	&	In situ halo	\\
 G 66$-$22    	&	 1159108770069883136 	&	5297	&	4.46	&	0.78	&	$-$0.88	&	0.02	&	0.01	&	$-$0.43	&	$-$0.44	&	0.10	&	$-$0.53	&	GSE	\\
 G 57$-$07    	&	 3916257626962267136 	&	5755	&	4.33	&	0.99	&	$-$0.48	&	0.29	&	0.23	&	$-$0.13	&	$-$0.36	&	0.15	&	$-$0.50	&	In situ halo	\\
 G 18$-$28    	&	 2728314787225296000 	&	5443	&	4.49	&	0.88	&	$-$0.85	&	0.31	&	$-$0.07	&	$-$0.40	&	$-$0.33	&	0.11	&	$-$0.43	&	In situ halo	\\
 G 170$-$56   	&	 4550762289589343488 	&	6112	&	4.11	&	1.39	&	$-$0.94	&	0.23	&	$-$0.10	&	$-$0.42	&	$-$0.33	&	0.17	&	$-$0.49	&	GSE	\\
 G 159$-$50   	&	 2494759795723713408 	&	5713	&	4.44	&	1.03	&	$-$0.94	&	0.36	&	$-$0.14	&	$-$0.57	&	$-$0.43	&	0.15	&	$-$0.57	&	In situ halo	\\
 G 31$-$55    	&	 2540830947835865728 	&	5731	&	4.35	&	1.26	&	$-$1.12	&	0.28	&	$-$0.40	&	$-$0.68	&	$-$0.28	&	0.16	&	$-$0.43	&	In situ halo	\\
 G 82$-$05    	&	 3202470247468181632 	&	5338	&	4.51	&	0.80	&	$-$0.78	&	0.00	&	0.16	&	$-$0.30	&	$-$0.46	&	0.10	&	$-$0.55	&	GSE	\\
 G 20$-$15    	&	 4165370682239910144 	&	6162	&	4.32	&	1.50	&	$-$1.50	&	0.27	&	$-$0.37	&	$-$0.90	&	$-$0.53	&	0.11	&	$-$0.63	&	GSE	\\
 CD$-$33 3337 	&	 5579891291853909504 	&	6112	&	3.86	&	1.56	&	$-$1.37	&	0.44	&	$-$0.50	&	$-$0.87	&	$-$0.37	&	0.12	&	$-$0.48	&	TD	\\
 CD$-$43 6810 	&	 5388634989413677440 	&	6059	&	4.32	&	1.24	&	$-$0.44	&	0.35	&	0.28	&	$-$0.01	&	$-$0.29	&	0.10	&	$-$0.38	&	In situ halo	\\
 CD$-$57 1633 	&	 5486881507314450816 	&	5981	&	4.29	&	1.08	&	$-$0.91	&	0.06	&	$-$0.08	&	$-$0.45	&	$-$0.38	&	0.10	&	$-$0.47	&	GSE	\\
 G 24$-$25    	&	 4228176122142169600 	&	5828	&	3.86	&	1.20	&	$-$1.40	&	0.34	&	$-$0.26	&	$-$0.55	&	$-$0.29	&	0.13	&	$-$0.41	&	GSE	\\
 G 63$-$26    	&	 3939515154043771392 	&	6175	&	4.17	&	1.65	&	$-$1.58	&	0.44	&	$-$0.64	&	$-$1.30	&	$-$0.66	&	0.16	&	$-$0.81	&	In situ halo	\\
 G 75$-$31    	&	 2502689198705422848 	&	6135	&	4.02	&	1.28	&	$-$1.04	&	0.24	&	$-$0.15	&	$-$0.50	&	$-$0.35	&	0.19	&	$-$0.53	&	GSE	\\
 HD 103723  	&	 3494135429225414016 	&	6050	&	4.20	&	1.11	&	$-$0.81	&	0.17	&	0.03	&	$-$0.30	&	$-$0.33	&	0.11	&	$-$0.43	&	GSE	\\
 HD 105004  	&	 3486853672952159872 	&	5852	&	4.35	&	1.09	&	$-$0.83	&	0.16	&	$-$0.07	&	$-$0.50	&	$-$0.43	&	0.15	&	$-$0.57	&	GSE	\\
 HD 106516  	&	 3581001280225616128 	&	6327	&	4.43	&	1.18	&	$-$0.69	&	0.46	&	0.09	&	$-$0.30	&	$-$0.39	&	0.10	&	$-$0.48	&	TD	\\
 HD 111980  	&	 3510294882898890880 	&	5878	&	3.98	&	1.39	&	$-$1.09	&	0.43	&	$-$0.31	&	$-$0.69	&	$-$0.38	&	0.10	&	$-$0.47	&	In situ halo	\\
 HD 113679  	&	 6141991322784963072 	&	5761	&	4.05	&	1.37	&	$-$0.66	&	0.40	&	0.11	&	$-$0.33	&	$-$0.44	&	0.13	&	$-$0.56	&	In situ halo	\\
 HD 120559  	&	 5871764831114179968 	&	5486	&	4.58	&	1.05	&	$-$0.91	&	0.30	&	$-$0.09	&	$-$0.45	&	$-$0.36	&	0.11	&	$-$0.46	&	TD	\\
 HD 121004  	&	 6095425184286680064 	&	5755	&	4.43	&	1.16	&	$-$0.71	&	0.36	&	0.14	&	$-$0.35	&	$-$0.49	&	0.10	&	$-$0.58	&	In situ halo	\\
 HD 132475  	&	 6232043867720079616 	&	5750	&	3.77	&	1.37	&	$-$1.51	&	0.46	&	$-$0.675	&	$-$1.05	&	$-$0.38	&	0.17	&	$-$0.54	&	In situ halo	\\
 HD 148816  	&	 4433711099194059904 	&	5923	&	4.17	&	1.33	&	$-$0.74	&	0.38	&	0.03	&	$-$0.30	&	$-$0.33	&	0.14	&	$-$0.46	&	In situ halo	\\
 HD 159482  	&	 4485972467412021376 	&	5829	&	4.37	&	1.21	&	$-$0.74	&	0.33	&	0.05	&	$-$0.35	&	$-$0.4	&	0.13	&	$-$0.52	&	In situ halo	\\
 HD 163810  	&	 4150413750729448320 	&	5592	&	4.61	&	1.17	&	$-$1.22	&	0.11	&	$-$0.225	&	$-$0.51	&	$-$0.29	&	0.15	&	$-$0.43	&	GSE	\\
 HD 177095  	&	 4082356695359119616 	&	5349	&	4.39	&	0.90	&	$-$0.74	&	0.38	&	0.02	&	$-$0.50	&	$-$0.52	&	0.13	&	$-$0.64	&	In situ halo	\\
 HD 179626  	&	 4263841431819084032 	&	5925	&	4.14	&	1.49	&	$-$1.06	&	0.41	&	$-$0.32	&	$-$0.59	&	$-$0.27	&	0.16	&	$-$0.42	&	In situ halo	\\
 HD 189558  	&	 4189078626829509248 	&	5707	&	3.83	&	1.29	&	$-$1.14	&	0.41	&	$-$0.33	&	$-$0.67	&	$-$0.34	&	0.11	&	$-$0.44	&	TD	\\
 HD 193901  	&	 6859076107589173120 	&	5729	&	4.43	&	1.31	&	$-$1.11	&	0.10	&	$-$0.18	&	$-$0.62	&	$-$0.44	&	0.12	&	$-$0.55	&	GSE	\\
 HD 194598  	&	 1752459532807704704 	&	6018	&	4.34	&	1.40	&	$-$1.11	&	0.21	&	$-$0.25	&	$-$0.67	&	$-$0.42	&	0.14	&	$-$0.55	&	GSE	\\
 HD 199289  	&	 6481426917515214464 	&	5915	&	4.30	&	1.21	&	$-$1.05	&	0.39	&	$-$0.28	&	$-$0.59	&	$-$0.32	&	0.11	&	$-$0.42	&	TD	\\
 HD 205650  	&	 6810862213470371072 	&	5793	&	4.35	&	1.17	&	$-$1.19	&	0.36	&	$-$0.36	&	$-$0.82	&	$-$0.46	&	0.12	&	$-$0.57	&	TD	\\
 HD 219617  	&	 4263841431819084032 	&	5983	&	4.28	&	1.42	&	$-$1.46	&	0.29	&	$-$0.56	&	$-$0.95	&	$-$0.39	&	0.11	&	$-$0.49	&	In situ halo	\\
 HD 22879   	&	 3250489115709122560 	&	5859	&	4.29	&	1.20	&	$-$0.86	&	0.36	&	$-$0.05	&	$-$0.45	&	$-$0.41	&	0.10	&	$-$0.50	&	TD	\\
 HD 284248  	&	 52624073910834176   	&	6271	&	4.21	&	1.51	&	$-$1.59	&	0.31	&	$-$0.65	&	$-$0.67	&	$-$0.02	&	0.18	&	$-$0.19	&	GSE	\\
 CD$-$51 4628 	&	 5358179685572496384 	&	6296	&	4.29	&	1.31	&	$-$1.32	&	0.22	&	$-$0.30	&	$-$0.75	&	$-$0.46	&	0.17	&	$-$0.62	&	GSE	\\
 HD 3567    	&	 2427069874188580480 	&	6180	&	4.01	&	1.40	&	$-$1.17	&	0.25	&	$-$0.07	&	$-$0.42	&	$-$0.35	&	0.18	&	$-$0.52	&	GSE	\\
 HD 51754   	&	 3112836479029139200 	&	5857	&	4.35	&	1.30	&	$-$0.58	&	0.35	&	0.20	&	$-$0.12	&	$-$0.32	&	0.10	&	$-$0.41	&	In situ halo	\\
 HD 59392   	&	 5586241315104190848 	&	6137	&	3.88	&	1.73	&	$-$1.62	&	0.40	&	$-$0.47	&	$-$1.00	&	$-$0.53	&	0.18	&	$-$0.70	&	GSE	\\
 HD 76932   	&	 5730371484020807808 	&	5977	&	4.17	&	1.30	&	$-$0.87	&	0.42	&	0.02	&	$-$0.55	&	$-$0.57	&	0.10	&	$-$0.66	&	TD	\\
 G 16$-$20    	&	 4412702764879270656 	&	5625	&	3.64	&	1.50	&	$-$1.42	&	0.22	&	$-$0.48	&	$-$0.82	&	$-$0.34	&	0.13	&	$-$0.46	&	GSE	\\
 CD$-$61 282 	&	 4715919175280799616 	&	5869	&	4.34	&	1.19	&	$-$1.25	&	0.21	&	$-$0.27	&	$-$0.77	&	$-$0.50	&	0.12	&	$-$0.61	&	GSE	\\
 \hline
             &                           &           &           &           &           & Archival data  &      &           &           &           &           &\\
\hline
 G 87$-$13     	&	 940306024863622400  	&	6217	&	4.11	&	1.42	&	$-$1.11	&	0.23	&	$-$0.10	&	$-$0.17	&	$-$0.07	&	0.15	&	$-$0.21	&	GSE	\\
 G 98$-$53     	&	 3440045641893430912 	&	5954	&	4.26	&	1.20	&	$-$0.89	&	0.25	&	$-$0.03	&	$-$0.35	&	$-$0.32	&	0.19	&	$-$0.50	&	GSE	\\
 G 161$-$73    	&	 3822808140853060224 	&	6108	&	3.99	&	1.26	&	$-$1.01	&	0.20	&	$-$0.20	&	$-$0.46	&	$-$0.26	&	0.17	&	$-$0.42	&	GSE	\\
 G 53$-$41     	&	 3855475490307708032 	&	5956	&	4.34	&	1.35	&	$-$1.21	&	0.22	&	$-$0.25	&	$-$0.52	&	$-$0.27	&	0.18	&	$-$0.44	&	GSE	\\
 G 188$-$22  	&	  1800518563283501440 	&	6116	&	4.20	&	1.42	&	$-$1.33	&	0.48	&	$-$0.54	&	$-$0.72	&	$-$0.18	&	0.20	&	$-$0.37	&	In situ halo	\\
 HD 25704  	&	  4682867168555426944 	&	5974	&	4.30	&	1.30	&	$-$0.86	&	0.36	&	$-$0.02	&	$-$0.45	&	$-$0.43	&	0.15	&	$-$0.57	&	TD	\\
\hline
\end{tabular} 
} \\
\vspace{0.05in}
Note: $^{\dagger}$ LTE values, NLTE corrections are negligible for the parameter range of the sample. \\
$^{\ast}$ $\rm \Delta_{NLTE}$(Th) $\sim$ 0.01 dex for the parameter range of the sample. \\ 
\end{table*} 

\begin{table*}[htp]
\centering
\caption{\small {Data for the literature sample}.} \label{Table_literature_sample} 
\resizebox{1.0\textwidth}{!}{
\begin{tabular}{lHccccccccccc} 
\hline                       
ID          & Gaia DR3 ID   & T$_{\rm eff}$    & log g     & [Fe/H]     & [Mg/Fe]$_{\rm NLTE}^{\dagger}$  & $ \rm \log \epsilon (Eu)_{LTE}$  & $\rm \log \epsilon (Th)_{LTE}$ & $ \rm \log \epsilon (Th/Eu)_{LTE}$   
& $\Delta_{\rm NLTE}$(Eu)     & $\Delta_{\rm NLTE}$(Th)    & $ \rm \log \epsilon (Th/Eu)_{NLTE}$      & Ref.    \\  
            &               & (K)          & (dex)     & (dex)      & (dex)                         & (dex)            & (dex)           & (dex)                
& (dex)                     & (dex)                    & (dex)                    &         \\
\hline
HE 2327$-$5642	&	6495850379767072128	&	5050	&	2.34	&	$-$2.78	&	0.14	&	$-$1.29	&	$-$1.67	&	$-$0.38	&	0.16	&	0.05	&	$-$0.49	&	1	\\
2MASS J09544277+5246414	&	828438619475671936	&	4410	&	0.61	&	$-$2.96	&	0.69	&	$-$1.16	&	$-$1.79	&	$-$0.63	&	0.14	&	0.09	&	$-$0.68	&	2	\\
HE 2252$-$4225	&	6542909786672462336	&	4710	&	1.65	&	$-$2.63	&	0.20	&	$-$1.3	&	$-$1.63	&	$-$0.33	&	0.14	&	0.08	&	$-$0.39	&	3	\\
BPS CS 29491$-$069	&	6600971319243174144	&	5300	&	2.80	&	$-$2.51	&	0.37	&	$-$1.03	&	$-$1.43	&	$-$0.40	&	0.11	&	0.05	&	$-$0.46	&	4	\\
HE 1219$-$0312	&	3693297117190538368	&	5060	&	2.30	&	$-$2.97	&	0.50	&	$-$0.98	&	$-$1.14	&	$-$0.16	&	0.11	&	0.06	&	$-$0.21	&	5	\\
BPS CS 30315$-$029	&	2332019357414106624	&	4570	&	0.99	&	$-$3.40	&	0.71	&	$-$2.24	&	$-$2.45	&	$-$0.21	&	0.10	&	0.12	&	$-$0.19	&	6	\\
LAMOST J080452.84+574019.6	&	1081668723016932736	&	4585	&	1.41	&	$-$2.38	&	0.42	&	$-$1.06	&	$-$1.28	&	$-$0.22	&	0.06	&	0.07	&	$-$0.21	&	7	\\
BPS CS 22953$-$003	&	4710594687144052096	&	4860	&	1.65	&	$-$3.00	&	0.42	&	$-$1.69	&	$-$1.92	&	$-$0.23	&	0.06	&	0.09	&	$-$0.20	&	8	\\
BD$-$15 5781	&	6887351938881045504	&	4550	&	0.70	&	$-$2.87	&	0.54	&	$-$2.28	&	$-$2.73	&	$-$0.45	&	0.27	&	0.16	&	$-$0.56	&	8	\\
BD+10 2495	&	3734341989334114304	&	4890	&	1.85	&	$-$2.14	&	0.41	&	$-$1.34	&	$-$1.77	&	$-$0.43	&	0.12	&	0.08	&	$-$0.47	&	8	\\
HD 26297	&	5097726306014470016	&	4400	&	1.10	&	$-$1.72	&	0.23	&	$-$1.30	&	$-$1.73	&	$-$0.43	&	0.12	&	0.11	&	$-$0.44	&	8	\\
HD 45282	&	3125113282265023488	&	5230	&	2.90	&	$-$1.73	&	0.47	&	$-$0.70	&	$-$1.08	&	$-$0.38	&	0.10	&	0.04	&	$-$0.44	&	8	\\
HD 175305	&	2268048503896398720	&	4920	&	2.30	&	$-$1.56	&	0.30	&	$-$0.64	&	$-$1.02	&	$-$0.38	&	0.10	&	0.06	&	$-$0.42	&	8	\\
2MASS J14244532$-$2542471	&	6271613367058424064	&	4762	&	1.58	&	$-$3.39	&	0.83	&	$-$1.25	&	$-$1.21	&	0.04	&	0.10	&	0.09	&	0.03	&	9	\\
BPS CS 29497$-$004	&	2322729725405593728	&	5090	&	2.40	&	$-$2.64	&	0.38	&	$-$0.45	&	$-$0.96	&	$-$0.51	&	0.11	&	0.05	&	$-$0.57	&	10	\\
BD+17 3248	&	4553184509407224576	&	5200	&	1.80	&	$-$2.00	&	0.49	&	$-$0.67	&	$-$1.18	&	$-$0.51	&	0.11	&	0.08	&	$-$0.54	&	11	\\
BPS CS 22892$-$052	&	6826025986350385280	&	4800	&	1.50	&	$-$3.10	&	0.44	&	$-$0.95	&	$-$1.57	&	$-$0.62	&	0.11	&	0.09	&	$-$0.64	&	12	\\
HE 1523$-$0901	&	6317828550897175936	&	4630	&	1.00	&	$-$2.95	&	0.63	&	$-$0.62	&	$-$1.20	&	$-$0.58	&	0.11	&	0.13	&	$-$0.56	&	13	\\
HD 221170	&	2869759781250083200	&	4510	&	1.00	&	$-$2.19	&	0.46	&	$-$0.86	&	$-$1.46	&	$-$0.60	&	0.10	&	0.12	&	$-$0.58	&	14	\\
BPS CS 31082$-$001	&	2451773941958712192	&	4825	&	1.50	&	$-$2.90	&	0.50	&	$-$0.76	&	$-$0.98	&	$-$0.22	&	0.06	&	0.07	&	$-$0.21	&	15	\\
HD 108577	&	3907872613852996864	&	4900	&	1.10	&	$-$2.20	&	0.40	&	$-$1.49	&	$-$2.00	&	$-$0.51	&	0.16	&	0.16	&	$-$0.51	&	16	\\
HD 115444	&	1474455748663044736	&	4500	&	0.70	&	$-$3.00	&	0.62	&	$-$1.82	&	$-$2.37	&	$-$0.55	&	0.23	&	0.15	&	$-$0.63	&	16	\\
BD+08 2856	&	1175731289937614080	&	4550	&	0.70	&	$-$2.00	&	0.59	&	$-$1.15	&	$-$1.65	&	$-$0.50	&	0.11	&	0.15	&	$-$0.46	&	16	\\
HD 6268	&	5033960575837265920	&	4600	&	1.00	&	$-$2.63	&	0.57	&	$-$1.56	&	$-$1.93	&	$-$0.37	&	0.19	&	0.13	&	$-$0.43	&	17	\\
HD 110184	&	3710786979933623424	&	4240	&	0.30	&	$-$2.52	&	0.23	&	$-$1.91	&	$-$2.50	&	$-$0.59	&	0.22	&	0.20	&	$-$0.61	&	17	\\
HD 186478	&	6871465473332110464	&	4720	&	1.60	&	$-$2.5	&	0.50	&	$-$1.34	&	$-$1.85	&	$-$0.51	&	0.13	&	0.08	&	$-$0.56	&	17	\\
BPS CS 30306$-$132	&	1163455203958063872	&	5110	&	2.50	&	$-$2.42	&	0.43	&	$-$1.02	&	$-$1.12	&	$-$0.10	&	0.10	&	0.04	&	$-$0.16	&	17	\\
BPS CS 31078$-$018	&	7189878332862720	&	5257	&	2.75	&	$-$2.85	&	0.55	&	$-$1.17	&	$-$1.35	&	$-$0.18	&	0.10	&	0.05	&	$-$0.23	&	18	\\
RAVE J005327.8$-$025317	&	2529476566134035200	&	4370	&	0.56	&	$-$2.16	&	0.51	&	$-$1.25	&	$-$2.14	&	$-$0.89	&	0.13	&	0.17	&	$-$0.85	&	19	\\
RAVE J024620.1$-$151842	&	5157037223094770944	&	4948	&	1.93	&	$-$2.45	&	0.26	&	$-$0.64	&	$-$1.40	&	$-$0.76	&	0.11	&	0.08	&	$-$0.79	&	19	\\
RAVE J031337.3$-$102056	&	5165907189554147584	&	4757	&	1.47	&	$-$2.05	&	0.42	&	$-$1.11	&	$-$1.30	&	$-$0.19	&	0.11	&	0.09	&	$-$0.21	&	19	\\
RAVE J034346.3$-$092417	&	5164146429057330816	&	4380	&	0.46	&	$-$1.92	&	0.52	&	$-$1.02	&	$-$1.52	&	$-$0.50	&	0.11	&	0.18	&	$-$0.43	&	19	\\
RAVE J141015.9$-$034355	&	3646021450089606912	&	4931	&	2.00	&	$-$2.06	&	0.40	&	$-$0.87	&	$-$1.59	&	$-$0.72	&	0.10	&	0.07	&	$-$0.75	&	19	\\
2MASS J15211026$-$0607566	&	6322811163932641152	&	4707	&	1.33	&	$-$2.00	&	0.39	&	$-$0.55	&	$-$1.36	&	$-$0.81	&	0.07	&	0.12	&	$-$0.76	&	19	\\
RAVE J153830.9$-$180424	&	6255142030043852928	&	4752	&	1.63	&	$-$2.09	&	0.43	&	$-$0.30	&	$-$0.95	&	$-$0.65	&	0.06	&	0.08	&	$-$0.63	&	19	\\
2MASS J22562536$-$0719562	&	2610720889058877696	&	4558	&	1.20	&	$-$2.26	&	0.48	&	$-$0.64	&	$-$1.66	&	$-$1.02	&	0.08	&	0.10	&	$-$1.00	&	19	\\
HD 20	&	2333756864959936000	&	5220	&	1.89	&	$-$1.78	&	0.60	&	$-$0.35	&	$-$0.85	&	$-$0.50	&	0.10	&	0.07	&	$-$0.53	&	20	\\
SMSS J200322.54$-$114203.3	&	4190620966764303488	&	5175	&	2.44	&	$-$3.50	&	0.32	&	$-$1.21	&	$-$1.31	&	$-$0.10	&	0.12	&	0.05	&	$-$0.17	&	21	\\
HD 222925	&	6487799171512458624	&	5636	&	2.54	&	$-$1.47	&	0.44	&	0.38	&	$-$0.06	&	$-$0.44	&	0.08	&	0.06	&	$-$0.46	&	22	\\
2MASS J22132050$-$5137385	&	6511387028823212544	&	5509	&	2.28	&	$-$2.20	&	0.72	&	0.77	&	0.22	&	$-$0.55	&	0.11	&	0.05	&	$-$0.61	&	23	\\
BD+29 2356	&	1464266230811360256	&	4760	&	1.30	&	$-$1.55	&	0.38	&	$-$0.69	&	$-$1.07	&	$-$0.38	&	0.11	&	0.12	&	$-$0.37	&	24	\\
BD+30 2611	&	1275876252107941888	&	4330	&	0.60	&	$-$1.45	&	0.28	&	$-$0.36	&	$-$0.98	&	$-$0.62	&	0.09	&	0.16	&	$-$0.55	&	24	\\
CD$-$68 1881	&	5810465477272078208	&	4780	&	1.50	&	$-$1.63	&	0.41	&	$-$0.63	&	$-$1.19	&	$-$0.56	&	0.10	&	0.09	&	$-$0.57	&	25	\\
HE 0050$-$0918	&	2474070010546861056	&	5310	&	2.00	&	$-$1.37	&	0.37	&	$-$0.14	&	$-$0.71	&	$-$0.57	&	0.13	&	0.11	&	$-$0.59	&	25	\\
HE 0324$-$0122	&	3267948604442696448	&	5145	&	2.45	&	$-$2.41	&	0.51	&	$-$0.77	&	$-$1.32	&	$-$0.55	&	0.10	&	0.05	&	$-$0.60	&	25	\\
CD$-$37 2066	&	4820462183959704448	&	4875	&	1.75	&	$-$2.02	&	0.51	&	$-$1.17	&	$-$1.75	&	$-$0.58	&	0.11	&	0.09	&	$-$0.60	&	25	\\
CD$-$22 7696	&	5660878702718068864	&	4580	&	0.95	&	$-$2.48	&	0.35	&	$-$1.69	&	$-$2.26	&	$-$0.57	&	0.19	&	0.12	&	$-$0.64	&	25	\\
HE 0007$-$1752	&	2366585258508216320	&	5215	&	2.65	&	$-$2.36	&	$-$0.19	&	$-$0.6	&	$-$1.16	&	$-$0.56	&	0.10	&	0.04	&	$-$0.62	&	25	\\
TYC 3431$-$689$-$1	&	1019002298308911360	&	4850	&	1.40	&	$-$2.05	&	0.51	&	$-$0.73	&	$-$1.34	&	$-$0.61	&	0.09	&	0.09	&	$-$0.61	&	26	\\
TYC 1191$-$918$-$1	&	2795305423661094016	&	4500	&	1.20	&	$-$2.15	&	0.71	&	$-$0.83	&	$-$1.24	&	$-$0.41	&	0.09	&	0.09	&	$-$0.41	&	26	\\
LAMOST J112456.61+453531.3	&	785388513000524160	&	5180	&	2.70	&	$-$1.27	&	$-$0.20	&	0.38	&	$-$0.19	&	$-$0.57	&	0.07	&	0.05	&	$-$0.59	&	27	\\
BPS CS 29529$-$0089	&	4678520146255821952	&	5014	&	2.16	&	$-$2.46	&	0.43	&	$-$0.15	&	$-$0.55	&	$-$0.40	&	0.10	&	0.06	&	$-$0.44	&	28	\\
HD 10837	&	2467368555894065664	&	5100	&	2.67	&	$-$2.37	&	0.45	&	$-$1.37	&	$-$1.99	&	$-$0.62	&	0.11	&	0.04	&	$-$0.69	&	29	\\
2MASS J14325334$-$4125494	&	6103374133661596416	&	4900	&	1.70	&	$-$2.97	&	0.66	&	$-$1.01	&	$-$1.49	&	$-$0.48	&	0.11	&	0.09	&	$-$0.50	&	30	\\
2MASS J20050670$-$3057445	&	6749102714060650880	&	4430	&	0.60	&	$-$3.03	&	0.58	&	$-$1.57	&	$-$2.18	&	$-$0.61	&	0.19	&	0.17	&	$-$0.63	&	30	\\
2MASS J15213995$-$3538094	&	6199959495651403776	&	5850	&	2.10	&	$-$2.80	&	0.55	&	$-$0.04	&	$-$0.60	&	$-$0.56	&	0.18	&	0.14	&	$-$0.60	&	31	\\
2MASS J05383296$-$5904280	&	4759914999195512064	&	5819	&	2.09	&	$-$2.55	&	0.66	&	$-$0.78	&	$-$1.16	&	$-$0.38	&	0.16	&	0.14	&	$-$0.40	&	32	\\
RAVE J094921.8$-$161722	&	5684624825278191744	&	4855	&	1.60	&	$-$2.22	&	0.54	&	$-$1.09	&	$-$1.70	&	$-$0.61	&	0.11	&	0.09	&	$-$0.63	&	33	\\
LAMOST J020632.21+494127.9	&	358364512036089472	&	4130	&	1.52	&	$-$0.54	&	0.17	&	1.18	&	0.48	&	$-$0.70	&	0.12	&	0.04	&	$-$0.78	&	34	\\
RAVE J203843.2$-$002333	&	4226556850750887168	&	4630	&	1.20	&	$-$2.91	&	0.48	&	$-$0.75	&	$-$1.24	&	$-$0.49	&	0.10	&	0.11	&	$-$0.48	&	35	\\
\hline
\end{tabular} 
} \\
\vspace{0.05in}
\noindent Note: $^{\dagger}$The NLTE corrections are calculated using the NLiTE tool by adopting the grid of \cite{Lind_2022}. \\
\vspace{0.05in}
\noindent References: 1. \cite{Mashonkina_2010}, 2. \cite{Shah_2023}, 3. \cite{Mashonkina_2014}, 4. \cite{Hayek_2009}, 5. \cite{Roederer_2009}, 
6. \cite{SiqueiraMello_2014}, 7. \cite{Lin2025}, 8. \cite{Roederer_2014}, 9. \cite{Placco_2023}, 10. \cite{Christlieb_2004}, 11. \cite{Cowan_2002}, 
12. \cite{Sneden2003}, 13. \cite{Frebel_2007}, 14. \cite{Ivans_2006}, 15. \cite{Hill2002}, 16. \cite{Johnson_2001},	17. \cite{Honda_2004}, 18. \cite{Lai_2008}, 
19. \cite{Sakari_2018}, 20. \cite{Hanke_2020}, 21. \cite{Yong_2021}, 22. \cite{Roederer_2018}, 23. \cite{Roederer_2024}, 24. \cite{Roederer_2010}, 
25. \cite{Gull_2021}, 26. \cite{Saraf_2023}, 27. \cite{Xing_2024}, 28. \cite{dasilva_2025}, 29. \cite{Roederer_2012}, 30. \cite{Cain_2018}, 31. \cite{Cain_2020}, 32. \cite{Hansen_2025}, 33. \cite{Gull_2018}, 34.\cite{Xie_2024}, 35. \cite{Placco_2017} 
\end{table*}

\longtab[3]{
\begin{landscape}
\begin{longtable}{lccccccccc}
\caption{Kinematics of the combined sample} \label{Table_kinematics} \\
\hline
\hline
ID      & Gaia DR3 ID     & Energy                 & L$_{z}$             & Eccentricity   & z$\rm_{max}$          & U$_{\rm LSR}$       & V$_{\rm LSR}$      & W$_{\rm LSR}$      & V$_{\rm spa}$  \\ 
        &                 & (km$^{2}$ s$^{-2}$)    & (km s$^{-1}$ kpc)   &                & (kpc)   & (km s$^{-1}$)   & (km s$^{-1}$)  & (km s$^{-1}$)  & (km s$^{-1}$) \\ 
\hline
\endfirsthead
\caption{continued}\\
\hline
ID      & Gaia DR3 ID     & Energy                 & L$_{z}$             & Eccentricity   & z$\rm_{max}$          & U$_{\rm LSR}$       & V$_{\rm LSR}$      & W$_{\rm LSR}$      & V$_{\rm spa}$  \\ 
        &                 & (km$^{2}$ s$^{-2}$)    & (km s$^{-1}$ kpc)   &                & (kpc)   & (km s$^{-1}$)   & (km s$^{-1}$)  & (km s$^{-1}$)  & (km s$^{-1}$) \\ 
\hline
\endhead
\hline
\endfoot
\hline
\endlastfoot
HE 2327-5642	&	6495850379767072128	&	-26994.70	&	-699.00	&	0.66	&	14.65	&	-65.32	&	-354.25	&	-235.14	&	430.18	\\
2MASS J09544277+5246414	&	828438619475671936	&	-23790.37	&	-1710.38	&	0.40	&	9.89	&	-146.20	&	-385.83	&	-109.34	&	426.84	\\
HE 2252-4225	&	6542909786672462336	&	7169.84	&	262.35	&	0.81	&	50.20	&	-116.23	&	-162.99	&	323.73	&	380.63	\\
BPS CS 29491-069	&	6600971319243174144	&	3655.79	&	-1440.49	&	0.73	&	36.11	&	-167.99	&	-422.83	&	276.14	&	532.22	\\
HE 1219-0312	&	3693297117190538368	&	--	&	--	&	0.12	&	7.54	&	--	&	--	&	--	&	--	\\
BPS CS 30315-029	&	2332019357414106624	&	-22074.91	&	-195.89	&	0.89	&	18.36	&	143.10	&	-287.39	&	175.00	&	365.64	\\
LAMOST J080452.84+574019.6	&	1081668723016932736	&	-16409.20	&	43.28	&	0.99	&	11.54	&	245.79	&	-237.30	&	-104.85	&	357.38	\\
BPS CS 22953-003	&	4710594687144052096	&	-24918.04	&	-748.28	&	0.77	&	12.17	&	278.02	&	-249.80	&	3.43	&	373.77	\\
BD-15 5781	&	6887351938881045504	&	-70825.33	&	355.43	&	0.76	&	1.01	&	-3.72	&	-167.33	&	-25.39	&	169.29	\\
BD+10 2495	&	3734341989334114304	&	--	&	--	&	0.08	&	1.28	&	--	&	--	&	--	&	--	\\
HD 26297	&	5097726306014470016	&	-51048.37	&	1119.67	&	0.38	&	2.71	&	-34.16	&	-87.88	&	81.59	&	124.69	\\
HD 45282	&	3125113282265023488	&	-35306.81	&	438.73	&	0.89	&	1.05	&	-238.06	&	-167.54	&	-34.89	&	293.18	\\
HD 175305	&	2268048503896398720	&	-14917.68	&	1179.85	&	0.48	&	18.56	&	-53.41	&	-72.34	&	-278.96	&	293.10	\\
2MASS J14244532-2542471	&	6271613367058424064	&	-46579.45	&	745.91	&	0.24	&	7.04	&	-83.96	&	-18.82	&	56.14	&	102.74	\\
BPS CS 29497-004	&	2322729725405593728	&	-30973.84	&	1170.19	&	0.61	&	7.58	&	-138.86	&	-64.28	&	-111.44	&	189.30	\\
BD+17 3248	&	4553184509407224576	&	-66644.32	&	-29.25	&	0.98	&	1.62	&	-39.67	&	-221.48	&	63.26	&	233.73	\\
BPS CS 22892-052	&	6826025986350385280	&	-27857.93	&	-167.49	&	0.93	&	14.68	&	-169.96	&	-186.18	&	-212.62	&	329.78	\\
HE 1523-0901	&	6317828550897175936	&	-33972.75	&	-1278.19	&	0.44	&	7.62	&	-130.40	&	-443.88	&	-134.26	&	481.72	\\
HD 221170	&	2869759781250083200	&	-51668.09	&	825.96	&	0.65	&	0.92	&	134.64	&	-125.77	&	-33.27	&	187.23	\\
BPS CS 31082-001	&	2451773941958712192	&	-22842.87	&	-1288.42	&	0.39	&	13.74	&	125.51	&	-374.03	&	-191.06	&	438.36	\\
HD 108577	&	3907872613852996864	&	-31510.47	&	-438.70	&	0.65	&	14.14	&	-181.05	&	-280.57	&	-177.54	&	378.18	\\
HD 115444	&	1474455748663044736	&	-50203.79	&	430.82	&	0.83	&	1.55	&	166.65	&	-169.43	&	16.20	&	238.20	\\
BD+08 2856	&	1175731289937614080	&	-40369.55	&	1066.61	&	0.62	&	2.69	&	184.92	&	-69.36	&	-17.41	&	198.26	\\
HD 6268	&	5033960575837265920	&	-35195.92	&	1562.19	&	0.48	&	1.95	&	147.53	&	-24.87	&	-35.18	&	153.69	\\
HD 110184	&	3710786979933623424	&	-55578.49	&	611.97	&	0.54	&	5.09	&	-25.29	&	-143.01	&	112.97	&	183.99	\\
HD 186478	&	6871465473332110464	&	-51469.25	&	-660.47	&	0.72	&	1.72	&	163.41	&	-318.58	&	-56.00	&	362.39	\\
BPS CS 30306-132	&	1163455203958063872	&	-52110.67	&	883.08	&	0.50	&	3.49	&	-132.37	&	-70.18	&	-2.61	&	149.84	\\
BPS CS 31078-018	&	7189878332862720	&	-49125.45	&	1014.35	&	0.56	&	1.36	&	-85.08	&	-105.85	&	-32.35	&	139.60	\\
RAVE J005327.8-025317	&	2529476566134035200	&	-45614.64	&	-1137.30	&	0.27	&	4.90	&	-27.54	&	-348.82	&	94.64	&	362.48	\\
RAVE J024620.1-151842	&	5157037223094770944	&	-13536.07	&	-1168.62	&	0.79	&	11.67	&	-220.75	&	-353.70	&	-133.65	&	437.83	\\
RAVE J031337.3-102056	&	5165907189554147584	&	-34291.81	&	-588.62	&	0.84	&	4.22	&	-184.58	&	-289.08	&	-73.15	&	350.70	\\
RAVE J034346.3-092417	&	5164146429057330816	&	-24901.73	&	620.48	&	0.86	&	6.60	&	-209.32	&	-174.77	&	-88.47	&	286.68	\\
RAVE J141015.9-034355	&	3646021450089606912	&	-55279.86	&	714.40	&	0.66	&	1.21	&	129.34	&	-118.33	&	-22.92	&	176.79	\\
2MASS J15211026-0607566	&	6322811163932641152	&	-50957.35	&	-749.49	&	0.51	&	5.32	&	104.35	&	-333.92	&	-147.20	&	379.56	\\
RAVE J153830.9-180424	&	6255142030043852928	&	-62652.43	&	-492.78	&	0.67	&	3.16	&	60.31	&	-287.50	&	95.04	&	308.75	\\
2MASS J22562536-0719562	&	2610720889058877696	&	-56216.61	&	499.86	&	0.71	&	3.33	&	108.77	&	-176.64	&	40.42	&	211.34	\\
HD 20	&	2333756864959936000	&	-41220.77	&	13.56	&	1.00	&	0.98	&	-223.83	&	-217.10	&	7.42	&	311.91	\\
SMSS J200322.54-114203.3	&	4190620966764303488	&	-21463.31	&	-1190.35	&	0.55	&	14.66	&	96.89	&	-429.77	&	-239.94	&	501.66	\\
HD 222925	&	6487799171512458624	&	-33367.02	&	-407.75	&	0.90	&	4.04	&	-244.11	&	-278.05	&	67.92	&	376.18	\\
2MASS J22132050-5137385	&	6511387028823212544	&	-66232.94	&	237.59	&	0.81	&	3.72	&	33.88	&	-183.35	&	-81.28	&	203.40	\\
BD+29 2356	&	1464266230811360256	&	-29876.24	&	-463.07	&	0.77	&	13.95	&	156.22	&	-281.16	&	-185.88	&	371.49	\\
BD+30 2611	&	1275876252107941888	&	-13487.36	&	1202.92	&	0.53	&	20.61	&	10.70	&	-60.23	&	-286.32	&	292.78	\\
CD-68 1881	&	5810465477272078208	&	-11503.92	&	820.58	&	0.60	&	24.24	&	-47.11	&	-103.35	&	-331.00	&	349.94	\\
HE 0050-0918	&	2474070010546861056	&	-10250.79	&	782.07	&	0.86	&	18.24	&	262.40	&	-139.50	&	171.07	&	342.90	\\
HE 0324-0122	&	3267948604442696448	&	-15679.57	&	1221.28	&	0.42	&	18.46	&	18.37	&	-80.54	&	-268.13	&	280.56	\\
CD-37 2066	&	4820462183959704448	&	-24397.31	&	-43.73	&	0.43	&	16.53	&	84.81	&	-218.66	&	267.92	&	356.07	\\
CD-22 7696	&	5660878702718068864	&	-47950.66	&	-66.13	&	0.54	&	9.17	&	-9.02	&	-230.15	&	-168.23	&	285.22	\\
HE 0007-1752	&	2366585258508216320	&	-38011.53	&	479.73	&	0.06	&	8.49	&	12.97	&	-160.02	&	-228.15	&	278.98	\\
TYC 3431-689-1	&	1019002298308911360	&	-41165.93	&	-777.26	&	0.32	&	8.50	&	-46.88	&	-298.09	&	-141.51	&	333.29	\\
TYC 1191-918-1	&	2795305423661094016	&	-51899.61	&	490.40	&	0.80	&	1.63	&	130.14	&	-175.70	&	44.28	&	223.08	\\
LAMOST J112456.61+453531.3	&	785388513000524160	&	-32286.56	&	-103.48	&	0.46	&	14.05	&	34.07	&	-232.75	&	121.53	&	264.77	\\
BPS CS 29529-0089	&	4678520146255821952	&	-50862.60	&	971.76	&	0.46	&	3.32	&	-100.13	&	-130.26	&	33.91	&	167.77	\\
HD 10837	&	2467368555894065664	&	--	&	--	&	0.08	&	1.14	&	--	&	--	&	--	&	--	\\
2MASS J14325334-4125494	&	6103374133661596416	&	2587.65	&	858.14	&	0.92	&	15.71	&	-357.03	&	-131.38	&	-107.48	&	395.33	\\
2MASS J20050670-3057445	&	6749102714060650880	&	-37429.04	&	-1183.14	&	0.60	&	3.54	&	-216.16	&	-418.40	&	6.63	&	470.98	\\
2MASS J15213995-3538094	&	6199959495651403776	&	-9042.92	&	1944.83	&	0.64	&	13.98	&	73.47	&	81.37	&	188.14	&	217.75	\\
2MASS J05383296-5904280	&	4759914999195512064	&	-51333.23	&	750.07	&	0.54	&	5.12	&	-75.78	&	-135.87	&	-122.34	&	197.91	\\
RAVE J094921.8-161722	&	5684624825278191744	&	-38665.58	&	-1261.40	&	0.52	&	3.90	&	-74.35	&	-384.59	&	81.56	&	400.11	\\
LAMOST J020632.21+494127.9	&	358364512036089472	&	-31644.95	&	2183.74	&	0.04	&	0.66	&	31.32	&	-11.04	&	9.54	&	34.55	\\
RAVE J203843.2-002333	&	4226556850750887168	&	-14974.23	&	-1432.30	&	0.65	&	15.67	&	-26.12	&	-535.77	&	-136.82	&	553.58	\\
BD+00 2058A	&	3085891537839267328	&	-16298.49	&	1022.68	&	0.61	&	18.40	&	137.48	&	-92.50	&	-252.47	&	301.99	\\
BD+02 2541	&	3700341138433848832	&	-53019.50	&	652.76	&	0.69	&	3.19	&	109.08	&	-137.87	&	92.34	&	198.58	\\
BD+06 2932	&	1159108770069883136	&	-25629.13	&	55.17	&	0.99	&	1.08	&	-284.14	&	-213.12	&	30.65	&	356.51	\\
BD+11 2369	&	3916257626962267136	&	-61025.68	&	372.56	&	0.81	&	1.38	&	-73.16	&	-174.05	&	-54.56	&	196.52	\\
BD+11 4725	&	2728314787225296000	&	-66205.54	&	-108.68	&	0.94	&	0.09	&	-21.67	&	-233.51	&	6.35	&	234.60	\\
BD+18 3423	&	4550762289589343488	&	-63193.17	&	-397.49	&	0.79	&	0.70	&	-60.80	&	-269.66	&	-36.43	&	278.82	\\
BD-01 306	&	2494759795723713408	&	-44944.79	&	262.14	&	0.91	&	2.23	&	-192.64	&	-187.18	&	64.48	&	276.23	\\
BD-03 56	&	2540830947835865728	&	-64392.56	&	332.72	&	0.82	&	0.85	&	-19.40	&	-178.42	&	-41.84	&	184.29	\\
BD-06 855	&	3202470247468181632	&	-5168.78	&	385.20	&	0.96	&	2.74	&	-342.79	&	-173.21	&	39.35	&	386.08	\\
BD-08 4501	&	4165370682239910144	&	-22706.04	&	1334.41	&	0.59	&	11.31	&	158.64	&	-51.65	&	-188.57	&	251.78	\\
CD-33 3337	&	5579891291853909504	&	-40544.93	&	1441.36	&	0.14	&	4.29	&	27.93	&	-40.68	&	-135.29	&	144.01	\\
CD-43 6810	&	5388634989413677440	&	-56749.96	&	413.12	&	0.80	&	2.55	&	102.43	&	-166.68	&	-81.28	&	211.85	\\
CD-57 1633	&	5486881507314450816	&	-21366.36	&	-141.27	&	0.98	&	1.23	&	-297.01	&	-240.90	&	-31.99	&	383.76	\\
G 24-25	&	4228176122142169600	&	-41417.29	&	-628.68	&	0.80	&	2.63	&	-200.72	&	-295.96	&	-74.54	&	365.29	\\
G 63-26	&	3939515154043771392	&	-52991.14	&	-1027.78	&	0.50	&	1.25	&	92.43	&	-349.42	&	44.55	&	364.17	\\
G 75-31	&	2502689198705422848	&	-30373.15	&	185.65	&	0.96	&	6.35	&	-243.51	&	-196.41	&	101.80	&	328.99	\\
HD 103723	&	3494135429225414016	&	-62409.14	&	179.90	&	0.91	&	1.33	&	-69.06	&	-198.36	&	55.33	&	217.20	\\
HD 105004	&	3486853672952159872	&	-62901.48	&	-65.13	&	0.96	&	2.36	&	-34.73	&	-228.64	&	-78.53	&	244.23	\\
HD 106516    	&	3581001280225616128 	&	-50000.50	&	1289.25	&	0.33	&	1.23	&	64.50	&	-58.69	&	-53.79	&	102.46	\\
HD 111980	&	3510294882898890880	&	-24559.02	&	264.70	&	0.95	&	6.22	&	270.68	&	-185.27	&	-100.62	&	343.10	\\
HD 113679	&	6141991322784963072	&	-61185.97	&	-451.77	&	0.78	&	0.09	&	-88.77	&	-278.15	&	2.34	&	291.98	\\
HD 120559	&	5871764831114179968	&	-51143.20	&	1350.19	&	0.26	&	0.75	&	-24.33	&	-50.89	&	-40.79	&	69.61	\\
HD 121004	&	6095425184286680064	&	-58167.55	&	-252.71	&	0.87	&	4.15	&	65.33	&	-251.43	&	109.49	&	281.91	\\
HD 132475	&	6232043867720079616	&	-54174.56	&	-1093.87	&	0.41	&	1.42	&	53.11	&	-357.86	&	61.13	&	366.90	\\
HD 148816	&	4433711099194059904	&	-58095.90	&	-325.10	&	0.84	&	2.10	&	102.11	&	-261.00	&	-72.16	&	289.40	\\
HD 159482	&	4485972467412021376	&	-34499.59	&	1343.05	&	0.56	&	4.10	&	-159.12	&	-50.74	&	101.11	&	195.24	\\
HD 163810    	&	4150413750729448320	&	-31961.85	&	108.50	&	--	&	1.57	&	262.28	&	-206.95	&	35.68	&	336.00	\\
HD 177095	&	4082356695359119616	&	-53342.90	&	574.94	&	0.76	&	0.43	&	146.64	&	-147.94	&	-22.91	&	209.56	\\
HD 179626	&	4263841431819084032	&	-60101.04	&	-72.80	&	0.97	&	1.12	&	106.64	&	-229.98	&	49.14	&	258.22	\\
HD 189558	&	4189078626829509248	&	-56022.17	&	896.66	&	0.55	&	1.02	&	81.89	&	-107.53	&	47.85	&	143.38	\\
HD 193901	&	6859076107589173120	&	-52306.18	&	-252.08	&	0.89	&	2.69	&	-145.22	&	-251.41	&	-80.50	&	301.29	\\
HD 194598	&	1752459532807704704	&	-63394.91	&	-332.14	&	0.83	&	0.32	&	-65.64	&	-261.38	&	-20.09	&	270.24	\\
HD 199289	&	6481426917515214464	&	-52259.27	&	1329.26	&	0.28	&	0.24	&	-26.53	&	-53.11	&	-16.33	&	61.57	\\
HD 205650	&	6810862213470371072	&	-49936.34	&	1097.12	&	0.53	&	0.16	&	-119.23	&	-81.97	&	9.09	&	144.97	\\
HD 219617    	&	4263841431819084032	&	-14959.22	&	-242.84	&		&	2.65	&	317.85	&	-251.43	&	-41.18	&	407.36	\\
HD 22879	&	3250489115709122560	&	-50675.79	&	1134.42	&	0.48	&	0.69	&	-99.82	&	-78.61	&	-35.08	&	131.81	\\
HD 284248	&	52624073910834176	&	-1544.74	&	510.00	&	0.94	&	9.14	&	-345.87	&	-156.49	&	-68.04	&	385.68	\\
HD 298986	&	5358179685572496384	&	-17163.32	&	645.81	&	0.88	&	13.97	&	254.49	&	-134.90	&	165.23	&	332.06	\\
HD 3567	&	2427069874188580480	&	-51824.90	&	-234.07	&	0.91	&	1.28	&	162.04	&	-249.95	&	-46.62	&	301.50	\\
HD 51754	&	3112836479029139200	&	-36030.78	&	570.20	&	0.85	&	0.33	&	234.76	&	-148.06	&	13.50	&	277.88	\\
HD 59392	&	5586241315104190848	&	-51050.07	&	-796.98	&	0.68	&	0.45	&	143.58	&	-316.46	&	-23.13	&	348.27	\\
HD 76932	&	5730371484020807808	&	--	&	--	&	--	&	--	&	--	&	--	&	--	&	--	\\
Wolf 615	&	4412702764879270656	&	-33605.57	&	-268.94	&	0.94	&	2.46	&	253.16	&	-255.90	&	55.31	&	364.19	\\
CD-61 282	&	4715919175280799616	&	-23789.52	&	-480.36	&	0.91	&	0.17	&	287.56	&	-278.31	&	-0.45	&	400.19	\\
G 87-13	&	940306024863622400	&	-44257.05	&	-399.84	&	0.87	&	0.42	&	-196.37	&	-268.68	&	18.69	&	333.31	\\
G 98-53	&	3440045641893430912	&	-50096.18	&	-191.95	&	0.92	&	3.26	&	-149.68	&	-243.45	&	-85.92	&	298.42	\\
G 161-73	&	3822808140853060224	&	-38691.70	&	-243.46	&	0.93	&	1.73	&	227.81	&	-245.56	&	46.54	&	338.18	\\
G 53-41	&	3855475490307708032	&	-59975.66	&	-227.69	&	0.87	&	3.95	&	-25.68	&	-248.66	&	-103.69	&	270.63	\\
G 188-22	&	1800518563283501440	&	-42409.93	&	1003.06	&	0.65	&	2.05	&	170.71	&	-96.90	&	65.99	&	207.09	\\
HD 25704	&	4682867168555426944	&	-45565.05	&	1333.02	&	0.45	&	0.05	&	-118.40	&	-53.99	&	-3.06	&	130.16	\\	
\end{longtable}
\end{landscape}
}

\clearpage
\pagebreak

\twocolumn

\section{Statistical Analysis} \label{section_statistical_analysis}
To corroborate the robustness of the bimodality and quantify the populations, we fitted the 
$ \rm \log \epsilon (Th/Eu)$ distributions (both LTE and NLTE) with Gaussian Mixture Models (GMM), followed by a 
Bayesian posterior probability distribution analysis. 

First, we performed a GMM analysis for K = 1, 2, and 3 component models and the Bayesian Information Criterion (BIC) 
was evaluated to determine the optimal number of populations. In NLTE, the two-component model is decisively preferred over 
both the one- and three-component models. The two-component model minimizes the $\text{BIC}$ value with 
a $\Delta \text{BIC} = \text{BIC}_1 - \text{BIC}_2 = 22.38$, corresponding to 
a Bayes Factor $\ln(B_{21}) = 11.20$. A $\ln(B_{21}) > 5$ ($\Delta \text{BIC} > 10$) provides decisive 
statistical evidence in favor of two distinct populations over a single continuous population \citep{Kass_1995}. 
The LTE distribution is strongly favored by a unimodal distribution with a $\Delta \text{BIC} = -14.45$ 
and Bayes factor $\ln(B_{12}) = 7.23$ {\footnote{Bayes factor B$_{12}$ indicates evidence for hypothesis 1 against 2}}. 
The 2-component GMM identified two well-separated Gaussian distributions in NLTE with mean ($\mu$) and 
standard deviations ($\sigma$): $-$0.54$\pm$0.11 (standard peak; $\mu_{1}\pm\sigma_{1}$) comprising 89.1\% of the sample 
and $-$0.20$\pm$0.02 (actinide-boost peak; $\mu_{2}\pm\sigma_{2}$) comprising 10.9\% of the sample. 
The LTE distribution yielded a single Gaussian with $\mu\pm\sigma$ = $-$0.43$\pm$0.18. 

\begin{figure}[ht!]
\centering
\includegraphics[scale=0.45]{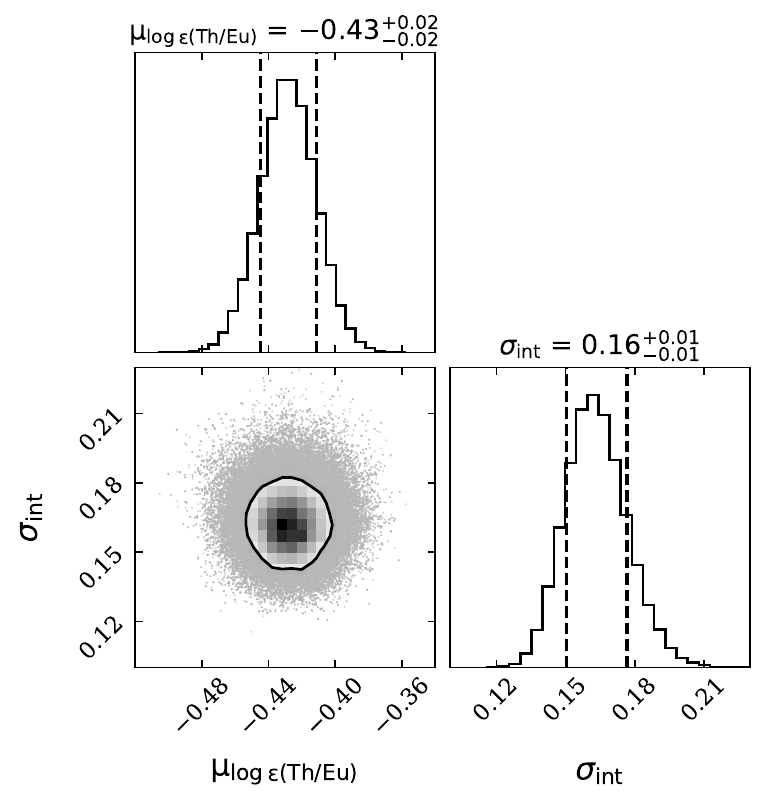}
\includegraphics[scale=0.45]{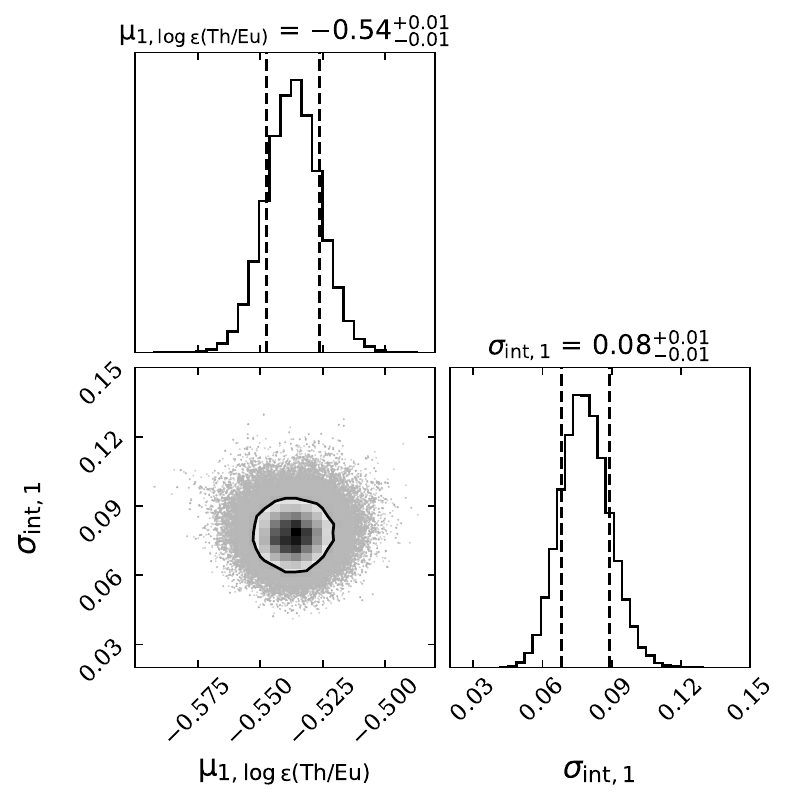}
\includegraphics[scale=0.45]{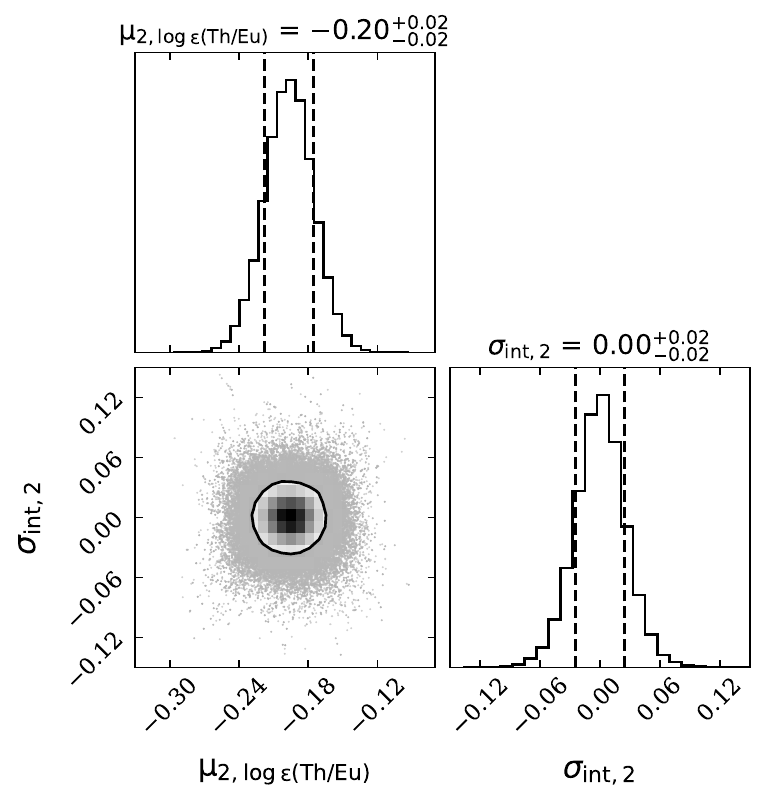}
\caption{Posterior probability distributions of $\mu$ and $\sigma_{\rm int}$ of $ \rm \log \epsilon (Th/Eu)$ distribution based on Bayesian analysis for LTE abundances, standard, and actinide-boost peaks (from top to bottom).}\label{figure_posteriors}
\end{figure}

The standard deviations obtained from the GMM are the observed widths of the $ \rm \log \epsilon (Th/Eu)$ distributions. To deconvolve 
the observational uncertainties ($\Delta$$\sim$0.07 dex) from this dispersion and to obtain the intrinsic (cosmic) scatter 
of the $ \rm \log \epsilon (Th/Eu)$ distribution, we performed a Bayesian Markov Chain Monte Carlo (MCMC) analysis using the 
parameters obtained from GMM as priors. The resulting posterior probability distributions for the 
intrinsic parameters, $\mu$ and $\sigma_{\rm int}$, are shown in Fig.~\ref{figure_posteriors}. 
In NLTE, intrinsic parameters, along with the 1$\sigma$ confidence level, obtained from the posterior distribution 
are $\mu_{1}$ = $-$0.54$\pm$0.01 with $\sigma_{\rm int, 1}$ = 0.08$\pm$0.01 for the standard peak and 
$\mu_{2}$ = $-$0.20$\pm$0.02 with $\sigma_{\rm int, 2}$ = 0.00$\pm$0.02 for the actinide-boost peak. 
The probability density in the observational gap $-$0.35 $<$ $ \rm \log \epsilon (Th/Eu)_{NLTE}$ $<$ $-$0.25 is zero at 
94.6\% (1.9$\sigma$) confidence level, as inferred from a posterior predictive distribution. 
The LTE distribution is represented by the intrinsic parameters 
$\mu$ = $-$0.43$\pm$0.02 and $\sigma_{\rm int}$ = 0.16$\pm$0.01. In all cases, the $\mu$ values are identical to 
those obtained from GMM despite the slightly lower values (by $\sim$ 0.02 dex) obtained for $\sigma_{\rm int}$.

\FloatBarrier
\clearpage

\end{appendix}
\end{document}